\documentclass[useAMS,usenatbib,onecolumn]{mn2e}
\usepackage{graphicx}
\usepackage{bm}
\usepackage{epsfig}
\usepackage{amssymb, ulem}
\usepackage{amsmath}
\usepackage{empheq}
\usepackage{amsfonts}
\usepackage{cleveref}
\usepackage{courier}
\usepackage{bm}
\usepackage{booktabs}
\usepackage{tabularx}
\usepackage[makeroom]{cancel}

\newcommand{\bea}{\begin{eqnarray}}
\newcommand{\eea}{\end{eqnarray}}

\newcommand{\beq}{\begin{equation}}
\newcommand{\eeq}{\end{equation}}

\newcommand{\simless}[0]{\mathbin{\lower 3pt\hbox
   {$\rlap{\raise 5pt\hbox{$\char'074$}}\mathchar"7218$}}}
\newcommand{\simgreat}[0]{\mathbin{\lower 3pt\hbox
   {$\rlap{\raise 5pt\hbox{$\char'076$}}\mathchar"7218$}}}

\newcommand{\figref}[1]{figure \ref{#1}}
\newcommand{\figrefs}[1]{figures \ref{#1}}
\newcommand{\figrefbare}[1]{\ref{#1}}
\newcommand{\capfigref}[1]{Figure \ref{#1}}
\newcommand{\capfigrefs}[1]{Figures \ref{#1}}
\newcommand{\eqnref}[1]{eq. (\ref{#1})} 
\newcommand{\eqnrefs}[1]{eqs. (\ref{#1})} 
\newcommand{\eqnrefbare}[1]{(\ref{#1})} 
\newcommand{\capeqnref}[1]{Equation (\ref{#1})}

\title[The Einstein-Boltzmann Equations Revisited]{The Einstein-Boltzmann Equations Revisited}
\author[Sharvari Nadkarni-Ghosh and Alexandre Refregier]{Sharvari Nadkarni-Ghosh$^{1}$\thanks{E-mail:
sharvari@iitk.ac.in} and Alexandre Refregier$^{2}$\thanks{E-mail:alexandre.refregier@phys.ethz.ch} \\
$^{1}$Department of Physics, I.I.T. Kanpur, Kanpur, U.P. 208016 India \\
$^{2}$Institute for Astronomy, Department of Physics, ETH Z\"{u}rich, Wolfgang-Pauli-Strasse 27, CH-8093 Z\"{u}rich, Switzerland  
}
\begin{document}
\date{}
\pagerange{\pageref{firstpage}--\pageref{lastpage}} \pubyear{}

\maketitle
\label{firstpage}

\begin{abstract}
The linear Einstein-Boltzmann equations describe the evolution of perturbations in the universe and its numerical solutions play a central role in cosmology. We revisit this system of differential equations and present a detailed investigation of its mathematical properties. For this purpose, we focus on a simplified set of equations aimed at describing the broad features of the matter power spectrum. We first perform an eigenvalue analysis and study the onset of oscillations in the system signalled by the transition from real to complex eigenvalues. We then provide a stability criterion of different numerical schemes for this linear system and estimate the associated step-size. We elucidate the stiffness property of the Einstein-Boltzmann system and show how it can be characterised in terms of the eigenvalues. While the parameters of the system are time dependent making it non-autonomous, we define an adiabatic regime where the parameters vary slowly enough for the system to be quasi-autonomous. We summarise the different regimes of the system for these different criteria 
as function of wave number $k$ and scale factor $a$. We also provide a compendium of analytic solutions for all perturbation variables in 6 limits on the $k$-$a$ plane and express them explicitly in terms of initial conditions. These results are aimed to help the further development and testing of numerical cosmological Boltzmann solvers.
\end{abstract}

\section{Introduction}
\label{sec:intro}
In the past few decades, observations of the cosmic microwave background (CMB) and of the large scale structure (LSS) have provided a wealth of information about the origin and evolution of our Universe (e.g., \citealt{planck_collaboration_planck_2016}; \citealt*{nicola_integrated_2016,nicola_integrated_2016part2};  \citealt{alam_clustering_2016}). These measurements suggest a standard model of cosmology: the universe consists primarily of dark matter and dark energy in addition to small amounts of baryons and radiation (photons and neutrinos) which evolve in a spatially flat background. The temperature anisotropies and the galaxy distribution are seeded by primordial fluctuations in the radiation and matter sectors respectively; these fluctuations were set up during the inflationary era and have a nearly scale invariant power spectrum. The parameters of this standard model of cosmology have been measured with percent level accuracy and current and future missions such as the Dark Energy Survey (DES\footnote{http://www.darkenergysurvey.org.}),   the Dark Energy Spectroscopic Instrument (DESI\footnote{http://desi.lbl.gov}), the 
Large Synoptic Survey Telescope (LSST\footnote{http://www.lsst.org.}), Euclid\footnote{http://sci.esa.int/euclid/.} and the Wide Field Infrared
Survey Telescope (WFIRST\footnote{http://wfirst.gsfc.nasa.gov.}) aim to push this limit even further. 

The increased precision in these measurements needs to be matched with precision in theoretical predictions for the observables. In particular, the dynamics of cosmological perturbations are governed by the coupled Boltzmann equations for radiative species, the fluid equations for the matter species and Einstein equations for the metric (see for e.g., \citealt{kodama_cosmological_1984,  sugiyama_gauge_1989, ma_cosmological_1995}). 
For CMB analyses, the relevant statistic is the angular power spectrum $C_l$ and linear perturbation theory is generally accurate enough. In the case of LSS data, it is usually necessary to compute the non-linear power spectrum. This is generally done by N-body codes or higher order perturbation schemes, which take as input linearly evolved matter variables. Thus, precision evolution of the linear Einstein-Boltzmann (E-B) system is required for both CMB as well as LSS data analyses.  

Numerical codes to solve this system have been developed since the nineties, starting from the pioneering work by \citet{ma_cosmological_1995} and the accompanying COSMICS package \citep{bertschinger_cosmics:_1995}. This was followed by CMBFAST which incorporated a novel method based line of sight integration \cite{seljak_line--sight_1996} thereby reducing the computation time by two orders of magnitude over traditional codes. Over the next three to four years, several effects were incorporated: CMB lensing \citep{seljak_gravitational_1996, zaldarriaga_gravitational_1998}, improved treatment of polarization \citep{seljak_measuring_1997} and extensions to closed geometries \citep{zaldarriaga_cmbfast_2000}. \citet{lewis_efficient_2000} then developed CAMB, a parallelized code based on CMBFAST. CMBEASY, a translation of CMBFAST in C++ was developed by \cite{doran_cmbeasy:_2005,doran_speeding_2005} to include gauge invariant perturbations and quintessence support, and Lesgourgues 
and collaborators have recently developed a new general code called CLASS \citep{lesgourgues_cosmic_2011-1}. Other authors have developed independent codes for example, Hu and co-workers \citep{hu_effect_1995,white_why_1996,hu_cmb_1997,hu_complete_1998} developed codes for general geometries, Sugiyama and collaborators \citep{sugiyama_perturbations_1992, sugiyama_cosmic_1995} developed a code using gauge invariant variables or more recently \cite{cyr-racine_photons_2011} improved on the tight-coupling approximation, but these were not available as documented packages (see \citet*{seljak_comparison_2003} for a comparative study of some earlier codes). Currently, CAMB and CLASS are the only two publicly available codes that are being maintained.  

Evolving the E-B system is a challenging task for several reasons. First, the equations are complicated because of the effect of various different physical processes with multiple time scales making it a stiff system. Certain variables can thus be highly oscillatory while others are very smooth in the same regime. Moreover, the system is a non-autonomous dynamical system, i.e., the parameters of the system are time dependent. Such systems are significantly more complicated to analyse than autonomous systems, as the information given by the eigenvalues of the jacobian can be incomplete
or even misleading. Also, the system generally has many perturbation variables due to the presence of the different physical components (dark matter, baryons, photons and neutrinos) and the multipole expansion for the radiation fields. Thus, although linear, the system is highly complex requiring advanced numerical treatment in the different regimes of evolution. 

Given the importance and complexity of the system, it is worth understanding its mathematical structure. We thus revisit the E-B system from a dynamical systems perspective and perform a detailed investigation of its mathematical properties. For this purpose, we focus on a simplified set of equations aimed at describing the broad features of the matter power spectrum while being analytically tractable. We first perform a detailed eigenvalue analysis of the linear system and study the onset of oscillations as well as the stiffness, numerical stability, and adiabaticity of the system in different regimes. We then provide a  compendium of analytical solutions to the system in six different asymptotic regimes, with the new feature that the analytic solutions are obtained for all perturbation variables and are given explicitly in terms of the initial conditions.  These results are aimed to aid the development of cosmological Boltzmann codes in terms of numerical design and testing. 

The paper is organized as follows. \S \ref{sec:setup} describes the simplified E-B system to be solved and the change of variables that further simplifies the equations. \S \ref{sec:eigenstruc} computes the eigenvalues and studies their structure. \S \ref{sec:adiabatic} gives precise definitions of the adiabaticity of the system based on time derivatives of the parameters and eigenvalues. \S \ref{sec:onset} examines the eigenvalue structure and predicts the onset of oscillations in the system. \S \ref{sec:stability} uses the eigenvalues to analyse the stability of various numerical solvers applied to the E-B system. \S \ref{sec:stiffness} examines the issue of stiffness of the E-B system. Typically in the CMB literature, the stiffness is attributed to the photon-baryon coupling term which is very large at early times (tight coupling regime). We demonstrate that even in the absence of baryons the system is stiff due to the high frequency oscillations of the photon moments at late epochs. We discuss the definition of stiffness and the parameter that can be used to quantify it. \S \ref{sec:limits} gives a summary of analytic solutions in six different regimes defined by various limits of the parameters. \S \ref{sec:summary} provides a discussion and conclusion. 
The paper has seven appendices. Appendix \ref{app:identities} derives various identities used throughout the paper. Sturm's theorem and Descartes' rule of signs, which are used to predict the onset of oscillations in \S \ref{sec:onset} are explained in appendix \ref{app:sturm}. Appendix \ref{app:freq} discusses the frequency of oscillations and explains why these are not visible for super-horizon modes. The general theory of stability of numerical schemes is reviewed in appendix \ref{app:stability}. Appendix \ref{app:limits} gives the details of the analytic solutions summarized in \S \ref{sec:limits}. Appendix \ref{app:withbaryons} shows the structure of the equations when baryons and neutrinos are included and \ref{app:4Dsystem} shows the eigenvalues of the system when the gravitational potential is not treated like a dynamical variable.

Throughout this paper, we consider a {\it flat} $\Lambda$CDM cosmology\footnote{$\Omega_{r,0}$ includes the contribution from the neutrino background although we do not include neutrino perturbations; see \citealt{dodelson}. The precise value of $\Omega_{r,0}$ does not affect this analysis.} with $\Omega_{m,0} = 0.3$, $\Omega_{r,0} = 4.15 \times 10^{-5} h^{-2}$, ${\rm H_0} = 100 h$ ${\rm km/s}$ ${\rm Mpc^{-1}}$ and $h=0.7$ and work in the conformal Newtonian gauge. 
 
\section{The Einstein-Boltzmann equations}
\label{sec:setup}
\subsection{Differential equations and initial conditions}
We are mainly interested the evolution of the dark matter power spectrum and hence it suffices to consider a reduced set of variables. The homogenous energy density of the radiation and matter are denoted by $\rho_r$ and $\rho_m$ respectively. The primary components of the photon distribution that affect the matter variables are the monopole and dipole moments denoted by $\Theta_0$ and $\Theta_1$ respectively. The matter fluctuations are characterised by the overdensity $\delta$ and the irrotational peculiar velocity $v$. We use the conformal
 Newtonian gauge and consider only scalar metric perturbations with no anisotropic stresses; thus the metric perturbations are characterised by only one scalar potential $\Phi$ \footnote{The form of the metric is $ds^2 = -(1+ 2 \Psi) dt^2 + a^2(1+ 2 \Phi) \delta_{ij} dx^i dx^j$. In the absence of anisotropic stresses, $\Phi = -\Psi$}.  For this simplified system, the coupled Boltzmann, fluid and Einstein equations become (e.g., \citealt{dodelson}) 
\begin{subequations}
\label{system}
\begin{align}
\frac{ d \Theta_0}{d\eta} + k \Theta_1 & =-\frac{d \Phi}{d\eta}, \\
\frac{d \Theta_1}{d\eta} -  \frac{k}{3} \Theta_0  &= - \frac{k}{3} \Phi \\
\frac{d \delta}{d\eta} + i k v & = - 3 \frac{d \Phi}{d\eta} \\
\frac{ d v}{d\eta}+ \frac{1}{a} \frac{d a}{d\eta}  &=  i k \Phi\\
k^2 \Phi + 3 \frac{1}{a} \frac{d a}{d\eta} \left(\frac{d \Phi}{d\eta} + \frac{1}{a} \frac{d a}{d\eta} \Phi \right) &= 4 \pi G a^2 \left[\rho_m \delta + 4 \rho_r\Theta_0\right] \label{oldeq5}. 
\end{align}
\end{subequations}
Here the time variable is the conformal time ($d\eta =  dt/a $, where $a$ is the scale factor) and $k$ is the comoving wavenumber.  
There are five variables and correspondingly five initial conditions which, in general, may be specified independently. However for adiabatic initial conditions given by standard single-field inflation the relations are
\begin{subequations}
\begin{align}
\nonumber \Theta_0(k, a_i) &= \frac{1}{2} \Phi(k,a_i)\\
\nonumber \Theta_1(k,a_i) &= -\frac{1}{6} \frac{k}{a_i H_i}  \Phi(k,a_i)\\
\nonumber \delta(k,a_i) &= 3 \Theta_0 = \frac{3}{2} \Phi(k,a_i) \\
\nonumber u(k,a_i) &= 3 \Theta_1 = -\frac{1}{2}\frac{k}{a_iH_i} \Phi(k,a_i),
\end{align}
\end{subequations}
where $a_i$ and $H_i$ are the initial values of the scale factor and Hubble parameter and $\Phi(k,a_i)$ is the initial potential. In order to further simplify the system, we introduce new variables 

\begin{subequations}
\begin{align}
y_1 &= \Theta_0 + \Phi\\ 
y_2 &= 3 \Theta_1\\
y_3 &= \delta + 3\Phi\\
y_4 &= iv\\
y_5 &= \Phi
\end{align}
\end{subequations}
and define  the parameter 
\beq
\epsilon \equiv \epsilon(k,a) = \frac{k}{Ha}.
\label{eq:epsdefn}
\eeq
Changing the time variable from $\eta$ to $ \ln a$, and noting that 
$\frac{d}{d \eta} =  (Ha) \frac{d}{d \ln a}$, the system given by \eqnref{system} can be re-written as 
\begin{subequations}
\label{newsystem}
\begin{align}
{\dot y}_1&=  -\frac{\epsilon(k,a)}{3} y_2 					\label{eq1}\\
{\dot y}_2&=  \epsilon(k,a)\left[y_1 - 2 y_5\right] 	\label{eq2}\\
{\dot y}_3  &= -\epsilon(k,a) y_4						 \label{eq3}\\
{\dot y}_4 &= -y_4 -\epsilon(k,a) y_5					\label{eq4}\\
{\dot y}_5  &= \frac{1}{2} \left[ \Omega_m(a) y_3 + 4 \Omega_r(a) y_1 - \left\{3 \Omega_m(a)  + 4 \Omega_r(a) + \frac{2}{3} \epsilon^2(k,a) +2 \right\} y_5\right],  			\label{eq5}
\end{align}
\end{subequations}
where the `dot' denotes derivative w.r.t. $\ln a$. The initial conditions become
\begin{subequations}
\label{newinit}
\begin{align}
\label{eq1init} y_1(k,a_i) &= \frac{3}{2} y_5(k,a_i)\\
\label{eq2init} y_2(k,a_i) &= -\frac{1}{2} \epsilon(k,a_i) y_5(k,a_i)\\
\label{eq3init} y_3(k,a_i) &= \frac{9}{2} y_5(k,a_i)\\
\label{eq4init} y_4(k,a_i) &= -\frac{1}{2} \epsilon(k,a_i)y_5(k,a_i). 
\end{align}
\end{subequations} 

\subsection{Parameters of the system}
There are three time-dependent dimensionless parameters in this system $\Omega_m$, $\Omega_r$ and $\epsilon$.  The first two, as usual, denote the fraction of radiation and matter density and are independent of $k$: 
\beq
\label{eq:omor}\Omega_m(a) = \frac{\Omega_{m,0} a_0^3 H_0^2}{a^3 H^2} \; \; \; \; \; {\rm and} \;\;\;\;
\Omega_r(a) = \frac{\Omega_{r,0} a_0^4 H_0^2}{a^4 H^2}.
\eeq
where `0' denotes the values of the parameters today i.e. at $a=a_0=1$.

The parameter $\epsilon$ has a dual interpretation. It is the ratio of two time scales: the Hubble time $H^{-1}$ and the time scale of oscillation $a k^{-1}$ of a photon mode of wavelength $k^{-1}$. It is also the ratio of two length scales: the comoving Hubble radius $(aH)^{-1}$ and the wavelength of a perturbation $k^{-1}$. $\epsilon$ is related to the conformal time or comoving horizon $\eta = \int \frac{dt}{a} $ by
\beq 
k \eta = \int \epsilon(k,a) d \ln a.
\eeq
If $H\sim a^{-n}$, then $\epsilon \sim (aH)^{-1} \sim a^{n-1}$ and $ k \eta = (n-1)^{-1}\epsilon$. Thus, in the {\it radiation-dominated} epoch, $n=2$, $\epsilon \sim a$ and the relation is $k \eta_{rad} = \epsilon_{rad}$. In the {\it matter-dominated} epoch, $n=3/2$, $\epsilon \sim a^{1/2}$ and the relation is $k \eta_{mat} = 2 \epsilon_{mat}$. By definition, $k \eta \ll 1$ denotes super-horizon modes whereas $k\eta \gg1$ denotes sub-horizon modes. However, since $k\eta$ and $\epsilon$ differ only by a factor of a few, in this work we will use $\epsilon\ll1$ and $\epsilon \gg 1 $ to denote super- and sub-horizon modes respectively. $\epsilon =1$ denotes the horizon crossing condition. In terms of time scales, for given $k$, $\epsilon\ll1$, implies that the time scale for oscillation is much larger than the age of the universe and $\epsilon\gg 1$ implies, fast oscillations, on time scales much smaller than the age of the universe. 

It is also useful to examine the rate at which the parameters $\epsilon$, $\Omega_m$ and $\Omega_r$ evolve. From the above definitions, their derivatives are (see appendix \ref{app:identities} for details)
\bea 
\label{epsder}\frac{\dot \epsilon}{\epsilon}&=& -\left[1-\frac{1}{2} \left(3 \Omega_m + 4 \Omega_r\right)\right],\\
 \label{omder}\frac{\dot \Omega_m}{ \Omega_m} &=& -\left[3-  \left(3 \Omega_m + 4 \Omega_r\right) \right],\\
\label{order} \frac{\dot \Omega_r }{ \Omega_r} &=& -\left[4 -\left(3 \Omega_m + 4 \Omega_r \right) \right].
\eea
\begin{figure}
\centering
\includegraphics[width=16cm]{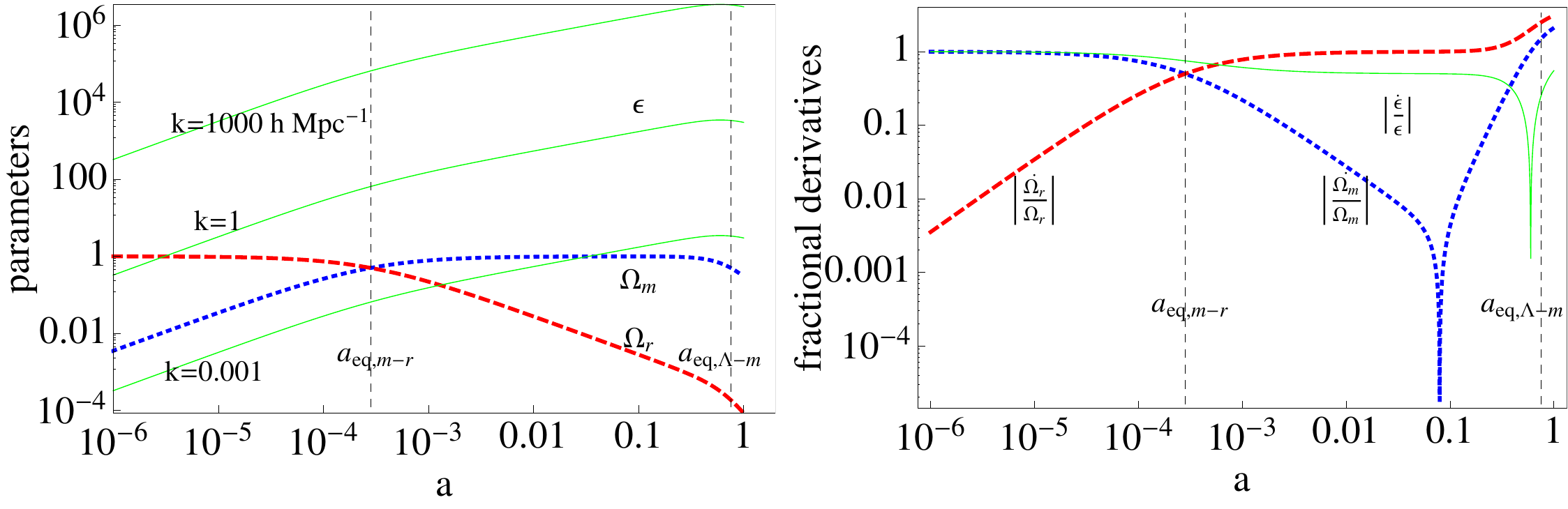}
\caption{Time dependent parameters in the system. The left panel shows $\Omega_m$, $\Omega_r$ and the k-dependent $\epsilon$ parameter. Note that at late times $\epsilon$ dominates the other parameters. The right panel shows the fractional derivatives of these quantities, w.r.t. the time variable i.e., $\ln a$. The fractional derivative of $\epsilon$ is the same for all $k$ values considered.  }
\label{parameters}
\end{figure}
\capfigref{parameters} shows the three parameters as a function of time (left panel) and their fractional derivatives (right panel). 
As can be seen, for a flat cosmology, the density parameters $\Omega_m$ and $\Omega_r$ stay bounded by unity, whereas, $\epsilon$ keeps increasing for any $k$. For large values of $k$ and/or late epochs, $\epsilon \gg  \Omega_m,\Omega_r$. Also, the upper bound on the fractional derivatives for all three parameters is of order unity. We will use these facts when we define the adiabatic condition in \S \ref{sec:adiabatic}.

There are three important epochs for this system: the epoch when a given mode crosses the horizon ($a_{hc}$) given by $\epsilon=1$, the epoch of 
of matter-radiation equality ($a_{eq, m-r}$) given by $\Omega_m = \Omega_r$ and the epoch of dark energy-matter equality ($a_{eq, \Lambda-m}$) given by $\Omega_{\Lambda} = \Omega_m$. These are given by
\bea 
a_{hc}(k) &=&\frac{k}{H},\\
a_{eq, m-r} &=& \frac{\Omega_{r,0}}{\Omega_{m,0}} \simeq2.8 \times 10^{-4}, \\
a_{eq, \Lambda-m} &=& \left(\frac{\Omega_{m,0}}{\Omega_{\Lambda, 0}}\right)^{1/3} \simeq 0.75, 
\eea
where the numerical values correspond to the cosmological parameters given in \S \ref{sec:intro}. 

\subsection{Algebraic equation for the potential}
\capeqnref{oldeq5} and its transformed version \eqnref{eq5} correspond to the time-time component of Einstein's equations in the absence of any anisotropic stress. Combining the time-time component with the time-space component gives an algebraic equation for $\Phi$ (e.g., \citealt{dodelson}): 
\beq
k^2 \Phi = 4 \pi G a^2 \left[\rho_m \delta + 4 \rho_r \Theta_0 + 3 \frac{a H}{k} \left(i\rho_m v + 4 \rho_r \Theta_1\right)\right].
\eeq
Converting to the $y$-variables defined by \cref{newsystem} and using the definitions of $\epsilon,\Omega_m$ and $\Omega_r$ given by 
\eqnrefs{eq:epsdefn} and \eqnrefbare{eq:omor}, we get 
\beq 
y_5 = B^{-1}\left[ 4 \Omega_r\left(y_1 + \frac{1}{\epsilon}y_2\right) + \Omega_m \left(y_3 + \frac{3}{\epsilon} y_4\right)\right],
 \label{eq5II}
 \eeq
 where, 
 \beq 
 B \equiv B(k,a) = 3 \Omega_m + 4 \Omega_r + \frac{2}{3} \epsilon^2. 
 \eeq 
By using \eqnrefs{epsder}, \eqnrefbare{omder} and \eqnrefbare{order} for the time derivatives of the parameters and \cref{eq1,eq2,eq3,eq4} for the time derivatives of the variables, it can be shown that the above form of $y_5$ satisfies \eqnref{eq5}. Thus, it forms a particular solution for $y_5$. The full solution is 
\beq 
y_5(k,a) = C e^{-\int_{a_i}^a \frac{B+2}{2} d\ln a} + B^{-1}\left[ 4 \Omega_r\left(y_1 + \frac{1}{\epsilon}y_2\right) + \Omega_m \left(y_3 + \frac{3}{\epsilon} y_4\right)\right], 
\eeq
where $C$ is set by the initial conditions. For adiabatic initial conditions of the form given by \cref{eq1init,eq2init,eq3init,eq4init} and assuming 
that $\epsilon \ll1$ gives 
\beq 
y_5(k,a_i) \simeq C + y_{5,i} \implies C\simeq0. 
\eeq
Furthermore, for sub-horizon modes, where $\epsilon \gg 1$, the homogenous term decays exponentially (in terms of the time variable $\ln a$). 

Using both the time-time and time-space components of Einstein's equations is redundant. In the original COSMICS code, one of them was used to check integration accuracy (see discussion in \citealt{ma_cosmological_1995}). Alternatively, it is also possible to substitute the algebraic solution for $y_5$ in \eqnrefs{eq1} to \eqnrefbare{eq4} to give a 4D dynamical system. Mathematically, this is possible because for adiabatic initial conditions, $C \sim 0$ and physically this means that for scalar perturbations, the Einstein equation is just a constraint equation that does not introduce new propagating degrees of freedom. In appendix \S \ref{app:4Dsystem} we compute the eigenvalues of this system. Applying the ideas presented in the rest of this paper, it seems possible that the 4D system may be numerically more stable than the 5D system. However, the advantage of using the algebraic equation for error control may still outweigh the advantage one gains by reducing the dimensionality of the system. A detailed analysis is required to comment more concretely on this issue and whether the results extend to the full Boltzmann system remains to be investigated. 

\section{Eigenvalue Structure}
\label{sec:eigenstruc}
\begin{table}
\centering
\begin{tabular}{|c|c|}
\hline
Before transition & After transition\\
\hline
\hline
$\lambda_1<0$ & $\lambda_1<0$\\
\hline
$\lambda_2<0$ & $\lambda_2<0$\\
\hline
$\lambda_3<0$ & $\lambda_3>0$\\
\hline
$\lambda_4<0$ & $Re[\lambda_4] < 0$\\
\hline
$\lambda_5>0$ & $Re[\lambda_5] <0$\\
\hline
\end{tabular}
\caption{Signs of eigenvalues of ${\mathcal A}$. At early times there are four negative roots and one positive root. These transition to two negative, one positive and two complex roots. \capfigref{eigenvalues} shows the magnitude of the roots and \figref{trans} shows the transition epoch.  } 
\label{eigensigns}
\end{table}

\begin{figure}
\centering
\includegraphics[width=18cm]{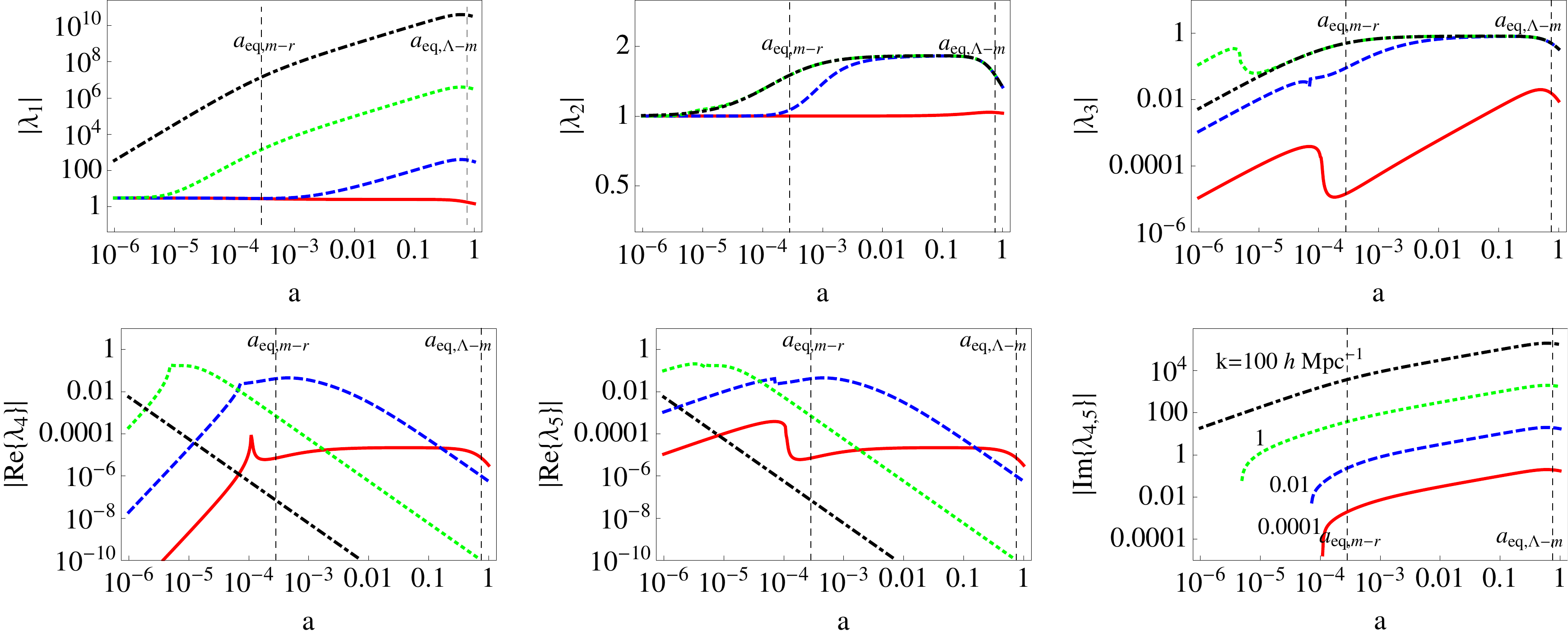}
\caption{Magnitude of eigenvalues of ${\mathcal A}$. The signs are shown in table \ref{eigensigns}. Four values of $k$ are chosen as shown in the last panel of the bottom row. Kinks in the plots occur at the transition points. $\lambda_1$ and $\lambda_2$ are smooth through the transition, but the other three change sign or form. The transition for the largest $k$, occurs before $a \sim 10^{-6}$. }
\label{eigenvalues}
\end{figure}  
\capeqnref{newsystem} can be written in compact form as 
\beq 
{\dot {\bf y}} ={\mathcal A} \cdot {\bf y}, 
\label{mainsystem}
\eeq
where the column vector ${\bf y} = \{y_1, y_2, y_3, y_4, y_5\}$ and the Jacobian matrix ${\mathcal A}$ is 
\beq
{\mathcal A} =\left( \begin{array}{ccccc}
0& -\frac{\epsilon}{3} & 0 &0 & 0\\ 
\epsilon &0 &0 & 0 &-2\epsilon \\
0& 0 & 0& -\epsilon  &0\\
0& 0 &  0 & -1  &  - \epsilon\\
 2 \Omega_r & 0 &   \frac{\Omega_m}{2} & 0  & - \frac{B+2}{2} \\
\end{array} \right)
\eeq
The matrix ${\mathcal A}$ has five eigenvalues which depend on $k$ and on time. Although ${\mathcal A}$ is a sparse matrix, it has no special symmetries. Hence calculating its eigenvalues analytically is rather cumbersome. Instead we compute them numerically. We find that for each value of $k$, all eigenvalues are real at a sufficiently early time: four of them are negative and one is positive. Eventually there is a transition after which three are real, of which two are negative and one is positive, and two are complex with negative real parts. The positive eigenvalue denotes a growing mode and a negative eigenvalue denotes a decaying mode. The eigenvalue structure is summarized in table \ref{eigensigns} and the temporal evolution of the magnitudes for four values of $k = 0.001,0.01,1$ and $100$ $h$ ${\rm Mpc}^{-1}$ is plotted in \figref{eigenvalues}. Note from the table that $\lambda_3$ stays real, but changes sign after the transition. $\lambda_4$ and $\lambda_5$ become complex; the real part of $\lambda_5$ changes sign. Thus, one eigenvalue is always positive throughout the evolution. This is expected since gravitational instability is inbuilt in the E-B system. In \figref{eigenvalues}, the kinks in the plots corresponding to $\lambda_3$ and real parts of $\lambda_4$ and $\lambda_5$ mark the epoch of transition. It is clear that the transition epoch is different for each value of $k$.

\section{The adiabatic conditions}
\label{sec:adiabatic}
The E-B system is non-autonomous because the matrix ${\mathcal A}$ is time-dependent. Non-autonomous systems are significantly more complicated because there the usual tools used to analyze autonomous systems cannot be applied. For example\footnote{Consider the linear system ${\dot x}_1 = -x_1 + x_2 e^{\gamma t}$ and ${\dot x}_2 = -x_2$. This has eigenvalues $\{-1,-1\}$ suggesting that both modes are stable, but directly solving the system shows that $x_2 \sim e^{-t}$ (stable) but $x_1 \sim e^(\gamma-1)t/t$, which is an unstable solution if ${\mathcal Re}(\gamma) >1$ and oscillatory if $\gamma$ is complex. This example has been adapted from the book by \citet{slotine}}, analyzing the eigenvalues for such systems to understand the stability can be mis-leading. However, it is always true that, for a linear \footnote{For a non-linear system, even for the autonomous case, eigenvalues of the linear system are useful only when ${\mathcal Re}(\lambda)\neq 0$ \citep{strogatz}} system, the presence of complex eigenvalues signals an oscillatory behaviour. In \S \ref{sec:onset} we will obtain an analytic prediction for this transition epoch. Another application of 
the eigenvalue analysis is to predict the stability of numerical schemes. In \S \ref{sec:stability} we investigate the stability of some popular numerical schemes applied to linear autonomous and non-autonomous systems. Dealing with the latter is significantly more involved. This  motivates the need to define the adiabatic regime, where the system's parameters vary slowly enough so that the system becomes `quasi-autonmous' and one can apply the results from the autonomous case. 

The E-B system as defined through \eqnref{mainsystem} consists of five dependent variables ($y_i$), the independent temporal variable ($\ln a$) and the Jacobian matrix ${\mathcal A}$ which is a function of three time-dependent parameters ($\epsilon, \Omega_m$ and $\Omega_r$). 
The system can be considered quasi-autonomous if the matrix ${\mathcal A}$ varies slowly as compared to the variables or alternately, the fractional change in the matrix in time $dt$ is small compared to the change in the dependent variables. The matrix ${\mathcal A}$ can be characterised either by the three parameters or by its five eigenvalues. 
The variation of the variables defines five time scales $y_i/{\dot y_i}$. If ${\mathcal A}$ is characterised by its parameters, the change in ${\mathcal A}$ gives three time scales $\epsilon/{\dot \epsilon}, \Omega_m/{\dot \Omega_m}$ and $\Omega_r/{\dot \Omega_r}$ whereas if it is characterised by its eigenvalues change in ${\mathcal A}$ gives five time scales $\lambda_i/{\dot \lambda_i}$. As we shall see below, the two descriptions give different adiabatic conditions.

\begin{enumerate}
\item {\it Adiabatic condition based on parameters}: We demand that for the system to be `quasi-autonomous', the fractional change in parameters in time $dt$ is small compared to the fractional change in the dependent variables. Refer to \eqnref{newsystem}. Assuming that all the dependent variables $y_i$ have the same order of magnitude, it is clear that when $\epsilon\ll1$, the fractional change ${\dot y}_i/y_i$ is of order unity (note that $\Omega_m, \Omega_r$ are always bounded by unity) and when $\epsilon \gg 1$, ${\dot y}/{y} \sim \epsilon$. Thus, to a good approximation ${\dot y}/y \sim \max\{1, \epsilon\}$. The adiabatic condition can thus be expressed as  
\beq 
\max\{1, \epsilon\}  \gg  \max\left\{ \left|\frac{\dot \epsilon}{\epsilon}\right|,  \left|\frac{\dot \Omega_m}{\Omega_m}\right|,  \left|\frac{\dot \Omega_r}{\Omega_r}\right|\right\}.
\label{adcond}
\eeq
The $\max$ on the r.h.s. is necessary to guarantee that {\it all} the parameters vary slower than the variables. 
We can thus define a first `adiabatic parameter' $p_1$ as the ratio 
\beq 
p_1(k,a) = \frac{\max\left\{ \left|\frac{\dot \epsilon}{\epsilon}\right|,  \left|\frac{\dot \Omega_m}{\Omega_m}\right|,  \left|\frac{\dot \Omega_r}{\Omega_r}\right|\right\}}{\max\{1, \epsilon\} }.
\label{adparam1}
\eeq
As a threshold value, we demand that the function varies at least ten times faster than the variation in the parameters i.e. $p =0.1$.

An alternate analytic expression for $p_1$ can be derived as follows. Referring back to \figref{parameters}, we see that the fractional change in the parameters for all epochs is roughly of order unity. Thus, when $\epsilon \ll1$, which happens at early times and/or for very small $k$, \eqnref{adcond} is never satisfied and the adiabatic condition cannot be implemented. When $\epsilon \gtrsim 1$, the denominator of \eqnref{adparam1} can be replaced by $\epsilon$. To estimate the numerator, note that,  until about matter-radiation equality, the fractional derivative of $\epsilon$ (which is equal to that of $\Omega_m$) dominates the $\Omega_r$ derivative. After the equality, $\Omega_r$ derivative starts to dominate. However, in this regime, the value of $\Omega_r$ is diminishing. Similarly, far into the matter dominated era, the matter derivative becomes larger, but in this regime $\epsilon$ is very large for most $k$ values of interest. Thus, in the regime where the $\Omega_m$ and $\Omega_r$ derivatives are dominant, the parameters themselves are sub-dominant and one can replace the numerator of \eqnref{adparam1} with ${\dot \epsilon}/{\epsilon}$. Using \eqnref{epsder} to substitute for ${\dot \epsilon}/\epsilon$ gives
\beq
p_{1,{\rm approx}}(k,a) \approx \frac{1}{\epsilon(k,a)}\left| 1-\frac{1}{2} \left(3 \Omega_m(a) + 4 \Omega_r(a)\right)\right|  \; \; \; \ {\rm when } \; \; \epsilon \gtrsim 1
\label{adcond2}
\eeq

\capfigref{adiabatic1} (left panel) shows the numerically evaluated parameter $p_1$ for five values of $k$ using \eqnref{adparam1}. The dashed line corresponds to $p_1 = 0.1$ For each $k$, the point where the curve intersects the dotted line denotes the epoch after which the adiabatic approximation is valid. For small wavenumbers ($\sim 10^{-4}$), the system is never adiabatic until the present epoch; as $k$ increases the range of epochs where the adiabatic approximation is valid increases. The right panel of the same figure shows the contour $p_{1,{\rm approx}}(k,a) =0.1$ (red dotted line) on the $k-a$ plane. The four coloured dots in both panels correspond to the points $p_1(k,a) =0.1$ numerically evaluated using \eqnref{adparam1}. It is clear that \eqnref{adcond2} forms a good approximation for $p(k,a)$. 
\begin{figure}
\centering 
 \includegraphics[width=8cm]{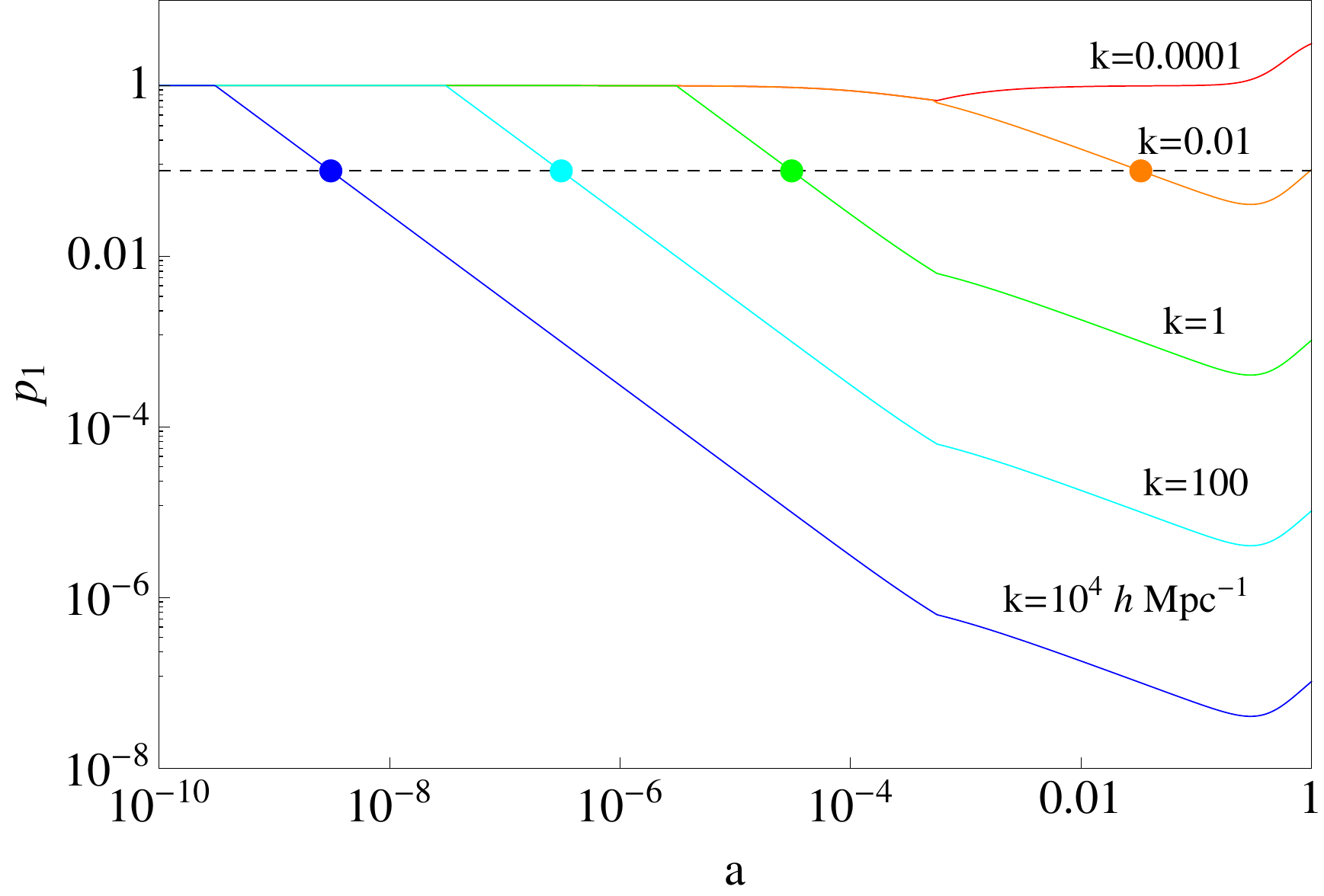} 
 \includegraphics[width=9cm]{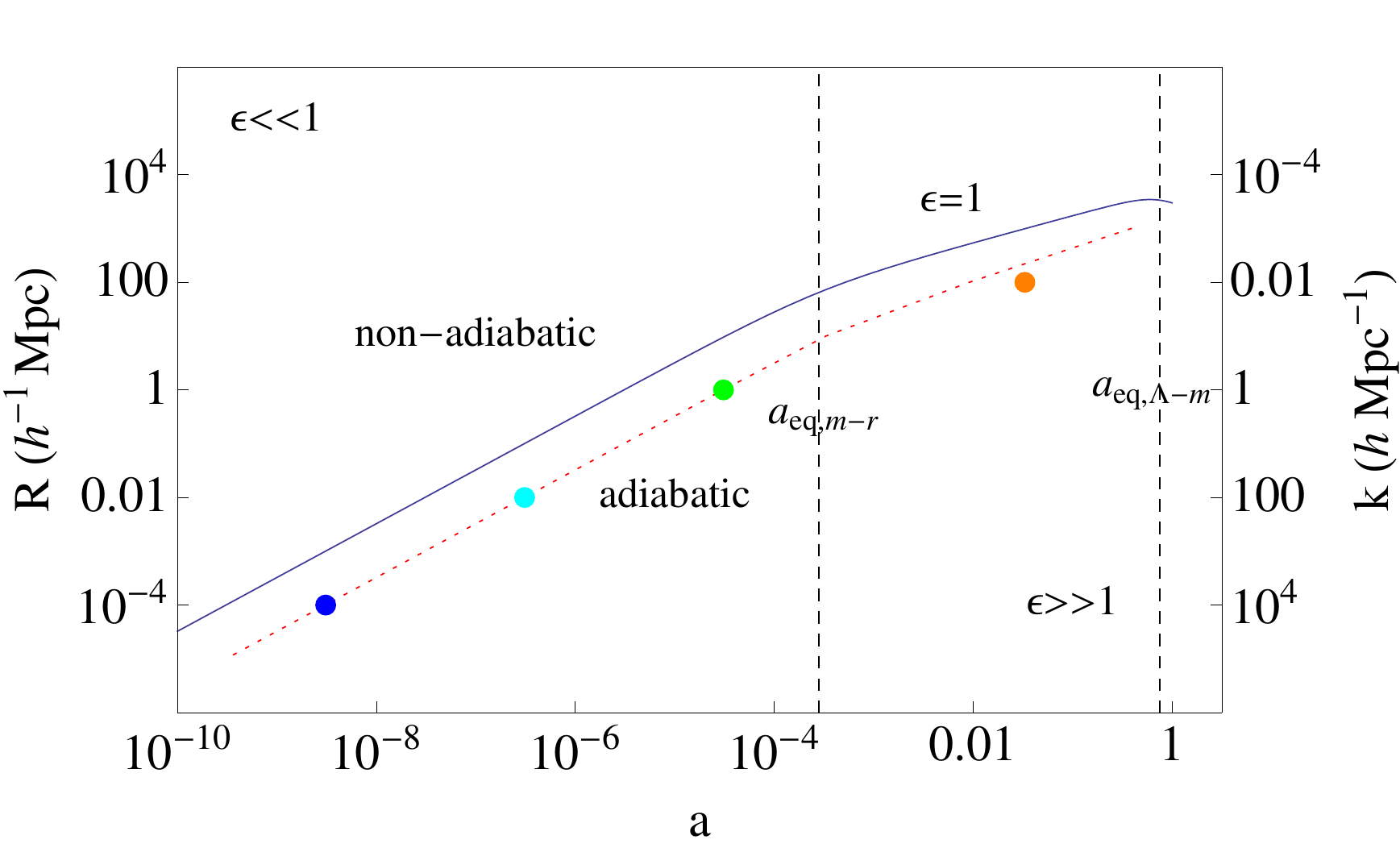}
\caption{Adiabatic condition based on parameters. The left panel shows the adiabatic parameter $p_1$ as defined in \eqnref{adparam1} for four values of $k$. The dotted line corresponds to our threshold value of $p_1 =0.1$ i.e., parameters should vary at least ten times slower than the variables. The resulting `adiabatic regime' lies to the left of the dotted line. The right panel shows the same condition on the $k-a$ plane (red dashed line), but uses the approximate analytic formula given in \eqnref{adcond2}. In both panels, the four points in orange, green, cyan and blue mark the condition $p_1(k,a) =0.1$, where $p_1$ is numerically evaluated using \eqnref{adparam1}. Thus, the formula of \eqnref{adcond2} is a fairly good approximation for $p_1$. The right panel also shows that imposing an `adiabatic regime' for super-horizon modes is not feasible. }
\label{adiabatic1}
\end{figure}

\item {\it Adiabatic condition based on eigenvalues}: 
In the eigenbasis 
\beq 
{\dot {\hat y}}_i= \lambda_i {\hat y}_i, 
\eeq
where the $\hat{}$ denotes eigenvectors. In this case, we demand that for the system to be `quasi-autonomous' the rate of fractional change in the eigenvalues is slow compared to the rate of fractional change in the eigenvectors. This condition can be implemented in two ways. For each eigen-direction one can demand ${\dot {\hat y}_i}/{\hat y}_i \gg  {\dot \lambda}_i/\lambda_i$. Alternatively, a more conservative way is to demand that the smallest time scale of change in eigenvalues $\min\{ |{\dot \lambda}_i/\lambda_i|^{-1}\}$ is larger than the largest time scale of change in eigenvectors $\max\{|{\dot {\hat y}_i}/{\hat y}_i|^{-1} \}$. 
Noting that, the rate of fractional change in the eigenvector is given by ${\dot y}/y \sim \lambda$, we  
define two other adiabatic parameters: 
\bea
\label{adparam2a}
p_{2,A}(k,a) &=&   \max \left\{\left |\frac{{\dot \lambda}_i}{\lambda_i^2}\right|\right\} \\
p_{2,B}(k,a) &=& \frac{\max |{\dot \lambda}_i|}{\min \{|\lambda_i|\}}, 
\label{adparam2b}
\eea 
where the latter is a more conservative definition. \capfigref{adiabatic2} shows the two parameters as a function of $k$ and $a$. The dashed line indicates the condition $p_{2,A/B} =0.1$. It is clear that these conditions are significantly more restrictive than the earlier condition of adiabaticity defined in terms of the parameters.

\begin{figure}
\centering
 \includegraphics[width=8cm]{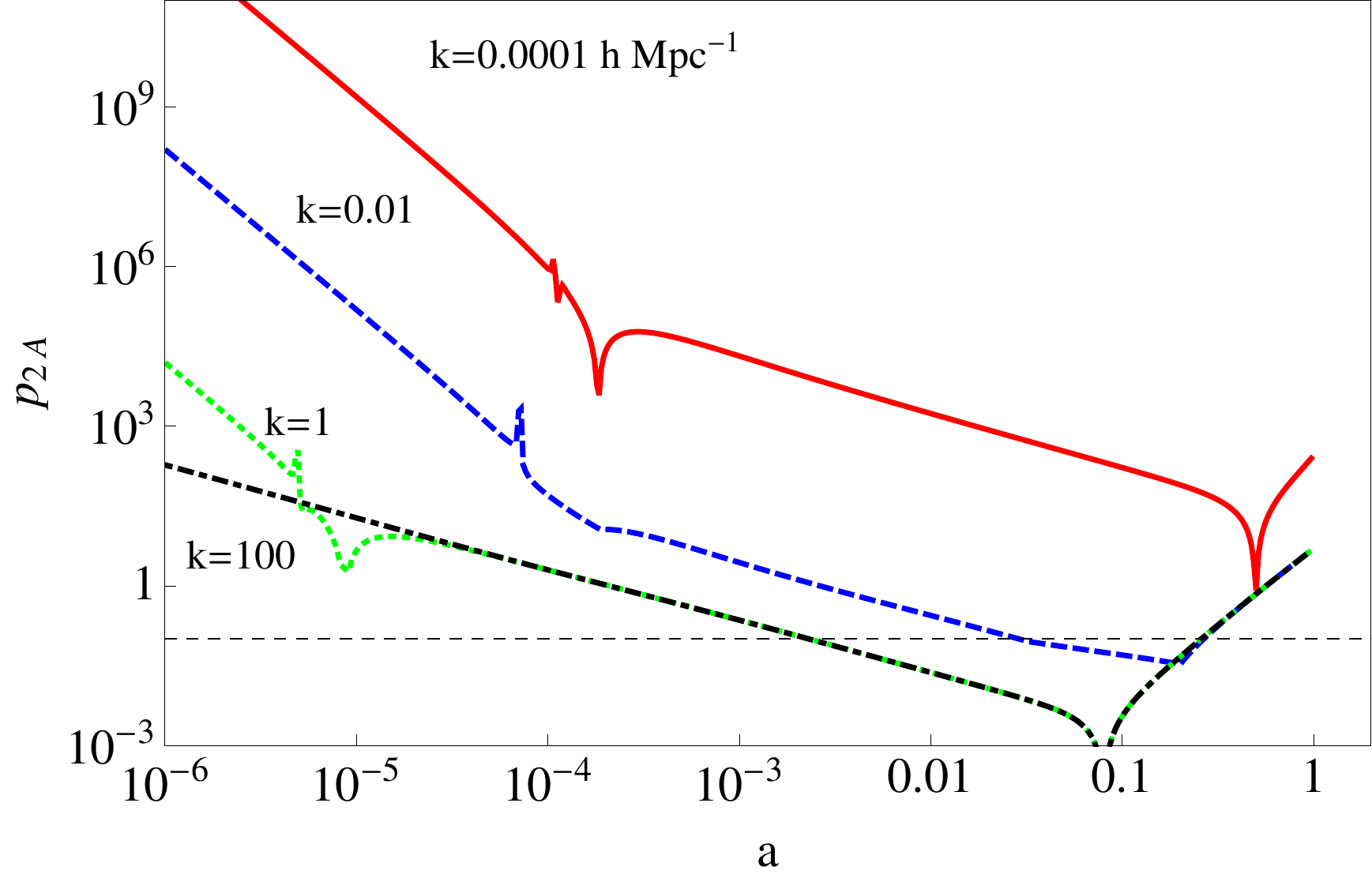} 
 \includegraphics[width=8cm]{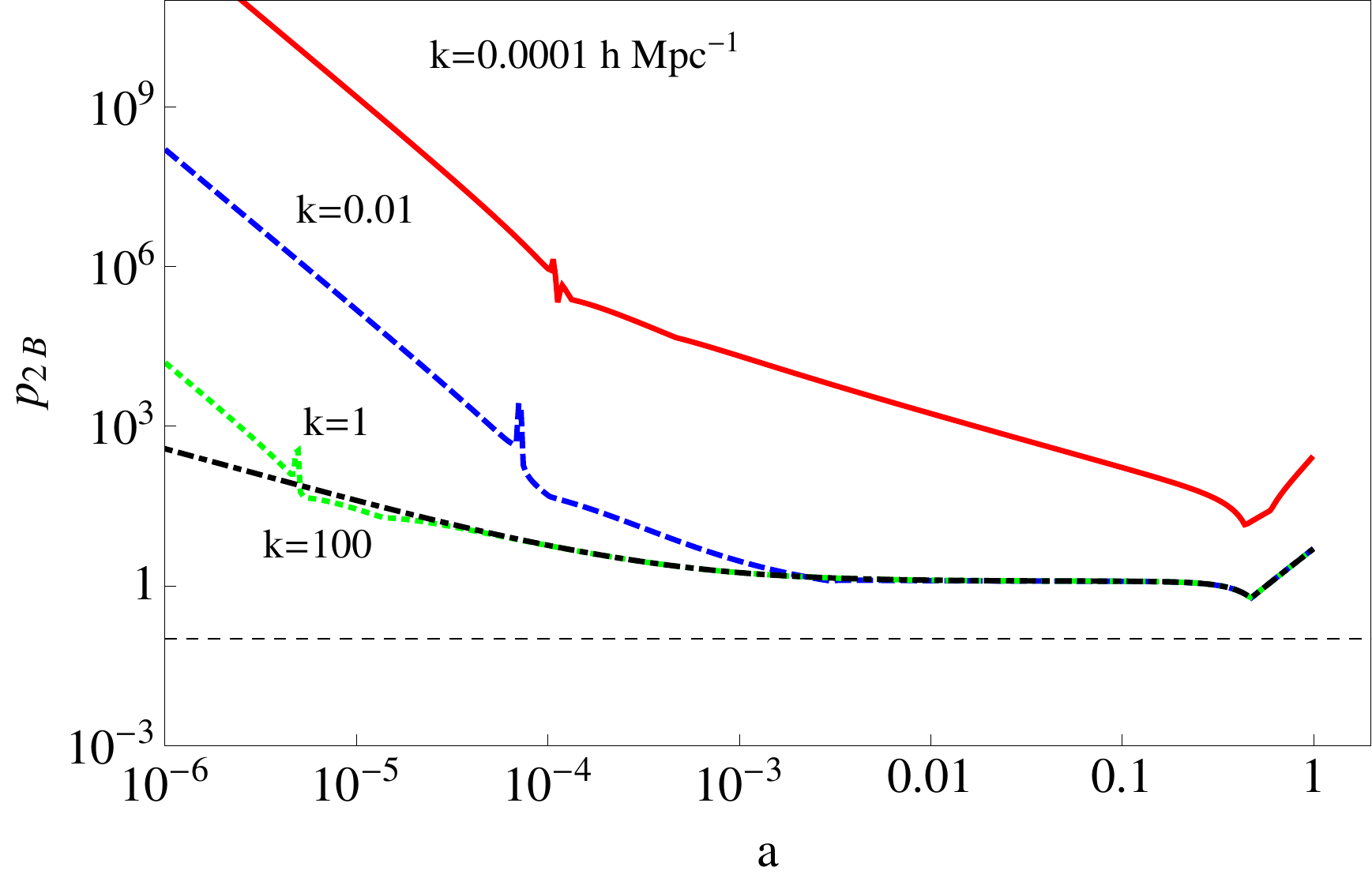}
\caption{Adiabatic conditions in the eigenbasis. The left and right panels show the adiabatic parameter as defined in \eqnrefs{adparam2a} and \eqnrefbare{adparam2b} respectively. The dashed line in both plots denoted the condition $p_{2,A/B} = 0.1$. While in the first case there is a small `adiabatic range', there is no appropriate range in the second definition. Overall, these are far more restrictive parameters than the adiabatic regime based on parameters. }
\label{adiabatic2}
\end{figure}
\end{enumerate}

Which definition of `adiabaticity' is appropriate depends upon the problem at hand. One can imagine evolving the E-B system by transforming to the eigenbasis. In this case, the evolution in the adiabatic regime will be given simply by the exponential of the diagonal matrix of eigenvalues. However, in this paper, in \S \ref{sec:limits}, we construct solutions in the basis defined by \eqnref{mainsystem}. Hence, in this paper, we will use the first adiabatic condition characterised in terms of the parameter $p_1$.

\section{Onset of oscillations}
\label{sec:onset}
It was shown in \S \ref{sec:eigenstruc} that the five eigenvalues of the matrix ${\mathcal A}$ undergo a transition from all real to three real and two complex. The epoch at which the transition takes place depends upon $k$. Although the eigenvalues are not known analytically, it is possible to analytically predict this epoch of transition. The method involves applying Sturm's theorem and Descartes' rule of sign to the characteristic polynomial of ${\mathcal A}$ and has been explained in detail in appendix \ref{app:sturm}. 
This analysis implies that the $k$-dependent transition epoch $a_{trans}$ is the solution of 
\beq 
9 \Omega_m(a_{trans}) + 2 \left[3 - 6 \Omega_r(a_{trans}) + \epsilon(a_{trans}, k)^2\right] = 0.
\eeq

\begin{figure}
\centering
\includegraphics[height=6cm]{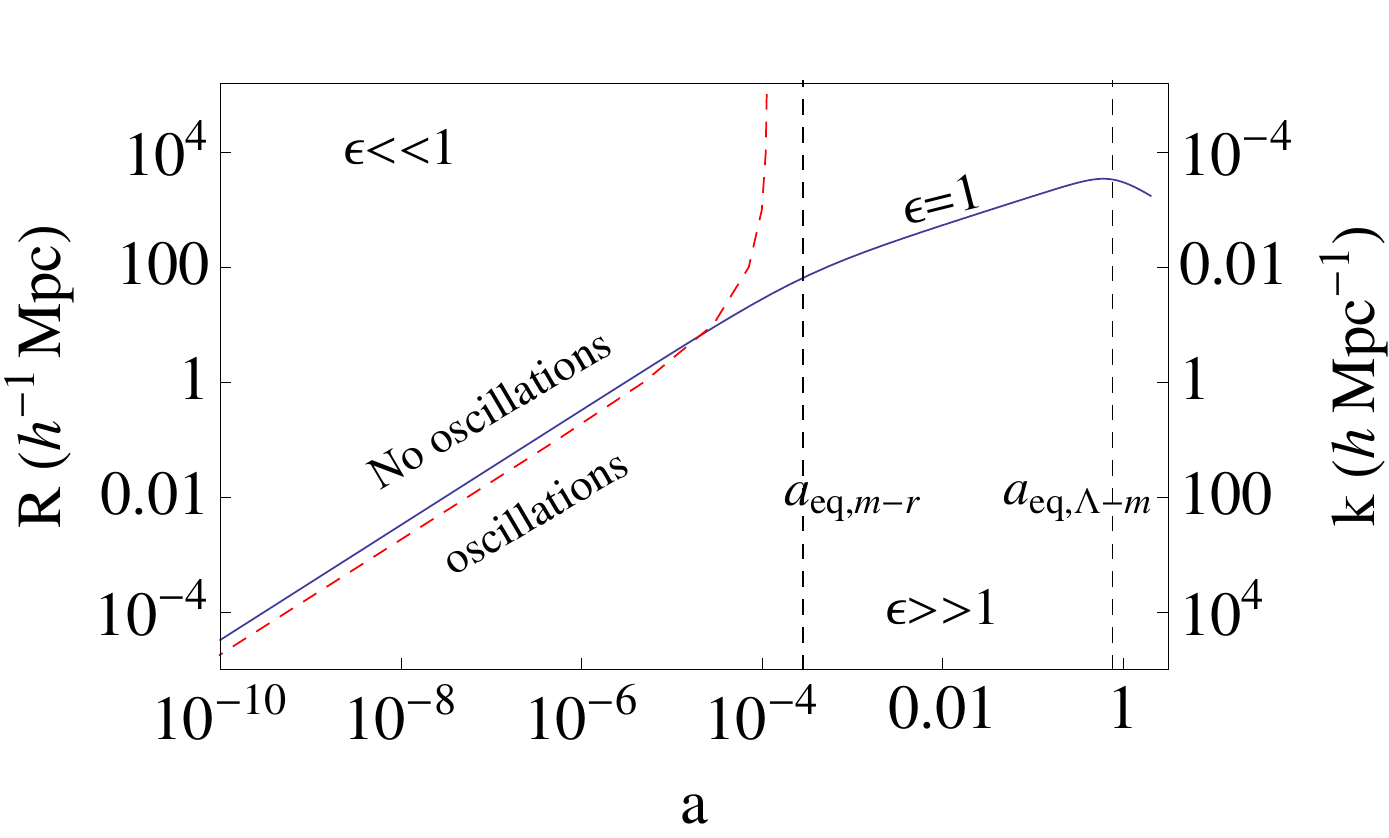}
\caption{Onset of oscillations: the transition from real to complex roots is denoted by the red dashed line. The blue solid line denotes the epoch of horizon crossing. It is clear that for most scales the onset of oscillations occurs very early, much before the epoch of matter-radiation equality.  }
\label{trans}
\end{figure}

\capfigref{trans} shows the scale that transitions to complex eigenvalues as a function of $a$. The figure suggests that, for all scales, the transition epoch occurs before the epoch of matter-radiation equality $a_{eq,m-r}$. This may seem counter-intuitive because, in general, no oscillations are expected for scales that enter the horizon after $a_{eq,m-r}$ i.e., for $k \ll k_{eq}$. However, note that Sturm's analysis does not predict the actual value of the frequency of oscillations. This is determined by the imaginary part of the eigenvalue, which in the large $\epsilon$ limit (late epochs) reduces to $\epsilon/\sqrt{3}$ and is smaller for earlier epochs (see \figref{oscfreq} in appendix \S \ref{app:freq}). For $k \sim 0.1 k_{eq}$, the average $\epsilon \sim 0.1$ in the interval from equality to today. In this interval, which corresponds to about 8 e-folds, one expects, $\sim 8 \times 0.1/\sqrt{3} \approx 0.5$ oscillations i.e., about half an oscillation. This issue is explained in greater detail in appendix \S \ref{app:freq}. For smaller values of $k$, this number will be even smaller and hence no oscillations are visible. It is also interesting to note that for modes with $k\gtrsim 0.1$ $h$ ${\rm Mpc}^{-1}$, the epoch of transition almost coincides with the epoch of horizon crossing. This means that, soon after the transition the oscillation frequency is of order unity or smaller: the regime of high frequency oscillations, which is numerically difficult to track, occurs well after the transition epoch.

As was discussed earlier, for a linear autonomous system complex eigenvalues imply oscillations and vice versa. For linear non-autonomous systems too, complex eigenvalues imply oscillations, however, the converse need not be true. Thus the transition to complex eigenvalues, in principle, indicates the presence of oscillations, not necessarily their onset. However, numerically, we do not find any evidence of oscillatory solutions before the epoch of transition for any value of $k$. Thus, we consider the transition epoch to denote the onset of oscillations.

\section{Numerical Stability}
\label{sec:stability}
Given an initial value problem to be solved numerically, there are many parameters that dictate the choice of an integration scheme. Apart from accuracy and available computing time, one important criterion is the stability of the numerical scheme. Loosely speaking, stability refers to the ability of the numerical solution to track the qualitative behaviour of the analytical solution. For example, if the analytic solution is bounded or converges to zero, the numerical approximation should also exhibit this behaviour. Mathematically, this property is characterised in terms of a function $r(z)$  called the {\it stability function} and the stability criterion is 
\beq 
|r(z)| \leq 1.
\label{stabcond} 	
\eeq 
For instance, for a one-dimensional differential equation of the type ${\dot y} = \lambda y$, the stability function depends only on the step size $h$ and the eigenvalue $\lambda$. Thus, given $\lambda$ and a particular method, one can estimate the allowed step size by applying the stability criterion.  It should be noted that such a stability criterion is relevant only when $\lambda$ is negative (or more generally, has negative real part). When $\lambda$ is positive the analytic solution grows exponentially and the choice of step size is dictated by how accurately the numerical solution is expected to track the analytic one. Similarly, if the eigenvalue is complex ($\lambda = i \omega$), the analytic solution oscillates with a frequency $\omega$ and numerically these oscillations can be fully resolved only if $h \leq 2 \omega^{-1}$. Thus, for a multi-dimensional system, like the E-B system considered here, the choice of step-size is dictated by a combination of accuracy requirements corresponding to the positive and complex eigenvalues and stability requirements related to the eigenvalues with negative real parts.

In the appendix, we review the basic stability theory \citep{butcher, harrier1, harrier2, petzold} for two classes of integration methods used to solve initial value problems: RK4 schemes and linear multistep methods, in particular the Backward Differentiation Formula (BDF) methods. RK4 schemes are single step schemes (the solution at the $n$-th step denoted by $y_n$ depends on the $n-1$-th step) but it can have multiple computations (called as stages) per step. In a linear multistep method with $k$ steps, $y_n$ depends linearly on $y_{n-1}, y_{n-2} \ldots y_{n-k}$ and/or the derivative at those points. Both RK4 and linear multi-step methods  have explicit schemes (where the solution for $y_n$ is computed directly from knowing the previous steps) and implicit schemes (where the solution for $y_n$ depends on solving a functional equation). Explicit schemes are easier to implement than implicit schemes, but tend to be less stable. Table \ref{stabtab} gives the stability functions for five commonly used methods: forward and backward Euler (these are the simplest first order explicit and implicit schemes), the popular fourth order Runge-Kutta solver (RK4), the trapezoidal rule (which is a second order implicit linear multistep method) and the second order BDF2 scheme. The formula for the third order BDF3 scheme is given in \eqnref{BDF3formula} in the appendix \ref{app:stability}. The order refers to how the error between the approximate and true solution scales as a function of the step size $h$.
 {\renewcommand{\arraystretch}{2}
\begin{table}
\centering
\begin{tabular}{|c|c|c|c|}
\hline
Method&Form& Order &Stability function ($z= h \lambda$)  \\
\hline
\hline
Forward Euler & $y_n = y_{n-1} + h f_{n-1}$ & 1&$r(z) = 1+z$\\
\hline
Backward Euler &$ y_n = y_{n-1} + h f_n $ & 1& $r(z) = \frac{1}{1-z}$\\
\hline
Standard Runge-Kutta &
$
\begin{array}{lcl}
y_n &= &y_{n-1} + \frac{h}{6}\left(k_1 + 2 k_2 + 2 k_3 + k_4\right)\\
k_1 &=& f(x_{n-1},y_{n-1})\\
k_2 &=& f(x_{n-1}+ \frac{h}{2}, y_{n-1} + \frac{h}{2} k_1)\\
k_3 &=& f(x_{n-1} + \frac{h}{2} , y_{n-1} + \frac{h}{2} k_2)\\
k_4 &=& f(x_{n-1} + h, y_{n-1} + h k_3). 
\end{array}
 $
& 4&$ r(z)=1+z+\frac{z^2}{2} + \frac{z^3}{6} + \frac{z^4}{24} $\\
\hline
Implicit Trapezoidal Rule &$y_n =y_{n-1} + \frac{h}{2} \left(f_n + f_{n-1}\right) $ &2& $r(z) = \frac{2+z}{2-z} $\\
\hline
BDF2 (implicit) &$y_n = \frac{4}{3} y_{n-1} -\frac{1}{3} y_{n-2} + \frac{2}{3} f_n0$ &2& $r(z) = \frac{2 + \sqrt{1 + 2 z}}{3- 2 z}  $\\
\hline
\hline
\end{tabular}
\caption{Stability functions for some commonly used numerical schemes. The function and its derivative at the $n$-th step are denoted by $y_n$ and $f_n$ respectively. For differential equations of the form ${\dot y} = \lambda y$, the stability condition is 
$|r(z= h \lambda)|=1$. Thus, given a $\lambda$, the step size can be determined. }
\label{stabtab}
\end{table}

\begin{figure}
\centering
\includegraphics[width=14cm]{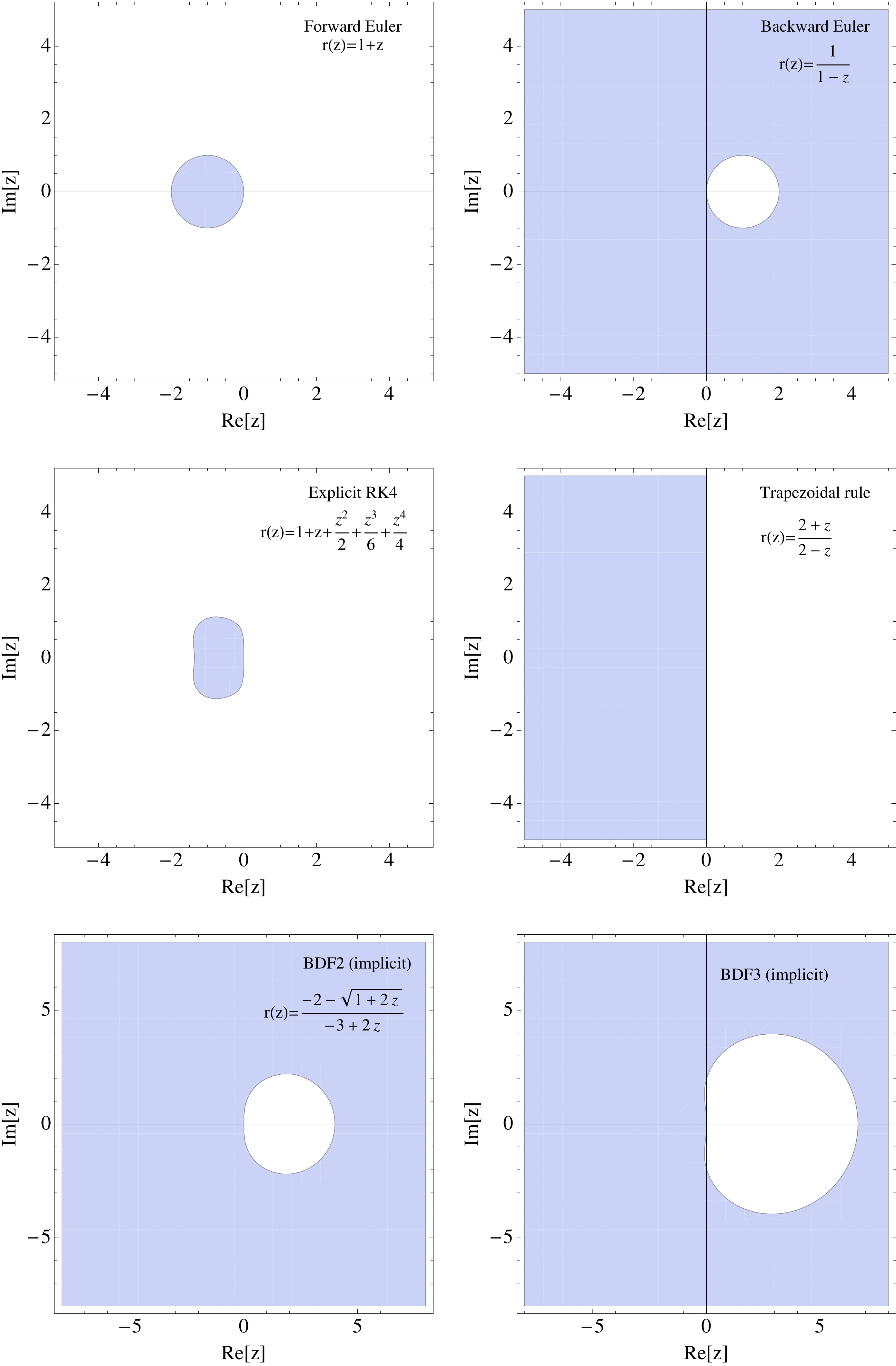}
\caption{Stability regions for some popular methods in the complex $z = \lambda h$ plane. Shaded regions satisfy the stability condition $|r(z)|\leq1$.  Note that the forward Euler and explicit RK4 are stable in a very small region. The BDF schemes are stable almost everywhere except for a small region in the $z$ plane. The BDF3 scheme has higher order than BDF2 so it is more efficient in terms of convergence, but it has a larger domain where it is unstable. If the eigenvalue $\lambda$ is real and positive, then the stability requirements for implicit methods imply a minimum step size. 
}
\label{numstab}
\end{figure}

\capfigref{numstab} shows the stability regions in the complex $z$ plane for each method. It is clear that implicit schemes have a greater region of stability than explicit schemes. For example, compare the forward and backward Euler schemes: in the first case $z=\lambda h$ must lie inside the shaded region in the left half-plane, whereas in the latter case $z=\lambda h$ can be anywhere {\it except} inside the non-shaded region in the right half-plane. Thus, for the particular schemes considered here, the stability requirement imposes a maximum allowed step size for explicit schemes and a minimum required step size for implicit schemes. Another feature to note is that sometimes there is a trade-off between order and stability. For example, the last panel implies that the BDF2 scheme has a greater region of stability than the higher order BDF3 scheme.   

\begin{figure}
\centering
\includegraphics[width=15cm]{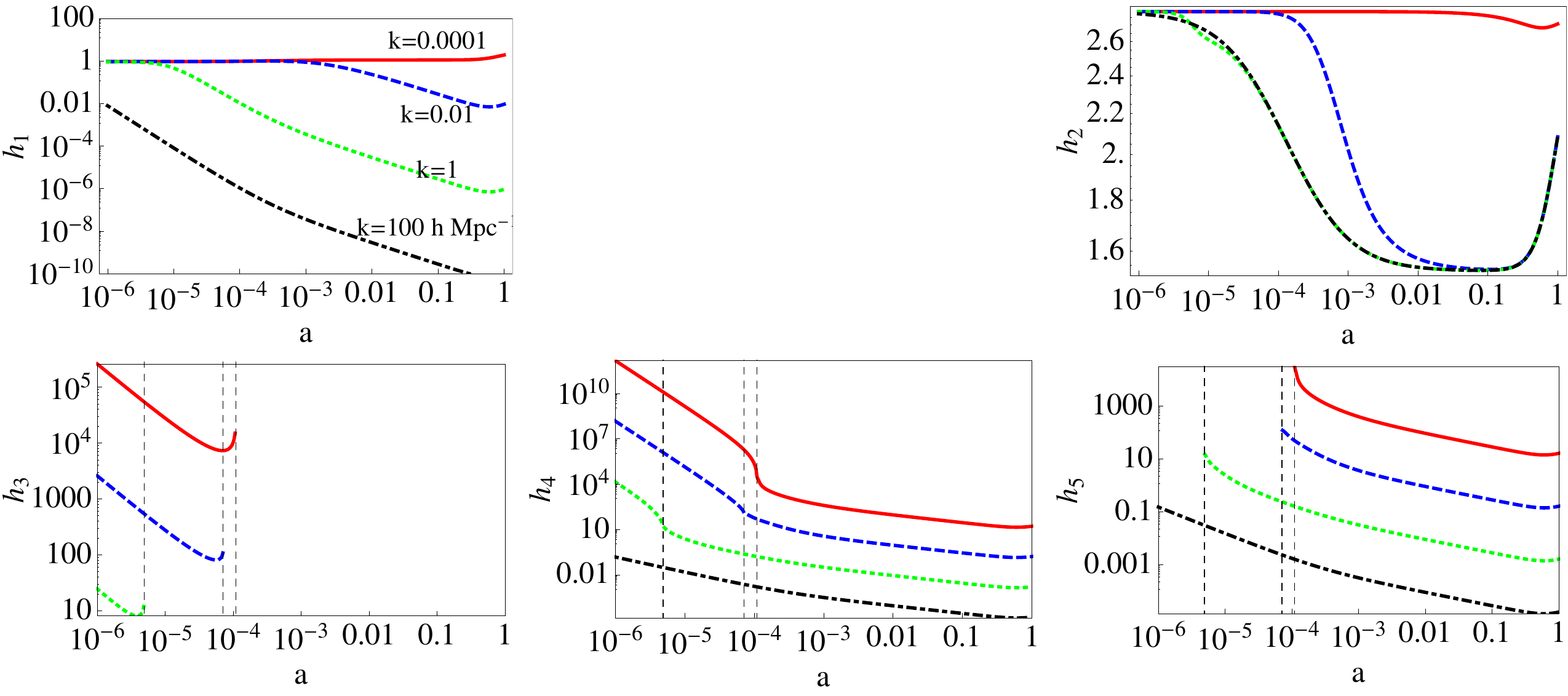}
\caption{The maximum allowed step size for a RK4 scheme applied to the E-B system. For simplicity, we assume that the system is autonomous and solve for $|r(z)|=1$ to get the minimum $h_i$ for each eigenvalue $\lambda_i$. Each resulting $h_i$ is a function of $k$ and $a$. The dashed vertical lines denote the transition from real to complex values. Refer to table \ref{eigensigns} for the signs of the eigenvalues. For positive eigenvalues, the maximum allowed value is zero i.e., the explicit scheme is never stable (third and fifth panel). For high negative values, the step size becomes vanishingly small as shown in the first panel.}
\label{hRK4}
\end{figure}
\begin{figure}
\centering
\includegraphics[width=15cm]{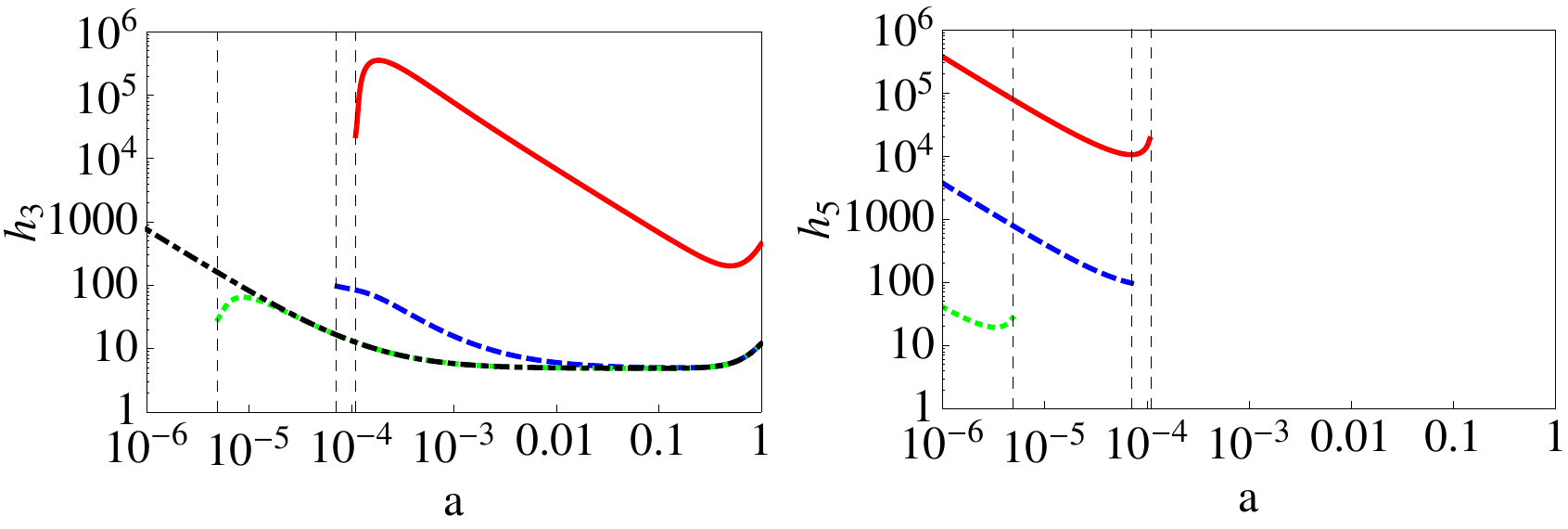}
\caption{The minimum required step size for the BDF2 scheme. The colour coding is the same as in \figref{hRK4}. For eigenvalues with negative real part, there is no minimum requirement. The step size can be arbitrary. Thus, only the step sizes $h_3$ and $h_5$ corresponding to $\lambda_3$ and $\lambda_5$ are shown here. }
\label{hBDF2}
\end{figure}
We now apply the stability conditions to the E-B system. There are five independent eigendirections \footnote{Without loss of generality, the stability analysis can be performed in the eigenbasis although the evolution need not be in this basis \citep{butcher}. }and the stability criterion can be applied separately for each of them. \capfigref{hRK4} shows the case for the explicit RK4 scheme for four $k$ values. For each eigenvalue $\lambda$, we numerically solve $|r(z=h \lambda)| = 1$ to get the maximum allowed step size $h$. Note that for the RK4 stability function, this gives only two real roots: $\lambda h = -2.785$ and $\lambda h =0$. Thus, if $\lambda$ is positive, the explicit RK4 scheme cannot work; the maximum allowed step size is zero. \capfigref{hRK4} can be understood in conjunction with table \ref{eigensigns}. $\lambda_1$, $\lambda_2$ are always real and negative and hence give a positive solution for $h$. $\lambda_3$ is negative until the transition point, after which it is positive. So $h_3$ is positive until the transition and zero thereafter. $\lambda_4$ is negative until the transition after which it is complex with a negative real part; thus for a real positive $h$, $z=\lambda h$ will lie in the second quadrant. Since the region of stability extends in this part of the plane, even complex eigenvalues give a real $h$. On the other hand, $\lambda_5$ is positive before the transition and hence the only solution in this part is $h_5=0$; after the transition, the solution for $h_5$ is identical to $h_4$ since the region of stability is symmetric about the $x$-axis. 

This can be contrasted with the behaviour of an implicit scheme applied to the same system. \capfigref{hBDF2} shows the minimum required step size from the stability condition $|r(z)| =1$ for the BDF2 scheme. Since $\lambda_1, \lambda_2<0$ and $Re(\lambda_4)<0$ for all epochs, there is no restriction on the step size, i.e., the scheme is always stable for evolution along these eigendirections. $\lambda_3$ ($\lambda_5$) is positive after (before) the transition which translates into a minimum required $h_3$ and $h_5$ in these regimes. Notice that the step-size is rather large; the total interval of integration (from $a \sim 10^{-8}$ to $a=1$) is about 18 e-folds, and thus much smaller than the minimum required step size. What does such a large stepsize mean ? This issue is also present in the case of explicit schemes, where stability requirements implied a zero step size for positive eigenvalues. This is expected. The stability criterion is derived from the condition that the numerical solution should decay when the analytic solution does so (see \S \ref{app:stability}). Thus, as noted earlier, it is relevant only to the sub-space of the system whose eigenvalues have negative real parts. The E-B system has gravitational instability encoded in it and some component of the solution is always growing which plausibly manifests as a positive eigenvalue.  In practice, for positive eigenvalues, the step size is dictated by accuracy considerations rather than stability, since the numerical as well as analytic solutions are unstable. 

\begin{figure}
\centering
\includegraphics[width=13cm]{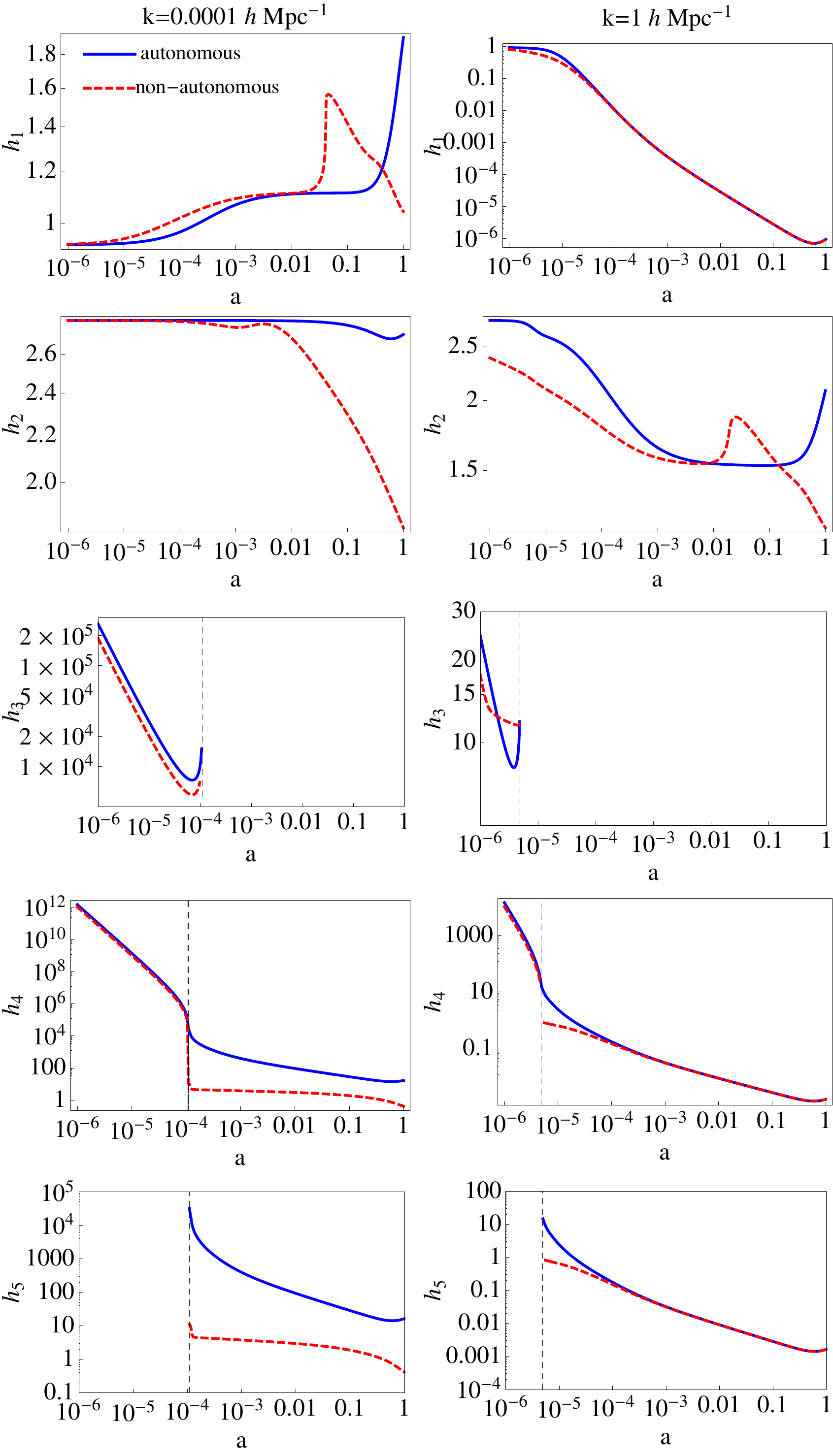}
\caption{Comparison of step sizes for a explicit RK4 scheme derived using the autonomous (blue solid) and non-autonomous (red dashed) stability condition for $k=0.0001 $ $h$ ${\rm Mpc}^{-1}$ (left) and $k=1$ $h$ ${\rm Mpc}^{-1}$ (right). The vertical dashed lines indicate the epoch of transition to complex values for each value of $k$.  }
\label{hRK4nonauto}
\end{figure}

As discussed earlier, the E-B system is non-autonomous. For a non-autonomous system the eigenvalue $\lambda$ is a function of the time variable and the stability function needs to be generalized. As an example, we consider the RK4 method applied to the non-autonomous system. If $x$ is the independent variable ($\ln a$ in the E-B case) then the stability function is \citep{burrage} 
 \beq
R(h,x) = 1 + \frac{z_1}{6} + \frac{z_2}{3} + \frac{z_1z_2}{6} + \frac{z_3}{3} + \frac{z_2 z_3}{6} + \frac{z_1z_2 z_3}{12} + \frac{z_4}{6} + \frac{z_3 z_4}{6} + \frac{z_2 z_3 z_4}{12} + \frac{z_1 z_2 z_3 z_4}{24}, 
\eeq
where the $z_i$s are functions of the eigenvalues evaluated at the sub-grid points defined by the nodes ($c$ values in the Butcher tableau): 
\bea
z_1 &=& h \lambda(x)\\
z_2 &=& h \lambda(x + \frac{h}{2})\\
z_3 &=& h \lambda(x + \frac{h}{2})\\
z_4 &=& h \lambda(x + h). 
\eea
The stability condition is unchanged:
\beq 
|R(h, x)|\leq1.
\eeq
\capfigref{hRK4nonauto} shows the step sizes derived using the autonomous (solid blue line) and non-autonomous (dashed red line) conditions  for two values $k = 0.0001$ (left column) and $k=1$ $h$ ${\rm Mpc}^{-1}$ (right column). We find that except when the eigenvalues are complex, the step size in the autonomous and non-autonomous cases are comparable. For example, for $k=0.0001$ $h$ ${\rm Mpc}^{-1}$, the plots of $h_1$ and $h_2$ in the two cases almost overlap for $a\lesssim 0.01$. They differ in the case of complex eigenvalues (for example $h_4$ and $h_5$ after the transition) but the difference is less for larger values of $k$. This can be understood from the adiabatic conditions discussed in \S \ref{sec:adiabatic}, where we show that in an appropriately defined `adiabatic' regime, the system can be considered autonomous. Larger values of $k$ satisfy the adiabatic condition for a longer range of epochs. Thus, in practice, it may be possible to use the stability analysis for autonomous systems to determine the step size and the non-autonomous nature can be accounted by making a conservative choice. For appropriate values of $k$, this strategy is further supported by the adiabatic conditions.

\section{Stiffness}
\label{sec:stiffness}
 A differential equation is generally considered to be stiff if there are two or more widely separated time scales in the problem. For the EB system without baryons the two scales are the oscillation time scale of a photon mode and the Hubble time. In terms of the conformal time these are $\eta_k = k^{-1}$ and $\eta_H \sim (a H)^{-1}$. In the presence of baryons, there is an additional time scale associated with the Thomson scattering of photons and baryons i.e., $\eta_c = (a n_e \sigma_T)^{-1}$, where $n_e$ is the electron density and $\sigma_T$ is the collisional cross section. 
In appendix \ref{app:withbaryons} we recast the system with $\ln a$ as the time variable and show that that the three time scales appear as two ratios: $\epsilon_k = \eta_H/\eta_k \sim k/aH$ and $\epsilon_c = \eta_H/\eta_c \sim (a n_e \sigma_T)/(aH)$ ($\epsilon_k = \epsilon$ in the rest of this text; we use the subscript only while considering the full system). For early epochs, before recombination, the Thomson opacity is large and the photons and baryons are tightly coupled i.e., $\eta_c<< \eta_H$ or $\epsilon_c \gg 1$. This regime is usually handled by invoking the tight coupling approximation. There are various implementations of this approximation (referred to as TCA; see \citealt*{blas_cosmic_2011}) all of which effectively re-write the coupling term such that the resulting equations are independent of $\eta_c^{-1}$. Another regime where the system becomes numerically difficult to track is when $\epsilon_k$ is large i.e., when the photon moments undergo rapid oscillations. However, for most modes of interest, this occurs at late epochs well into the matter or dark energy dominated phase and the radiation fields need not be tracked with high accuracy. Practically, the radiation streaming approximation (referred to as RSA in the second CLASS paper) is invoked wherein one substitutes approximate analytic expressions for the radiation fields thus circumventing the problem of numerically tracking them. This approximation has been discussed by \citealt{doran_speeding_2005} in the conformal Newtonian Gauge and by \citealt*{blas_cosmic_2011} in the synchronous gauge.

 In the CMB literature, only the tight coupling regime is usually referred to as the `stiff' regime. But, it is clear that $\eta_c << \eta_H$ and $\eta_k << \eta_H$ are both regimes where there are two widely separated time scales in the problem and the system is stiff. 
 Hence one of the aims in this paper is to understand better the definition of `stiffness' and discuss means to characterise it. Stiff systems have multiple time scales spanning a large dynamic range. To maintain stability of the solution, it is usually necessary that the step size be smaller than the smallest time scale in the system, even though accuracy requirements may allow a larger step size \citep{numrecipes}. Thus, a stiff system can be characterised as follows \citep{petzold}. An initial value problem is considered `stiff' over an interval, if the step size required to maintain stability of the forward Euler method is significantly smaller than the step size required to maintain accuracy. Hence, whether a problem is considered stiff or not depends upon: (1) the parameters of the differential equation (2) the accuracy criterion, (3) the length of the interval of integration and (4) the region of absolute stability of the method. For a stiff problem, the step size dictated by stability requirements of an explicit scheme becomes prohibitively small and typically an implicit method needs to be invoked.  

In the EB system without baryons there are two physical time scales and their ratio $\epsilon$ primarily governs the rates at which the variables evolve (coefficients of most terms are of the order of $\epsilon$). In the eigenbasis it has the form $ {\dot y}_i = \lambda_i y_i$, 
 and each eigenvalue has an associated time scale $\lambda_i^{-1}$. One way to characterise `stiffness' is to define the `stiffness ratio' $s$ as \citep{lambert}
\beq 
s = \frac{|{\mathcal Re}\{\lambda_{max}\}|}{|{\mathcal Re}\{\lambda_{min}\}|}\; \;   \;\; \lambda_{min, max}<0,
\label{stiffdef}
\eeq
where $\lambda_{max}$ and $\lambda_{min}$ denote the most and least negative eigenvalue respectively. Restricting to the subspace of negative eigenvalues is necessary since stability of the numerical solution is defined only in this subspace. \capfigref{stiffness} (left plot) shows the stiffness parameters defined above as a function of epoch for four values of $k$. It is clear from the figure that the E-B system has a large stiffness ratio for a wide range of $k$ and epochs. The right plot shows the stiffness parameter $s$ vs $\epsilon$. 
It is seen that the stiffness parameter is minimum when $\epsilon \sim 1$ i.e. when the time scale of oscillation and Hubble evolution are of the same order. \footnote{Note that, even for small values of $\epsilon$ the stiffness parameter is very large. This seems a bit counterintuitive since there are usually no numerical difficulties reported for super-horizon modes (small $\epsilon$). However, we checked (plot not shown) and found that indeed the explicit forward Euler fails to evolve a super-horizon mode (e.g., $k = 5 \times 10^{-5}$ $h$ ${\rm Mpc}^{-1}$) from $a=10^{-8}$ to $a=1$ although $\epsilon \ll1$ throughout this regime. Instead a higher order scheme such as the explicit Runge-Kutta was needed to evolve the system over the entire domain. }

\begin{figure}
\centering
\includegraphics[width=8cm]{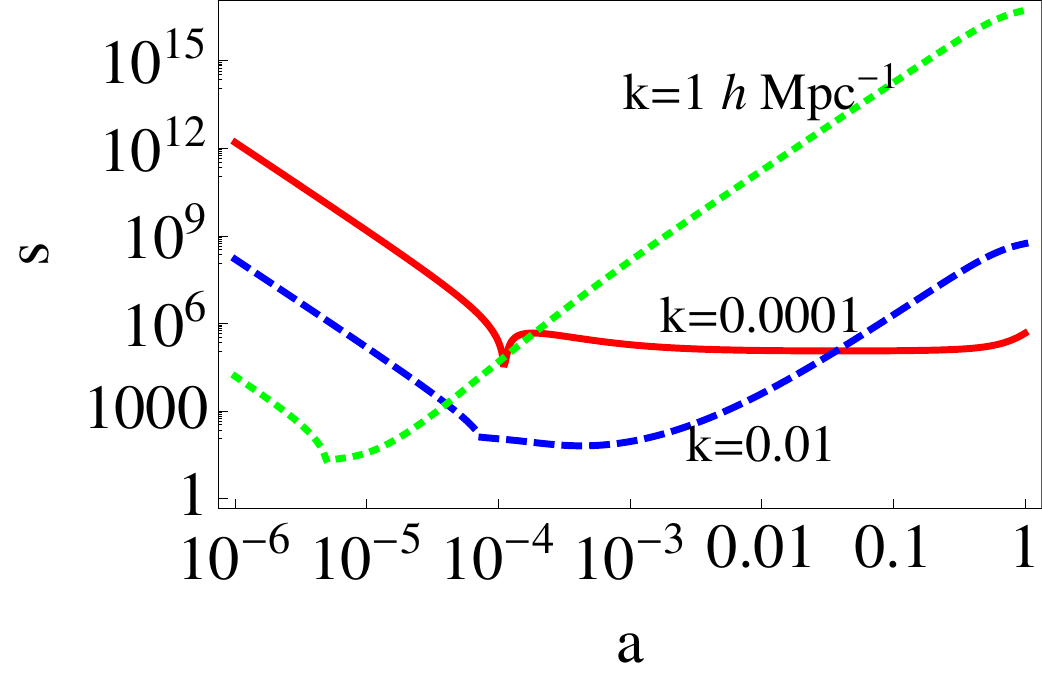}
\includegraphics[width=8cm]{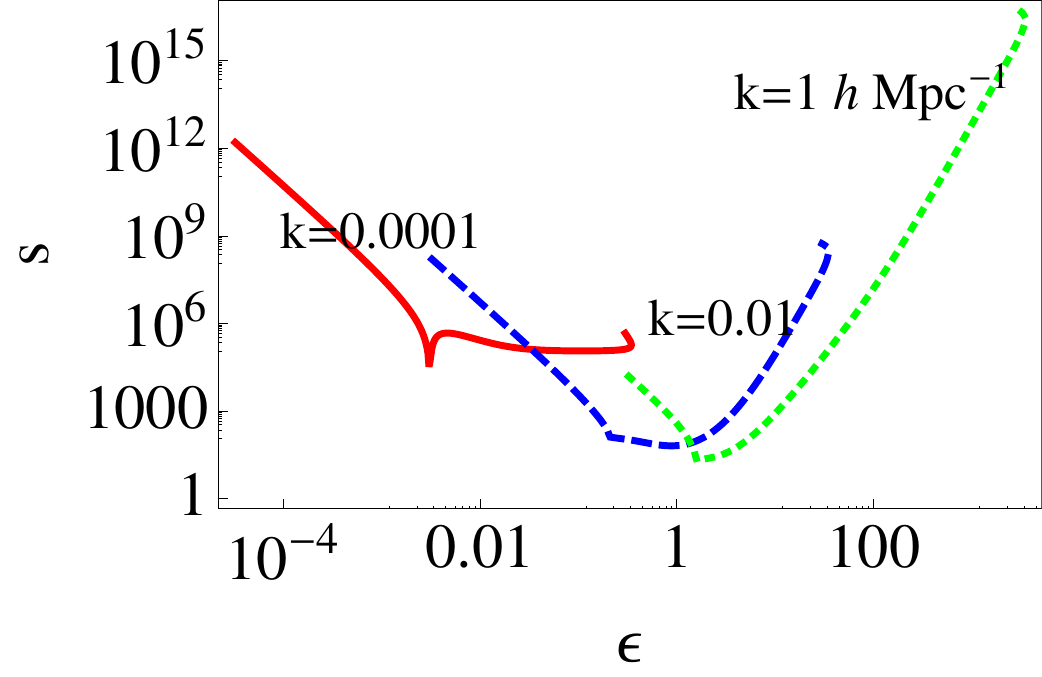}
\caption{The left plot shows the stiffness parameter defined in the text in \eqnref{stiffdef}. Large values of $s$ indicates that the E-B system is `stiff' at all epochs for all modes. Thus, implicit methods have to be employed since explicit methods are expected to be unstable for such systems. The right plot shows $s$ vs. $\epsilon$. Even for super-horizon modes with small values of $\epsilon$, the system is stiff.}
\label{stiffness}
\end{figure}

In the past implicit schemes have been advocated to evolve the full system E-B system including baryons, particularly to treat the tight coupling regime. For example, CLASS uses the solver \texttt{ndf15} which is a Numerical Differentiation Formula (NDF) very closely related to the BDF methods \mbox{\citep{Shampine}}. CAMB, on the other hand, uses the DVERK routine, which is based on higher order (adaptive) Runge-Kutta methods and is most efficient for non-stiff systems (see subroutines.f90 of the CAMB sourcecode). However, due to the various approximations (TCA and RSA) to treat the stiff regimes this does not prove to be prohibitive.

}

\section{Limits}
\label{sec:limits}

\begin{figure}
\centering
\begin{minipage}[]{0.85\textwidth}
\includegraphics[width=15cm]{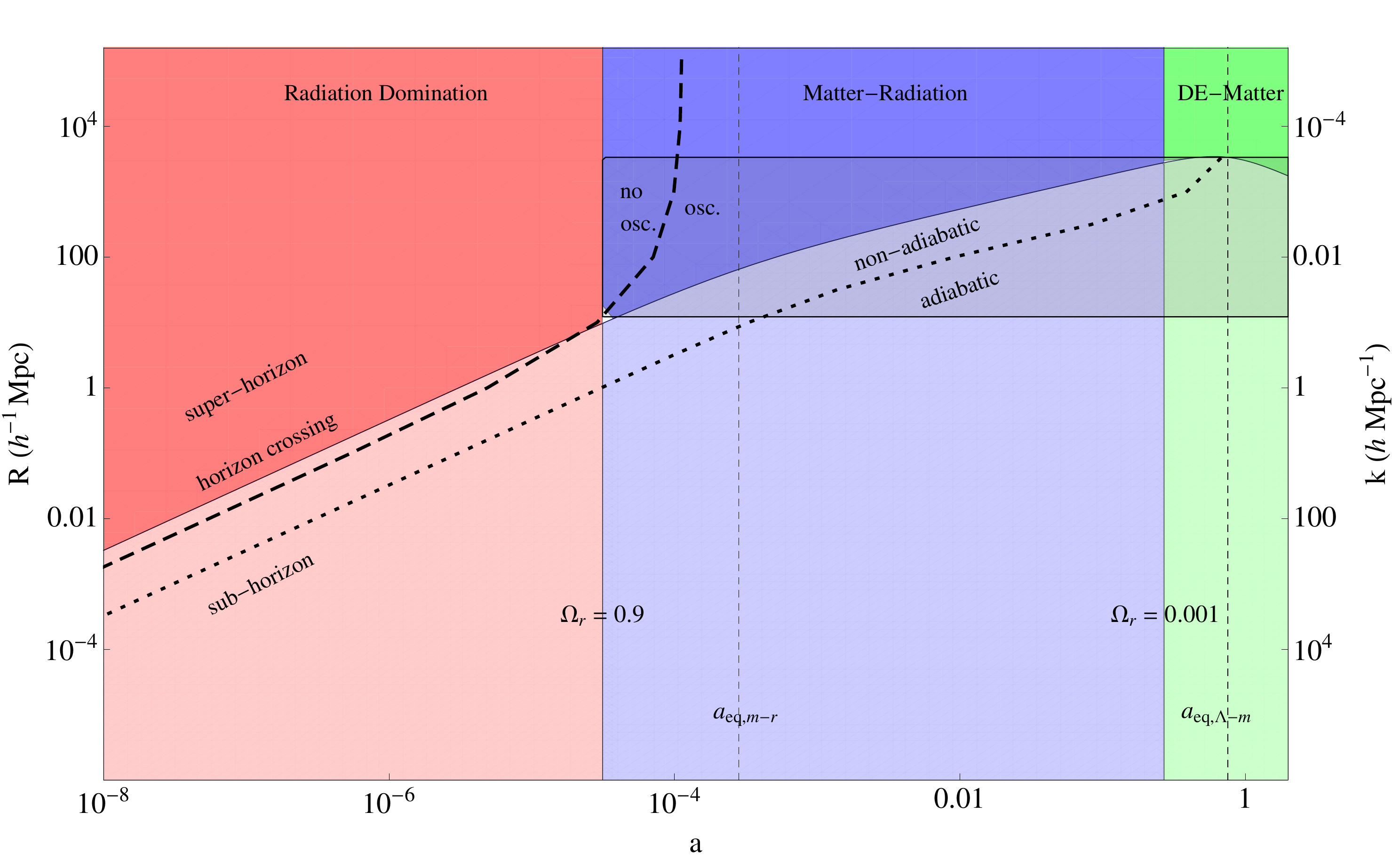}
\end{minipage}
\begin{minipage}[]{0.1\textwidth}
\includegraphics[width=2.5cm]{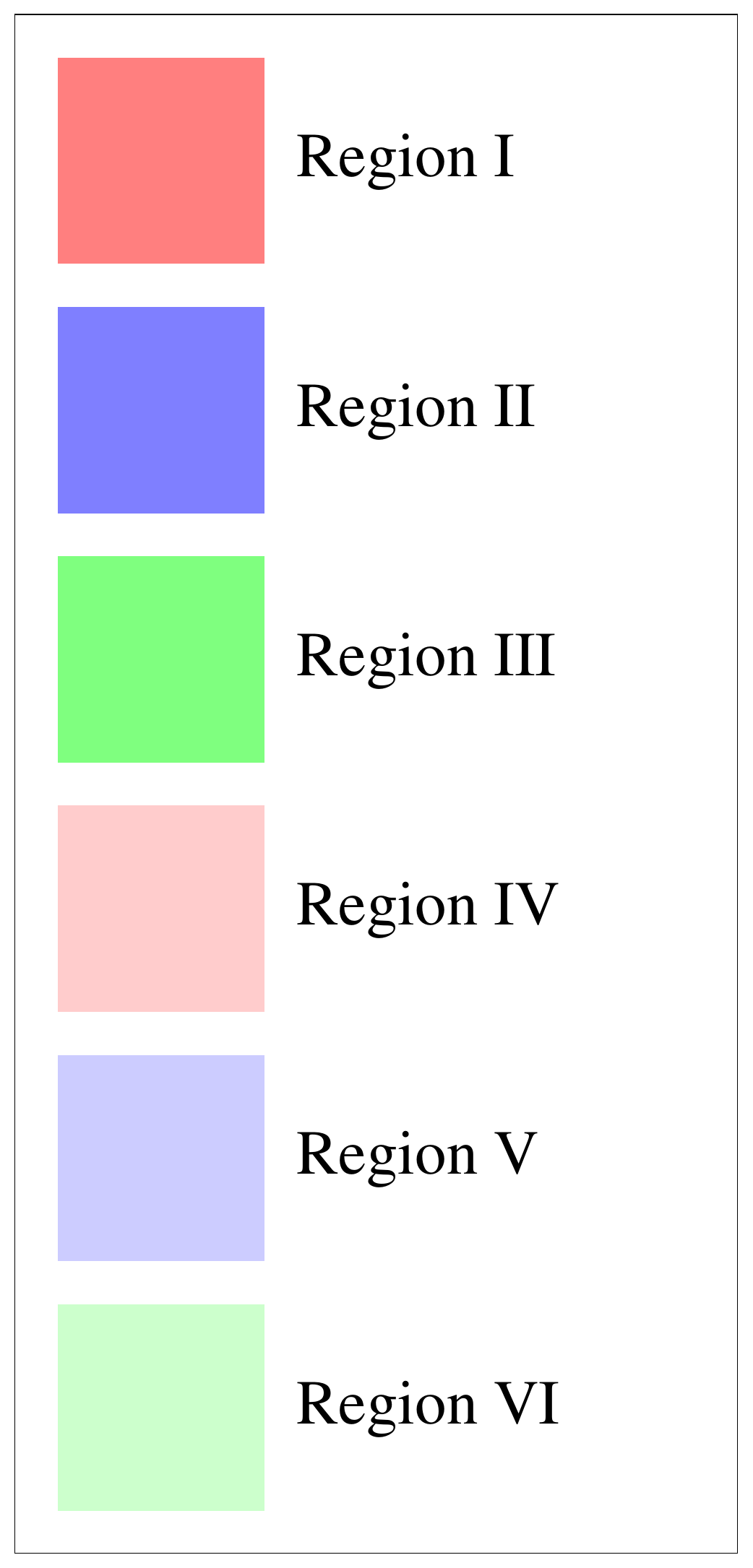}
\end{minipage}
\caption{Various regions on the `$k-a$' plane. The dot-dashed line separates the regions where the eigenvalues are real (no oscillations) from where they are complex (onset of oscillatory behaviour). The dashed line indicates the regions where the adiabatic condition holds and the system can be treated as an autonomous system. The regions marked I to VI indicate various limits where analytic solutions are available to test numeric codes.  We choose the end of radiation domination to be the epoch when $\Omega_r = 0.9$ and the beginning of the matter-dark energy era when 
$\Omega_r = 0.001$. The shaded rectangle indicates the range of $k$ values where a complete solution cannot be constructed. This range is from $k=0.1$ to $k= 3. \times 10^{-4}$ $h$ ${\rm Mpc}^{-1}$. 
}
\label{kaplot}
\end{figure}

There are three parameters in the E-B system: $\Omega_m$, $\Omega_r$ and $\epsilon$. 
Based on the value of $\epsilon$ there are two different regimes: superhorizon ($\epsilon \ll1$) and sub-horizon ($\epsilon \simgreat 1$ and $\epsilon \gg 1$). Based on the $\Omega$ parameters the evolution can be classified into three eras: the radiation domination when $\Omega_m\ll\Omega_r 0$, the matter-radiation era, when both $\Omega_m$ and $\Omega_r$ are non-zero and the matter-dark energy era when $\Omega_r\ll\Omega_m$ and $1-\Omega_m-\Omega_r = \Omega_\Lambda$. 
This classification defines six separate regions (denoted by the roman numerals `I' to `VI') where analytic forms can be obtained. \capfigref{kaplot}
shows these regions along with a summary of the various regimes given by the eigenvalue analysis discussed in earlier sections. 
We have chosen $\Omega_r=0.9$ to be the end of radiation domination and $\Omega_r=0.001$ to be the onset of the matter-dark energy era. The solutions in these different regions are listed below. The details involved in arriving at these forms are given in appendix \ref{app:limits}. The solutions depend on initial conditions denoted by subscript `$i$'. These initial conditions are in general different for each region \footnote{
The construction of these solutions (for the sub-horizon modes) assumes that $y_5$ satisfies the algebraic form of \eqnref{eq5II }which is valid when 
deep in the radiation era the system satisfies adiabatic initial conditions given by \cref{eq1init,eq2init,eq3init,eq4init}. However, at late epochs, these relations are not satisfied.}. This allows one to compare the solutions in each region independently.

\subsection{Analytic forms}
There has been extensive work in the past in terms of obtaining solutions to the Boltzmann system, for example, super-horizon solutions have been constructed by \cite{kodama_cosmological_1984} and extensive work has been done in the sub-horizon regime by Hu and collaborators (for e.g., \citealt{hu_anisotropies_1995,hu_small-scale_1996}). In this paper we obtain solutions in terms of the variables $y_1$ to $y_5$. In some cases we can reproduce earlier results, while in others there is a slight difference because of the change of variables.

\begin{itemize}
\item{\bf Region I: super-horizon modes in the radiation domination era}\\
To the lowest order in $\epsilon$, the first four variables in this region of the $k-a$ plane become 
\begin{subequations}
\begin{align}
\label{y1regIa} y_1(a) &\simeq y_{1,i}\\
\label{y2regIa} y_2(a) &\simeq y_{2,i}  \\
\label{y3regIa}y_3(a) &\simeq y_{3,i}\\
\label{y4regIa}y_4(a) & \simeq  \frac{y_{4,i} a_i}{a}
\end{align}
\end{subequations}
The solution for $y_5$ in regions I, II and III is derived assuming that $y_1$ and $y_3$ are constants. In both region I and II, the solution for $y_5$ is expressed in terms of the variable 
\beq 
x = \frac{a}{a_{eq,m-r}}, 
\label{x1eq}
\eeq
where $a_{eq,m-r}$ is the epoch of radiation-matter equality. 
For adiabatic initial conditions set at inflation, and assuming a very small $x_i\ll1$, the solution for $y_5$ is the well-known solution \citep{kodama_cosmological_1984}. 
\beq 
y_5(a) \equiv y_5(x)  = \frac{y_{5,i}}{10} \left(16 \frac{\sqrt{1+x}}{x^3} -\frac{16}{x^3} -\frac{8}{x^2} + \frac{2}{x} + 9 \right). 
\label{y5regI}
\eeq

It is possible to get refined approximations for $y_1$ to $y_4$ using the solutions given by \cref{y1regIa,y2regIa,y3regIa,y4regIa} and \eqnref{y5regI} and in the r.h.s. of \cref{eq1,eq2,eq3,eq4} and integrating the resulting system. 
This gives, 
\begin{subequations}
\begin{align}
\label{y1regIb} y_1(k,a) &= y_{1,i} - y_{2,i} \int_{a_i}^a \frac{\epsilon(k,a')}{3} d\ln a'\\
\label{y2regIb} y_2(k,a) &= y_{2,i} +  y_{1,i}  \int_{a_i}^a  \epsilon(k,a') d\ln a' - 2 \int_{a_i}^a y_5(a')  \epsilon(k,a') d\ln a' \\
\label{y3regIb}y_3(k,a) &= y_{3,i} -  y_{4,i} a_i \int_{a_i}^a  \frac{\epsilon(k,a')}{a'} d\ln a' \\
\label{y4regIb} y_4(k,a)& = \frac{y_{4,i} a_i}{a} -\frac{1}{a} \int_{a_i}^a a' \epsilon(k,a') y_5(a') d\ln a'.
\end{align}
\end{subequations}
Deep in the radiation dominated era, $\epsilon \sim a$ and the integrals can be evaluated analytically (see appendix \ref{app:limits}).

\item{\bf Region II: super-horizon modes in the radiation-matter era}\\
The main difference between region I and II is that the initial conditions in the latter are not necessarily those that are set by inflation. The super-horizon mode evolves through the radiation dominated era before entering region II and and variables change as a result of this evolution. The general solution for $y_5$ in terms of the variable $x$ defined in \eqnref{x1eq} is 
\beq
y_5(x) = \frac{\sqrt{1+x}}{x^3}\left( \frac{y_{3,i}}{5} \left.\left\{\frac{16+8x-2x^2+x^3}{\sqrt{1+x}} \right\}\right|_{x_i}^x + \frac{4 y_{1,i}}{3} \left.\left\{\frac{-8-4x+x^2}{\sqrt{1+x}} 
 \right\}\right|_{x_i}^x\right) +  y_{5,i}\left( \frac{x_i}{x}\right)^3 \sqrt{\frac{1+x}{1+x_i}}. 
\eeq
Note that this equation reduces to \eqnref{y5regI} for $x_i\ll1$ and initial conditions given by \cref{eq1init,eq2init,eq3init,eq4init}. 

The solutions for $y_1$ to $y_4$ have the same functional form as  \cref{y1regIb,y2regIb,y3regIb,y4regIb} but the integrals should be solved numerically with the appropriate values of the initial conditions. 
\item{\bf Region III: super-horizon modes in the matter-dark energy era.}\\
In this region, the functional form of the solutions for $y_1$ through $y_4$ are the same as in region I and II and given by \cref{y1regIb,y2regIb,y3regIb,y4regIb}. $y_5$ is solved in terms of the variable 
\beq 
x=a/a_{eq,\Lambda-m},  
\eeq
where $a_{eq, \Lambda-m}$ is the epoch of matter-dark energy equality. 
The solution for $y_5$ in region III in terms of $x$ and the initial value $x_i$ is given by a hypergeometric function (\citealt{arfken}):
\beq 
y_5(x,x_i) = C \frac{\sqrt{1+x^3}}{x^{5/2}} -\frac{y_{3,i}}{4} \frac{1}{x^{5/2}(1+x^3)^{1/6}} \; \; {}_2F_1\left(\frac{2}{3}, \frac{1}{6},\frac{5}{3}, \frac{1}{1+x^3}\right), 
\eeq
where 
\beq
C  = y_{5,i} \frac{x_i^{5/2}}{\sqrt{1+x_i^3}} + \frac{y_{3,i}}{4} \frac{1}{(1+x_i^3)^{2/3}} \; \; {}_2F_1\left(\frac{2}{3}, \frac{1}{6},\frac{5}{3}, \frac{1}{1+x_i^3}\right). 
\eeq
\item{\bf Region IV: sub-horizon modes in the radiation dominated era} \\
Let us define 
\beq x = \epsilon/\sqrt{3}.
\eeq 
In this region we then obtain
\begin{subequations}
\begin{align}
\label{y1regIV} y_1(x) &= - \frac{1}{2 x} \left[ (a_2 x -2 a_1 ) \sin x + (a_1 x + 2 a_2 ) \cos x \right]  \\
\label{y2regIV} y_2(x) &= -\frac{\sqrt{3}}{2 x^2} \left[ \left\{ 2  a_2 x + a_1 \left(x^2-2\right)\right\} \sin x + \left\{2 a_1 x-a_2   \left(x^2-2\right)\right\} \cos x\right]\\
\label{y3regIV}y_3(x) &= 3 a_1\left.\left(\frac{\sin x}{x} + \ln x \frac{\sin x_i }{x_i}  - \text{Ci}(x)\right)\right|_{x_i}^x
- 3 a_2 \left.\left( \frac{\cos x}{x} + \ln x \frac{\cos x_i }{x_i}  +  \text{Si}(x) \right)\right|_{x_i}^x -\sqrt{3} y_{4,i} x_i \ln \left(\frac{x}{x_i}\right) + y_{3,i}\\
\label{y4regIV} y_4(x) &= - \frac{\sqrt{3}}{x^2 x_i} \left[ a_1(x \sin x_i - x_i \sin x) + a_2 (x_i \cos x - x \cos x_i)\right]  + \frac{y_{4,i} x_i}{x}\\
\label{y5regIV} y_5(x) &= a_1 \left(\frac{\sin x - x \cos x}{x^3}\right) -a_2\left(\frac{x \sin x +  \cos x}{x^3}\right),
\end{align}
\end{subequations}
where 
\beq
\text{Ci}(x) =-\int_x^\infty \frac{\cos x'}{x'} dx' \; \; \; {\rm and } \;\;\;  \text{Si}(x) = \int_0^x \frac{\sin x'}{x'}dx'. 
\eeq
Here $a_1$ and $a_2$ are determined by initial conditions. Either one can determine them using the initial conditions of $y_1$ and $y_2$ in \eqnrefs{y1regIV} and \eqnrefbare{y2regIV}. Alternatively,  one can demand that at early times $y_5$ tends to a constant ($y_{5,i}$). For adiabatic initial conditions set up at a early time $a_i \sim 10^{-8}$, these two are equivalent and effectively give $a_1= 3 y_{5,i}$ and $a_2 =0$. In a more general case, the solution for $y_5$ contains a contribution from the homogenous part of the differential equation, \eqnref{eq5}, with an unknown constant. In this case, $a_1$, $a_2$ and the unknown constant are jointly set using the initial conditions on $y_1,y_2$ and $y_5$. 

 Note that in the radiation dominated era, $\epsilon = \eta$, $x=k\eta/\sqrt{3}$ and the solution for $y_5$ is the usual Bessel function solution for $\Phi$. The variable $y_3$ has a logarithmic dependence on $x$; this is similar to the logarithmic dependence of $\delta$ in the sub-horizon solution of \cite{hu_small-scale_1996}.

\item{\bf Region V: sub-horizon modes in the radiation-matter era}\\
In this region, the evolution for $y_3$ for most modes of interest happens to be scale-independent (i.e., does not depend on $\epsilon$). 
The remaining four variables, however, do depend on $k$. In solving for $y_3$, it is assumed that 
$y_5$ depends only on the matter variables, however, having solved for $y_1$ to $y_4$, a complete solution for $y_5$ can be constructed \eqnref{eq5II}. This `second order' solution that includes the effect of radiation on the potential is more accurate than the solution which ignores the radiation. 
 \bea
 y_1(k,a) &=& c_1 \cos I(k,a) + c_2 \sin I(k,a) + a_3 y_3(a)\\ 
y_2(k,a) &=& \sqrt{3}\left(c_1 \sin I(k,a) -c_2\cos I(k,a)\right) + b_4 y_4(k,a)\\
 y_3(a) &=& c_3 \left(x + \frac{2}{3} \right) + c_4 \left[\left(x + \frac{2}{3} \right) \log \left(\frac{\sqrt{1+x}+1}{\sqrt{1+x}-1}\right) - 2 \sqrt{1+x}\right] \; \; \; {\rm where} \; \; x = a/a_{eq,m-r} \\
 y_4(k,a) &=& -\frac{1}{\epsilon(k, a)} \left\{ c_3 x + c_4 \left[x \log \left(\frac{\sqrt{1+x}+1}{\sqrt{1+x}- 1}\right)  - \frac{2(1+3x)}{3\sqrt{1+x}} \right]  \right\}\\
y_5(k,a) &=&  \frac{1}{B(k,a)} \left[ 4 \Omega_r\left(y_1 + \frac{y_2}{\epsilon}\right) + \Omega_m \left(y_3+ \frac{3y_4}{\epsilon} \right)\right],
 \eea
where 
\beq 
I(k,a) = \int_{a_i}^{a} \frac{\epsilon(k,a')}{\sqrt{3}} d\ln a',
\eeq  
\beq
a_3 = \frac{2 \Omega_m}{B+ 3 \Omega_m}, \;\;\;b_4 = \frac{6 \Omega_m}{B + 3 \Omega_m}. 
\label{eqa3b4}
\eeq
and $c_1,c_2,c_3,c_4$ are set from initial conditions on $y_1, y_2,y_3$ and $y_4$.

\item{\bf Region VI: sub-horizon modes in the matter-dark energy era}\\
Here $y_1$ and $y_2$ and $y_5$ are the same functional form as those given above. However, $y_3$ and $y_4$ have a different solution.
\bea
y_3(a) &=& c_3 H + c_4 H \int \frac{da}{(aH)^3}
\label{y3mat}\\
y_4(a) &=& -\frac{1}{\epsilon} \left[ c_3  \frac{d H}{d\ln a}  + c_4 \left( \frac{1}{(aH)^2} + \frac{d H}{d \ln a} \int \frac{da}{(aH)^3} \right)\right].
\eea
Again $c_3$ and $c_4$ are set through the initial conditions on $y_3$ and $y_4$ respectively. 
\end{itemize}

\subsection{Numerical Comparison} 
\begin{figure}
\centering
\includegraphics[width=18cm]{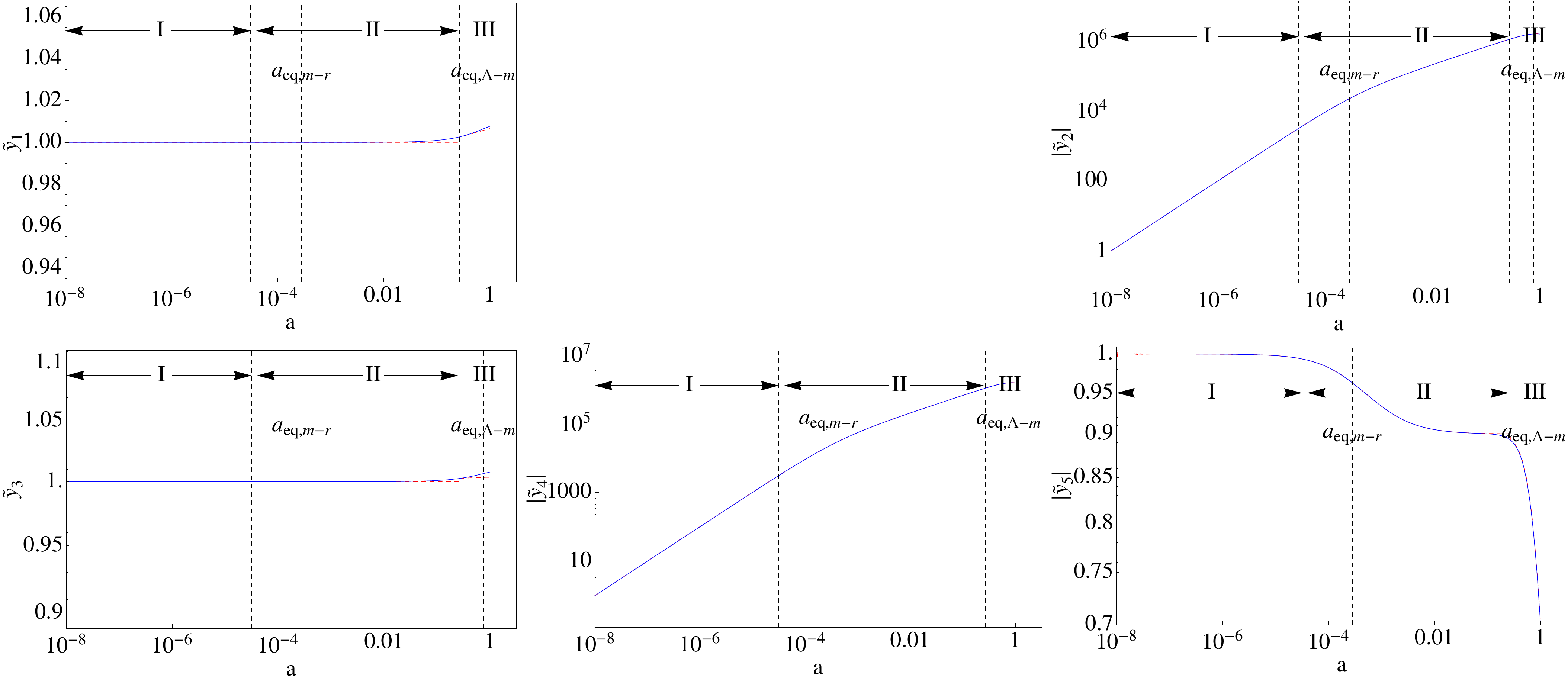}
\caption{Analytic approximations compared to numerical solutions for $k=5 \times 10^{-5}$ $h$ ${\rm Mpc}^{-1}$. This mode traverses regions I, II and III. The red dotted line corresponds to the analytic solution and the blue to the numerical one.    }
\label{kpt00005}
\end{figure}
\begin{figure}
\centering
\includegraphics[width=18cm]{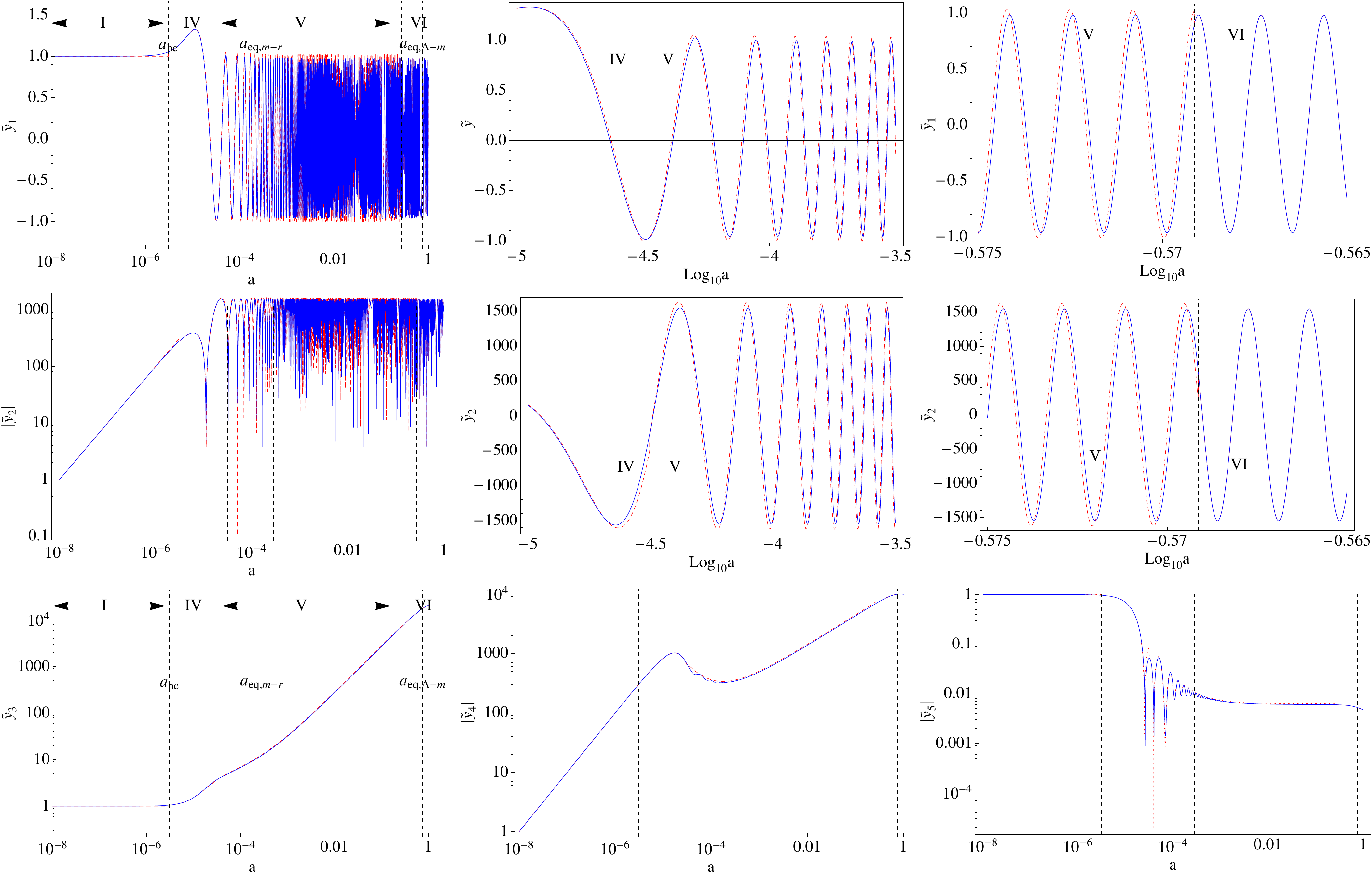}
\caption{Analytic approximations compared to numerical solutions for $k=1$ $h$ ${\rm Mpc}^{-1}$. This comparison serves as a check for regions I, IV, V and VI. The red dotted line corresponds to the analytic solution and the blue to the numerical one.   }
\label{k1}
\end{figure}
\begin{figure}
\centering
\includegraphics[width=18cm]{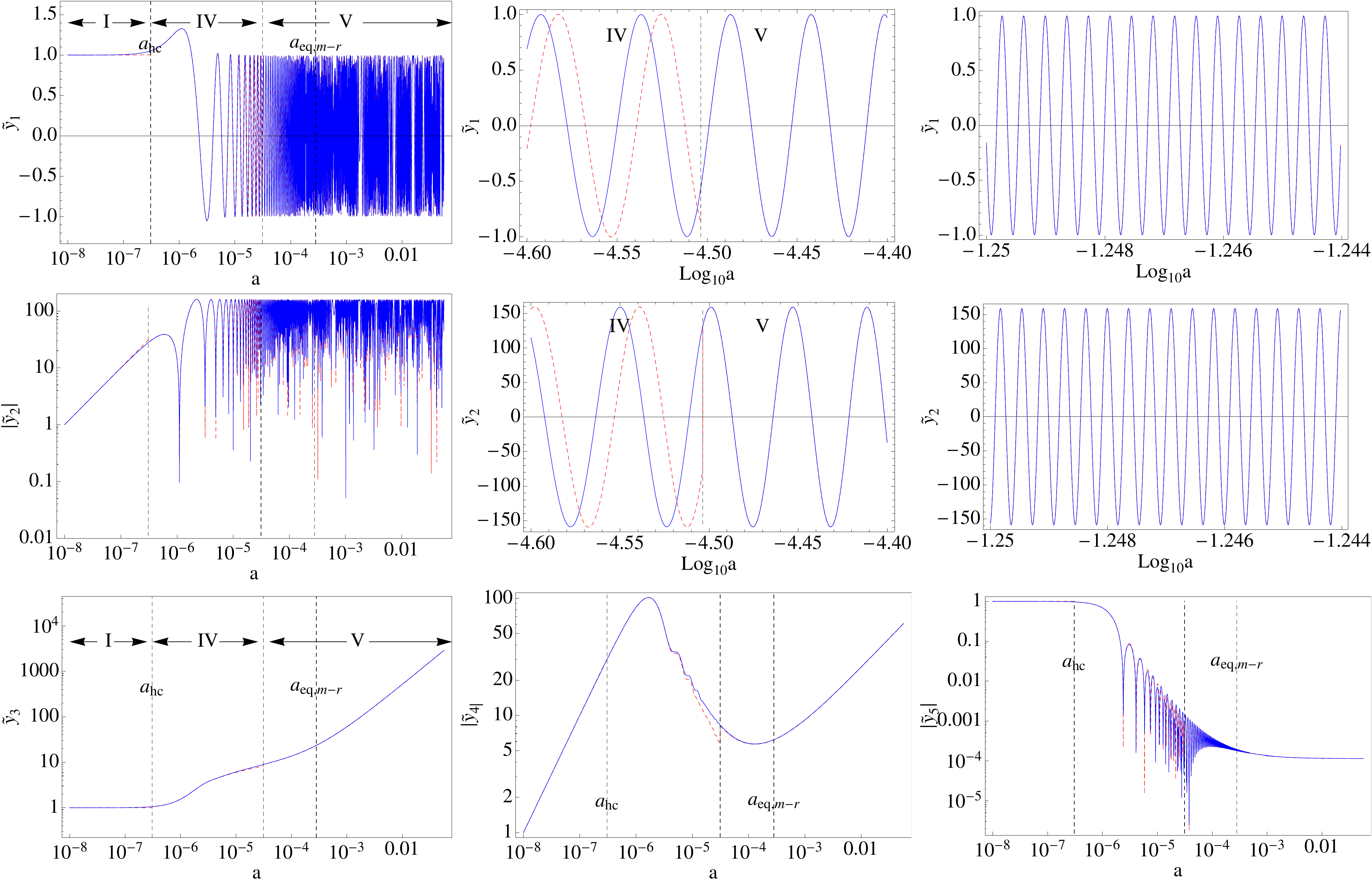}
\caption{Analytic approximations compared to numerical solutions for $k=10$ $h$ ${\rm Mpc}^{-1}$. The colour scheme is the same as that in \figref{k1}. Due to the large number of oscillations, the numerical solution breaks down at around $a \sim 0.2$ and thus the mode traverses regions I, IV, V.  }
\label{k10}
\end{figure}
We now compare the analytic forms given above with a numerical solution to the Boltzmann system. The numerical solution to \cref{eq1,eq2,eq3,eq4,eq5} was generated using an implicit Runge-Kutta solver inbuilt in the software package `Mathematica' \citep{mathematica}. The solution was evolved from $a_i=10^{-8}$ to $a_f=1$ with initial conditions given by \cref{eq1init,eq2init,eq3init,eq4init}. The value of $y_{5,i}$ was $0.01$. We considered three cases $k=5 \times 10^{-5}, 1$ and $10$ $h$ ${\rm Mpc}^{-1}$. The figures below plot the scaled variable
\beq 
{\tilde y_n}  = \frac{y_n}{y_{n,i}}, 
\eeq
where $y_n$, $n=1 \ldots 5$ is the variable of interest and $y_{n,i}$ is its initial value. 

\capfigrefs{kpt00005} shows the evolution of the mode $k= 5\times 10^{-5}$ $h$ ${\rm Mpc}^{-1}$ from $a=10^{-8}$ to $a=1$. This mode stays outside the horizon throughout its evolution until today and provides a check for the evolution in regions I, II and III. To construct the analytic solution initial conditions are needed at the starting epoch of each region. The initial conditions for region I are the same as those that the numerical solution is started with. To define the initial conditions for the other regions, the numerical solution is read off at two epochs: $a(\Omega_r=0.9)$ i.e., end of radiation domination and $a(\Omega_r=0.001)$ i.e., beginning of the matter-dark energy era. This provides initial conditions for regions II and III respectively. It can be seen from \figref{kpt00005} that the agreement between the analytic approximation and numerics is good. 

\capfigref{k1} shows the evolution of the mode $k=1$ $h$ ${\rm Mpc}^{-1}$. This mode crosses the horizon during the radiation domination era. Thus it starts in region I and covers regions IV, V and VI as it evolves. Here too the initial conditions for region I are those that the numerical solution started with. The interface between region I and IV is the epoch when the mode crosses the horizon denoted as $a_{hc}$. The numerical solution is read off at $a_{hc}$, at $a(\Omega_r=0.9)$ and at $a(\Omega_r=0.001)$ to provide initial conditions for regions IV, V and VI respectively. The agreement between the analytic answer and numerical solution is good. The first two rows of plots show the radiation variables $y_1$ and $y_2$. In each of these rows, the middle (right) plot shows a blown up interval around the transition between region IV (V) and region V (VI). 
Note that the approximation in region V worsens with time. This is because in constructing the radiation solutions in region V, we used the adiabatic approximation to obtain the coefficients $a_3$ and $b_4$ in \eqnref{eqa3b4}. This construction assumes that the parameter $\Omega_m$ is constant over the interval of integration. Although the time derivative of $\Omega_m$ is small, it is non-zero and the breakdown of this approximation causes the two solutions to deviate. For region VI, the initial conditions for the analytic approximations read-off from the numerical solutions; thus by construction the two solutions match at the junction between regions V and VI. The last row shows $y_3, y_4$ and $y_5$. Note that $y_5$ includes is constructed by using \eqnref{eq5II} and includes the contribution from the radiation variables. Thus we are able to reproduce the oscillations in $y_5$ to some extent.

\capfigref{k10} shows the evolution of the mode $k=10$ $h$ ${\rm Mpc}^{-1}$. This mode crosses the horizon during the radiation domination era, but the oscillations in the radiation variables are too large to allow evolution up until $a=1$ and the mode is evolved only until $a=0.27$. Thus it passes through regions I, IV and V. The initial conditions are set in a similar fashion to the $k=1$ mode. Note that the solution gets worse in region IV: this signals the breakdown of the radiation domination region. The natural question to ask is why this breakdown is not as prominent in the $k=1$ ${\rm Mpc}^{-1}$ case ? The reason is because in constructing the solutions, the starting point is to express $y_ 5$ only in terms of $y_1$ and $y_2$ through \eqnref{eq5II}; the $\Omega_m y_3$ and $\Omega_m y_4$ terms in this expression are ignored. When the radiation domination approximation breaks down the $\Omega_m$ term cannot be neglected, but it is weighted by $y_3$ and $y_4$. These are higher for higher values of $k$; thus the breakdown of the approximation occurs earlier for these values.

\section{Conclusions}
\label{sec:summary}
In this paper, we have investigated the mathematical properties of the E-B system using an eigenvalue analysis and computed analytical solutions in six different regimes of evolution. The main focus of our work was the dark matter power spectrum and hence it sufficed to consider a reduced set of variables. Thus, the photons are characterised only by their monopole and dipole moments $\Theta_0$ and $\Theta_1$ and dark matter is characterised by its density $\rho$ and irrotational peculiar velocity $v$ in the Newtonian gauge. The photon and matter sectors are indirectly coupled via the gravitational potential $\Phi$ and their dynamics is  dictated by the linearized E-B system. For mathematical simplicity, baryons and neutrinos were excluded from the system. 
Traditional analyses evolve the E-B system as a function of the conformal time $\eta$. Instead we chose the time variable as $\ln a$ and further simplified the system by making a change of coordinates. This allowed us to clearly define three parameters that dictate the evolution: the matter and radiation density parameters $\Omega_m$ and $\Omega_r$ and a parameter 
$\epsilon = k/(Ha)$, which is the ratio of the Hubble time to the oscillation time of a photon mode. These parameters are time-dependent, making this system non-autonomous, but we have defined appropriate `adiabatic conditions' when the parameters vary `slowly' and the system can be considered`quasi-autonomous'. 

The results from the investigation of the E-B system can be summarized in two parts. In the first part we perform an eigenvalue analysis which gives insight into various interesting properties of the system.  
\begin{enumerate}
\item {\it Onset of oscillations}: There are five eigenvalues for the system, which depend on $\epsilon$, $\Omega_m$ and $\Omega_r$. For every mode the eigenvalues  transition from being all real (four negative and one positive) to three real (two negative, one positive) and two complex (with negative real parts). The appearance of complex eigenvalues denotes the presence of oscillations. By applying Sturm's theorem and Decartes rule of sign, were able to analytically predict the transition epoch. We find that this epoch differs for each $k$; for a larger $k$ the transition occurs earlier. For $k\gtrsim 0.1$ $h$ ${\rm Mpc}^{-1}$, the transition occurs just after horizon crossing whereas for $k \lesssim 0.1$ $h$ ${\rm Mpc}^{-1}$, it occurs well before horizon crossing. We know from the analytical solutions and from the magnitude of the imaginary eigenvalue that the frequency of oscillations is of order $\epsilon$. This means that the oscillations are never visible for the super-horizon modes and that the high-frequency oscillations in sub-horizon modes occur well after the mode has crossed the horizon.

\item {\it Stability of a numerical solver}: We analyzed the stability properties of two classes of numerical schemes, namely, general Runge-Kutta methods and linear multistep methods. The stability of these schemes applied to a eigenvalue problem ${\dot y} = \lambda y$ is governed by the step size and eigenvalue. Applying the stability condition allowed us to estimate the step size given the eigenvalue. Typically, this condition imposes a maximum bound on the step size for explicit schemes and a minimum bound for implicit schemes.

\item {\it Stiffness}: In the literature, the `stiffness' of the E-B system is often attributed to the presence of baryons because the time scales of Thomson scattering are much smaller than the Hubble time.
We demonstrated that the late time regime of rapid oscillations also corresponds to the presence of two widely separated time scales making the system stiff. These rapid oscillations are present even in the absence of baryons. To better characterise stiffness, we plot the stiffness parameter defined as the ratio of the most and least negative eigenvalues (including eigenvalues with negative real parts) and find that this ratio is large for almost all modes and epochs of interest. 
\end{enumerate}

In the second part, we provide analytic solutions in six asymptotic regions of the $k-a$ plane which are defined in terms of the values of $\Omega_m$, $\Omega_r$ and $\epsilon$. Most of these limits have been discussed individually by various authors, primarily solving for the potential $\Phi$. Here we provide a comprehensive list of solutions for all five variables and give them explicitly in terms of the initial conditions. This allows an independent comparison in each region of the $k-a$ plane. The solutions for sub-horizon modes are constructed using the `adiabatic condition'. However, this condition is not satisfied by all sub-horizon modes.  Thus there is a range $k=0.1 - 0.0001$ $h$ ${\rm Mpc}^{-1}$ where solution for all five variables can be constructed only until the modes are super-horizon. We compared the analytic solutions for $k= 5 \times 10^{-5}, 1, 10$ $h$ ${\rm Mpc}^{-1}$  with the numerical solutions and found a good match. \capfigref{kaplot} summarizes our results for the different mathematical regimes of the system and the applicability
of the asymptotic analytical solutions. 

In this paper, we considered a restricted set of variables; those that influence the broad features of the dark matter power spectrum. The full system includes baryons and neutrinos and possibly other interacting species. New interactions will introduce new time scales in the problem. These may imply additional adiabatic conditions and it is conceivable that the system may not be adiabatic with respect to all parameters at the same time. Defining them appropriately will depend upon the problem at hand. Adiabaticity is an important criterion to satisfy because the stability analysis for autonomous systems is relatively simple. For non-autonomous system, using eigenvalues to analyze stability (of the system itself) or the numerical solver can sometimes give inaccurate results. The full system also consists of the whole hierarchy of multipoles resulting in a large number of perturbation variables. Numerically, computing the eigenvalues of the linear operator may be computationally intensive and cumbersome. In this work we have considered the equations only in the conformal Newtonian gauge; the equations could be recast in synchronous gauge or in terms of gauge independent variables. All these extensions potentially complicate the analysis, but the framework and results presented in this paper can still be applied to gain some insight into the mathematical structure of the system and thus help facilitate further code development and testing. 

\section{Acknowledgements}
SN would like to acknowledge the Science and Engineering Research Board (SERB) for the grant (YSS/2014/000526) and would like to thank Sayantani Bhattacharyya and Sagar Chakraborty for useful discussions. AR would like to thank Adam Amara and Lukas Gamper for useful discussions. In addition,  AR thanks the Tata Institute for Fundamental Research, where part of this work was done, for its hospitality and adjunct faculty programme support. We would like to thank Julien Lesgourgues and Antony Lewis for useful comments on the manuscript.

\appendix
\section{Useful identities}
\label{app:identities}
The three parameters are 
\beq
\epsilon = \frac{k}{Ha},  \Omega_m = \frac{\Omega_{m,0}  H_0^2a_0^3}{H^2a^3} \; \; {\rm and}  \; \; \Omega_r= \frac{\Omega_{m,0} H_0^2 a_0^4 }{H^2a^4}, 
\eeq
where 
\beq 
H^2= H_0^2\left[\frac{\Omega_{m,0} a_0^3}{a^3} +\frac{\Omega_{r,0} a_0^4}{a^4} +\Omega_{\Lambda,0}\right].
\eeq
Hence, 
\beq
2 H \frac{d H}{d a}  = \frac{H_0^2}{a} \left[-3\frac{\Omega_{m,0} a_0^3}{a^3} -4 \frac{\Omega_{r,0} a_0^4}{a^4} \right].
\eeq
Using the definitions of $\Omega_m$ and $\Omega_r$ gives
\beq
\frac{d \ln H} {d\ln a} = -\frac{1}{2} (3 \Omega_m + 4 \Omega_r). 
\label{Hderiv}
\eeq
Differentiating $\epsilon$, 
\beq
\frac{d \epsilon}{d a}  = -\frac{k}{a^2H} \left[ 1 + \frac{a}{H} \frac{dH}{da}\right].
\eeq
Substituting for the definition of $\epsilon$ and the derivative of $H$, gives 
\beq 
\frac{d \ln \epsilon}{d \ln a} = -\left[ 1 -\frac{1}{2} (3 \Omega_m + 4 \Omega_r)    \right].
\eeq
Differentiating $\Omega_m$, 
\beq 
\frac{d \Omega_m}{d a}  = -\frac{\Omega_{m,0} a_0^3}{a^3 H^2} \left[\frac{3}{a} + \frac{2}{H} \frac{d H}{da}\right]
\eeq
which, upon substituting for $dH/da$ and using the definition of $\Omega_m$ gives
\beq
\frac{d \ln \Omega_m}{d \ln a} = - [3 -(3 \Omega_m + 4 \Omega_r)].
\eeq
Similarly, the derivative for $\Omega_r$ is 
\beq
\frac{d \ln \Omega_r}{d \ln a} = - [4 -(3 \Omega_m + 4 \Omega_r)].
\eeq

\section{Sturm's theorem}
\label{app:sturm}
The characteristic polynomial of the Jacobian matrix ${\mathcal A}$ has five real roots a early epochs and three real roots at later epochs. The transition between these two structures denotes the onset of oscillations. Although the eigenvalues are not known analytically, it is possible to compute the number of real roots using Sturm's theorem and Descartes' rule of sign described below. 

Sturm's theorem \citep{collins,hook} gives the number of distinct real roots of a polynomial $p(x)$ in an interval $(a,b)$ by counting the number of changes of signs of the Sturm's sequence at the end points of the interval. Given a $n$-th order polynomial $p(x)$, the Sturm sequence is constructed as follows: 
\bea
p_0(x) &=& p(x)\\
p_1(x) &=& p'(x) \\
p_2(x) &=& -{\rm Rem}[p_0(x),p_1(x)]\\
p_3(x) &=& -{\rm Rem}[p_1(x),p_2(x)]\\
\vdots\\
0 &=& -{\rm Rem}[p_{n-1}(x),p_n(x)],
\eea
where ${\rm Rem}[p_i(x),p_{i+1}(x)]$ is the remainder of the polynomial division $p_ix)/p_{i+1}(x)$. The degree of each polynomial in the chain successively decreases and this sequence usually culminates in a constant. The minimum number of divisions is always less than or equal to the degree of the polynomial. The signs of each of these polynomials are recorded at the two end points, $a$ and $b$, in ascending order of the degree of the polynomial and the number of sign changes are noted at each end. Let this be $n_a$ and $n_b$ respectively. The number of real roots is then $|n_a-n_b|$. 

Using Sturm's theorem on the characteristic polynomial of ${\mathcal A}$, we first establish that there are two different root structures and the the transition point is evaluated by solving for the epoch at which the number of real roots changes from five (at early epochs) to three. The characteristic polynomial is 
\beq
 c_5 \lambda^5 +  c_4 \lambda^4 + c_3 \lambda^3  +  c_2 \lambda^2 +  c_1\lambda +c_0=0,
 \eeq
where
\bea
c_0 &=& \frac{\Omega_m \epsilon^4}{6} \\
c_1&=& -\frac{1}{18}\epsilon^2 \left[9 \Omega_m+ 2 (3 - 6 \Omega_r + \epsilon^2)\right]\\
c_2&=&  -\frac{1}{9} \epsilon^2 \left[6 - 6 \Omega_r + \epsilon^2\right]\\
c_3&=& -\frac{1}{6}\left[6 + 9 \Omega_m + 12 \Omega_r + 4 \epsilon^2 \right] \\
c_4&=& -\frac{1}{6} \left[9 \Omega_m + 2 (6 + 6 \Omega_r + \epsilon^2)\right] \\
c_5 &=& -1
\eea
The interval (a,b) in this case corresponds to $(-\infty, + \infty)$. This polynomial has five roots. Applying Sturm's theorem we find that there are two different root structures: at sufficiently early epochs, all roots are real; eventually, two complex roots are generated and three roots are real. The transition epoch depends upon $k$ and $a$ and is denoted as $a_{trans}(k,a)$. Sturm's theorem does not indicate the sign of the roots. The signs of the real roots can be estimated using Descartes' rule of sign. The rule states the following. Consider a single-variable polynomial ordered by descending variable exponents. Let $n$ be the number of sign changes between consecutive non-zero coefficients. Then the number of {\it positive} roots is equal to $n$ or less than $n$ by an even number. Similarly, the upper bound on the number of negative roots can be estimated by multiplying the coefficients of odd powers by minus one. Applying Descartes' rule of sign to our case, we note that the coefficients $c_2, c_3, c_4, c_5$ are always negative and $c_0$ is always positive. $c_1$ can be either positive or negative. This gives the constraint that the number of positive roots is less than or equal to one. In estimating the negative roots we change the signs of $c_1, c_3$ and $c_5$. Hence $c_3, c_5$ are always positive while $c_2, c_4$ are always negative for all epochs. $c_1$ changes sign, but in either cases, the rule gives four or two or zero negative roots at all times. We note that Descartes' rule of sign does not give the absolute number of real roots; just an upper bound. Combining with Sturm's theorem, we infer that when all five roots are real, one must be positive and four negative and when three are real, two have to be negative (all three cannot be negative by the rule of signs) and one has to be positive. 

In principle, to obtain the transition epoch analytically, one must compute the Sturm sequence, evaluate it at  $(-\infty, + \infty)$ and find the sign changes. But this is a cumbersome task. Instead we can guess the transition epoch from the rule of signs. Since $c_1$ is the only term in the coefficients that changes sign during the evolution, it is plausible that the transition epoch corresponds to the transition of the sign of this term. Thus, we postulate that the transition epoch satisfies the relation 
\beq 
(9 \Omega_m(a_{trans}) + 2 (3 - 6 \Omega_r(a_{trans}) + \epsilon(a_{trans}, k)^2)) = 0.
\eeq
We find that the analytical predictions of the transition epoch using the Descartes' rule of sign match the numerical prediction using Sturm's theorem. 

\section{Frequency of oscillations}
\label{app:freq}
 \begin{figure}
\includegraphics[width=6cm]{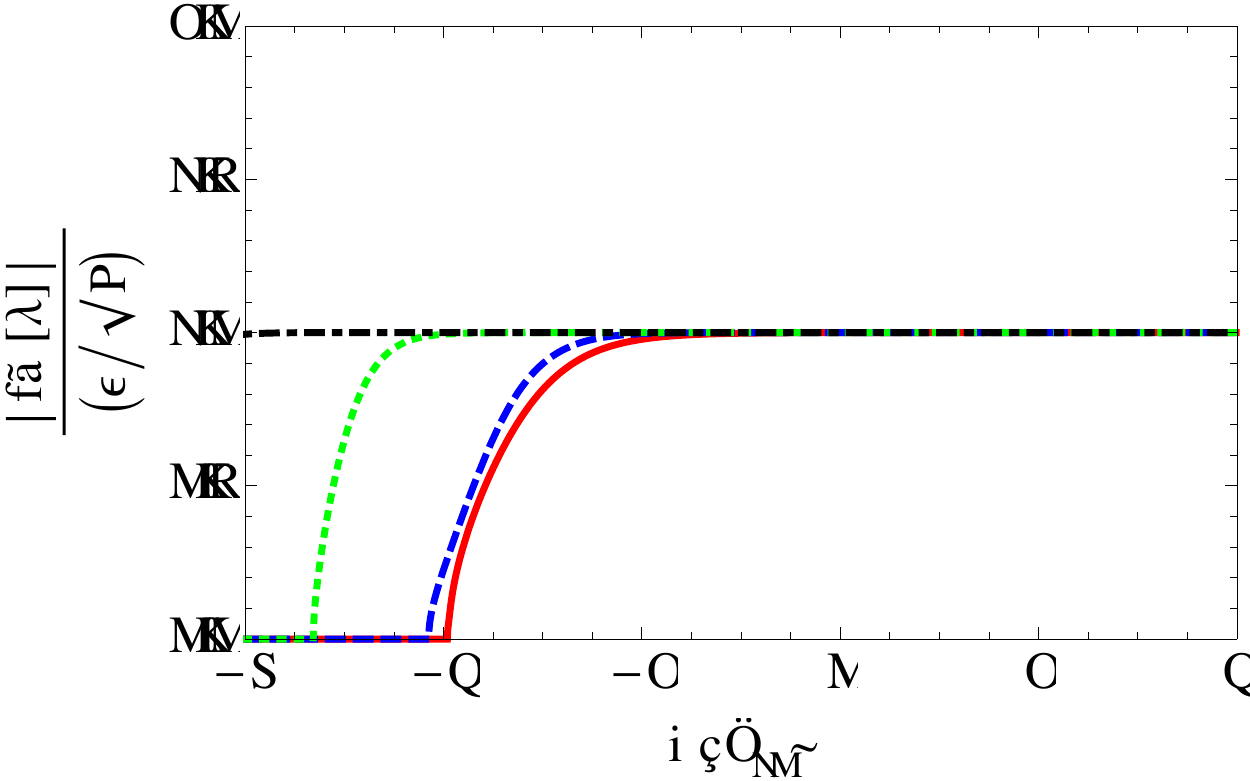}
\caption{Ratio of the imaginary part of the eigenvalue to $\epsilon(k,a)$. The colour coding is same as that of \figref{eigenvalues}. It is clear that at late times the imaginary part which determines the `instantaneous' frequency is proportional to $\epsilon$. We found that this figure does not change even when $\Omega_\Lambda$ is set to zero.  }
\label{oscfreq}
\end{figure}

\begin{figure}
\includegraphics[width=16cm]{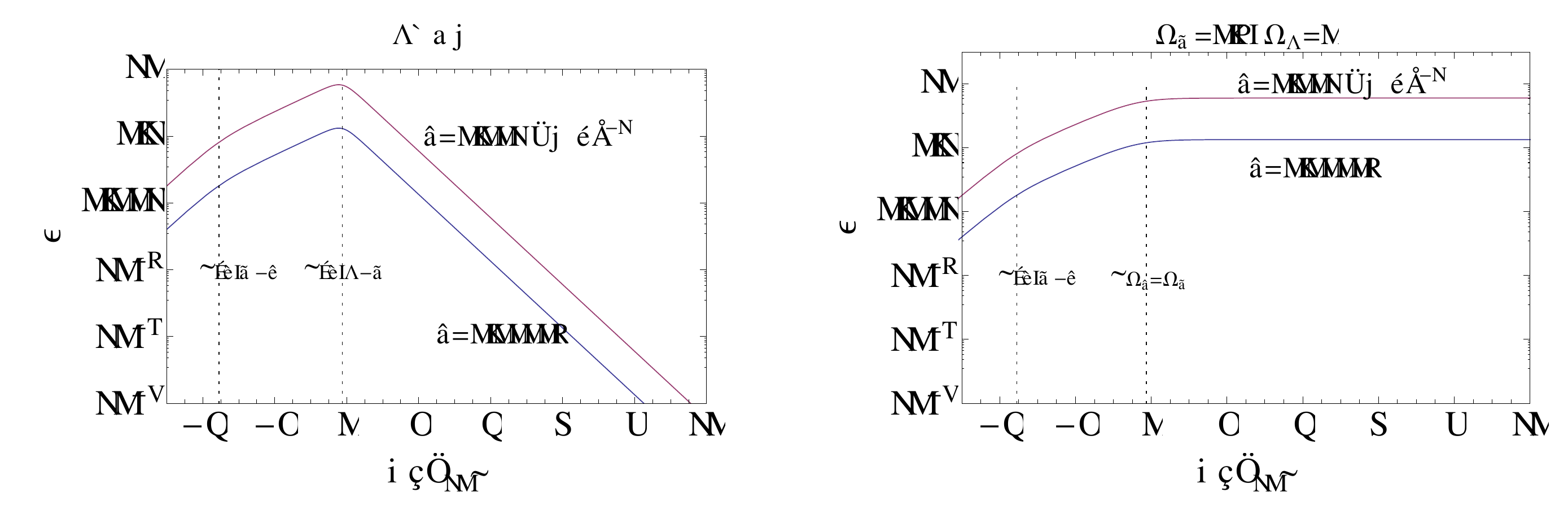}
\caption{The variation of $\epsilon$ for the $\Lambda$CDM cosmology and another open model with the same $\Omega_m$ and $\Omega_r$ as the $\Lambda$CDM model, but with $\Omega_{Lambda}=0$. For both models $k_{eq,m-r}$ = 
In the former case, $\epsilon$ stays small throughout the evolution for $k \ll k_{eq}$.}
\label{epscases}
\end{figure}

As is discussed in \S \ref{sec:onset}, Sturm's theorem predicts the transition to oscillations, by identifying the epoch when the eigenvalues of the system become imaginary. This transition epoch depends explicitly on $\Omega_m$ and $\Omega_r$ but not on $\Omega_\Lambda$. The $\Omega_\Lambda$ dependence is only through $\epsilon$. For small values $k<< k_{eq,m-r}$, the epoch occurs just before matter-radiation equality. However, as can be seen in \figref{kpt00005}, oscillations are not observed. This can be explained as follows. Sturm's theorem predicts the transition to complex eigenvalues but not their magnitude. Numerically, we find that the imaginary part of the eigenvalue which determines the `instantaneous' \footnote{instantaneous because the system is non-autonomous} frequency of oscillations is proportional to $\epsilon$ at late times (see \figref{oscfreq}).

\begin{figure}
\includegraphics[width=16cm]{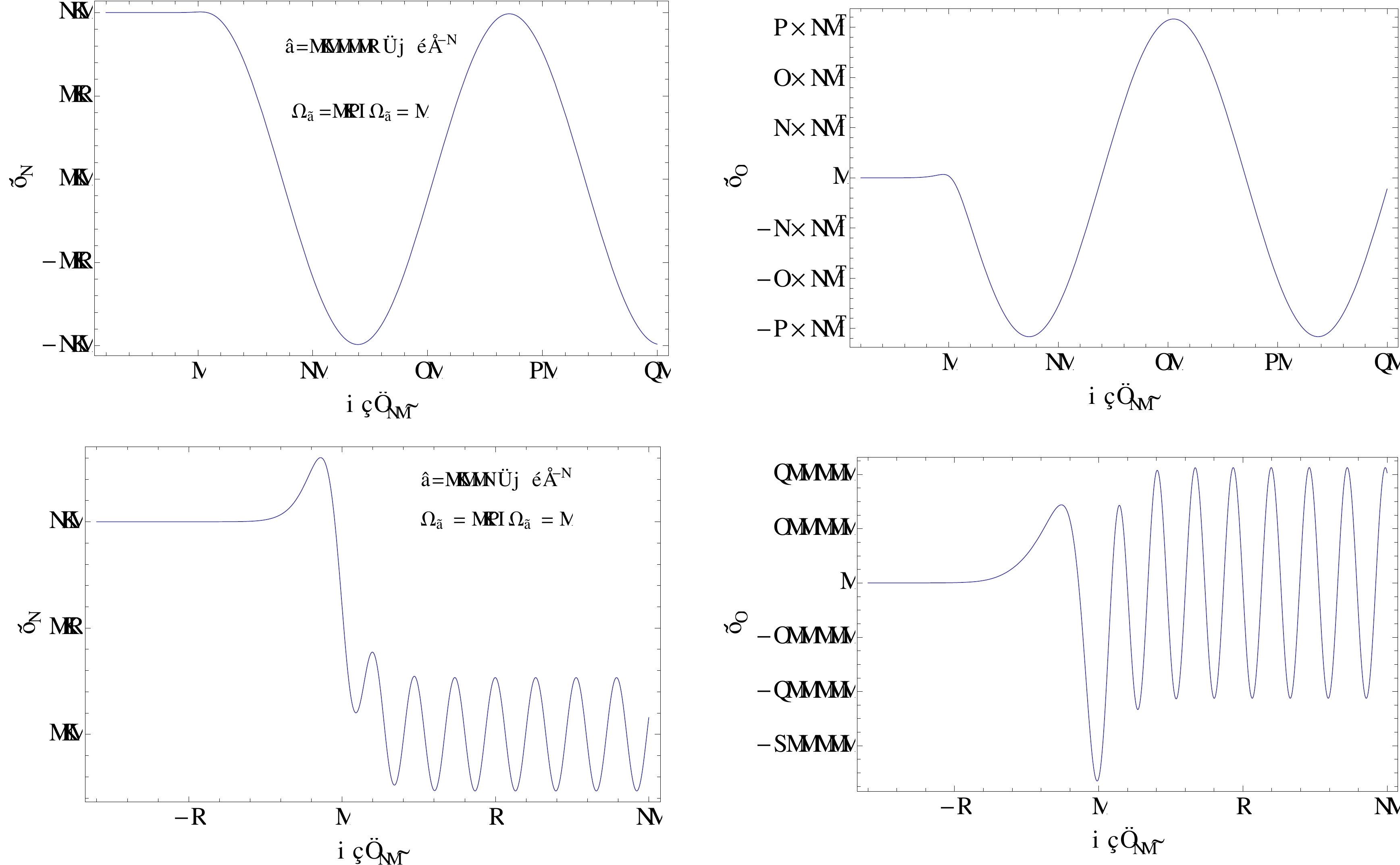}
\caption{Extended evolution for a cosmology with the same $\Omega_m$ and $\Omega_r$ as the $\Lambda$CDM case, but with $\Omega_{\Lambda}=0$. Oscillations are seen for long enough evolution times. }
\label{open}
\end{figure}

\begin{figure}
\includegraphics[width=16cm]{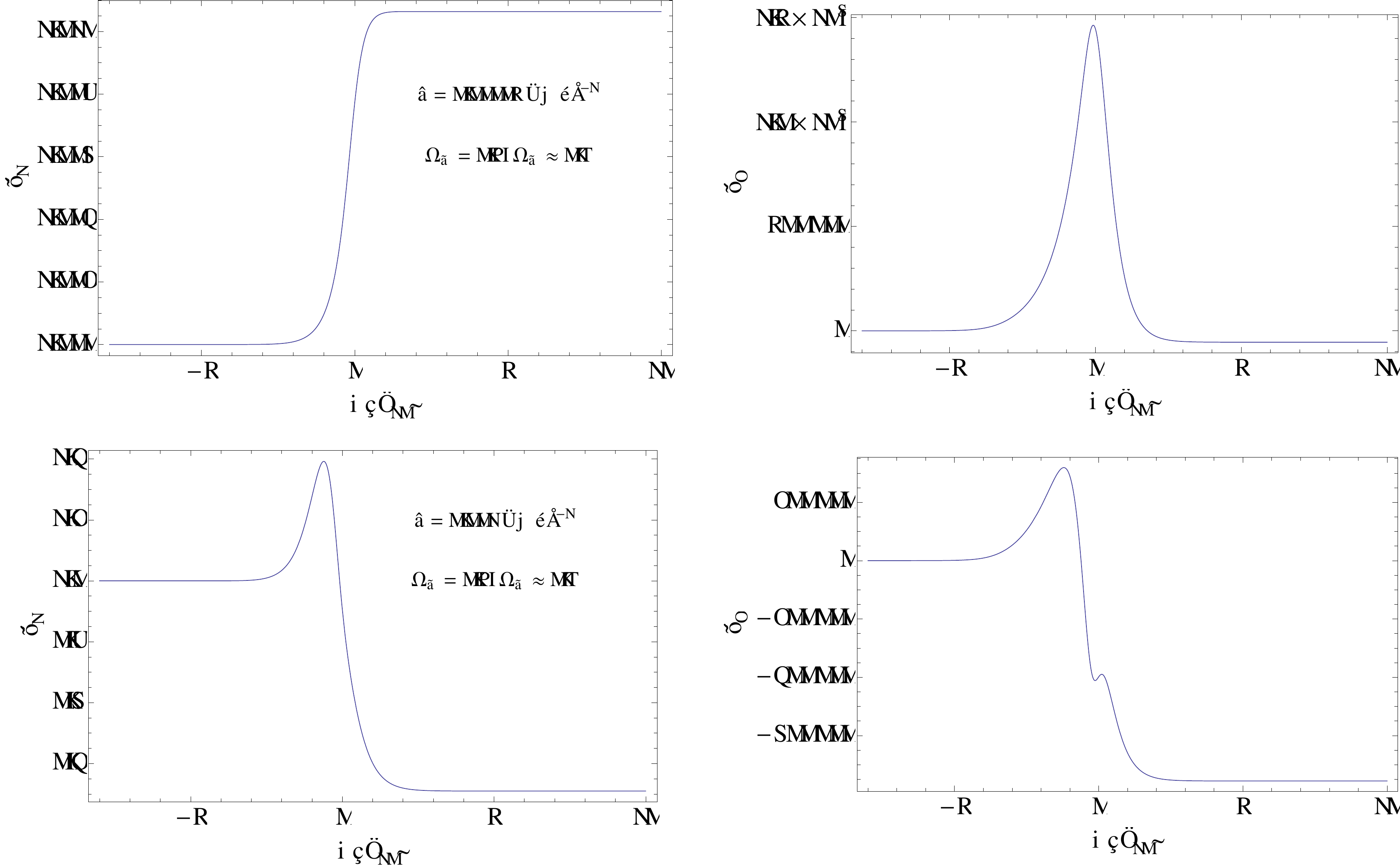}
\caption{Extended evolution for the $\Lambda$CDM cosmology. The `instantaneous' frequency is too small for noticeable oscillations. }
\label{flat}
\end{figure}

In the $\Lambda$CDM cosmology, $\epsilon$ starts small, reaches a maximum at $a=1$ and drops very sharply after the epoch of dark energy-matter equality $a_{eq, \Lambda-m}$ (see the left panel of \figref{epscases}). For small values of $k$ ($k\ll k_{eq,m-r}$), $\epsilon$ stays small throughout the evolution. For example $k = 5 \times 10^{-5}$ $h$ ${\rm Mpc}^{-1}$, $\epsilon \sim 0.1$ at $a \sim1$ and drops thereafter. Thus, no oscillations are visible because the oscillation frequency is very small. In roughly 8 e-folds (from $a \sim a_{eq,m-r}$ to $a \sim 1$) one expects to see around $8 \times Im[\lambda] $ oscillations. For $Im[\lambda] \sim 0.5 \epsilon $ and  $\epsilon \sim 0.1$ this corresponds to about half a oscillation. To understand this issue better we contrast it with another cosmology with the same $\Omega_m$ and $\Omega_r$ as $\Lambda$CDM but $\Omega_\Lambda=0$. These two cosmologies have practically the same $k_{eq,m-r}$ and Sturm's theorem predicts similar transition epochs. However, the $\epsilon$ variation is different. In the latter, $\epsilon$ tends to a constant because the curvature dominates (see right panel of \figref{epscases}). Thus, oscillations are seen for long enough evolution times. This is illustrated in \figrefs{open} (open cosmology with $\Omega_\Lambda=0$) and \figrefbare{flat} ($\Lambda$CDM). In each figure, the two rows correspond to two different values of $k$; each row shows the evolution of $y_1$ and $y_2$ for extended evolution times. Oscillations are clearly visible in the open cosmology. The oscillation frequency is higher for a higher value of $k$ since $\epsilon$ is higher \footnote{We find that \figref{oscfreq}, which shows that oscillation frequency is proportional to $\epsilon$ remains unchanged for the open case}. \capfigref{flat} shows that, for the $\Lambda$CDM case, oscillations cannot be sustained. We note that the oscillations in the system are primarily in the $y_1$ and $y_2$ (radiation) sector: since $\Omega_r$ is small after the transition epoch, these variables couple very weakly to the potential $y_5$ and there are no oscillations visible in the matter sector. Also, from a practical perspective, there are no visible oscillations before $a=1$ in any of the cases: again this is because the oscillation frequency is small in this domain.

\section{Stability of numerical schemes}
\label{app:stability}
Consider the general differential equation 
\beq 
\frac{d y}{d x} = f(x,y[x]),  
\label{testeq}
\eeq
with a given initial condition $y[x_0] = y_0$. If $f(x,y[x]) \equiv f(y[x])$, i.e., $f$ has no explicit time dependence, then the dynamical system is autonomous, else is it a non-autonomous system. To solve the equation numerically, the domain is discretized into finite number of grid points $\{x_0, x_1, \ldots\}$. The separation between the grid points gives the step size $h_n = x_n - x_{n-1}$. 
The values of the function at any point are given by $y({x_n}) = y_n$ and the derivatives are denoted as $f_n$.  
The simplest numerical scheme to solve this system is the explicit (or forward) Euler's method. In this scheme, 
\beq 
y_n = y_{n-1} + h f_{n-1}, \; \; \; n=1, 2 \ldots
\label{forwardEuler}
\eeq 
Although computationally simple, this scheme is not very stable and there are various extensions possible: (a) perform more number of computations (stages) within a step (RK schemes), (b) keep higher order derivatives in the Taylor expansion (Taylor series solutions), (c) use more past values in computing $y_n$ (multistep scheme) (d) use higher order derivatives of $f$ (Rosenbrock methods). See \cite{butcher}, section 214, for a nice schematic diagram showing these extensions. In this paper, we will consider (a) general RK schemes and (c) linear multistep schemes.

\subsection{Runge-Kutta (RK) schemes}
The commonly used 4th-order \footnote{An RK method has order $p$ if the local truncation error between the true solution and the approximation scales as $h^{p+1}$ (i.e., $|y(x_0 + h) - y_1|\leq K h^{p+1}$).} RK scheme is of the form 
\begin{subequations}
\label{rk4}
\begin{align}
y_n &= y_{n-1} + h\left(\frac{1}{6}k_1 + \frac{1}{3} k_2 + \frac{1}{3} k_3 + \frac{1}{6} k_4\right),\\
{\rm where} \; \; 
k_1 &= f(x_{n-1},y_{n-1})\\
k_2 &= f(x_{n-1}+ \frac{h}{2}, y_{n-1} + \frac{h}{2} k_1)\\
k_3 &= f(x_{n-1}+ \frac{h}{2}, y_{n-1} + \frac{h}{2} k_2)\\
k_4 &= f(x_{n-1} + h, y_{n-1} + h k_3). 
\end{align}
\end{subequations}
The forms listed above are explicit in the sense that the step at $y_n$ can be computed completely from the knowledge of the function at all points upto $y_{n-1}$. 
This scheme has four stages and it can be shown that the error between the numerical solution after $n$ steps ($y_n$) and the exact solution at the position of the $n$-th step ($y_{exact}[x_n])$) scales as ${\mathcal O}(h^4)$ (hence fourth order). This is an explicit scheme because the solution for the function after $n$ steps $y_n$ depends only on the solution at earlier time steps.
A general explicit RK scheme with $s$ stages has the form: 
\beq
y_n = y_{n-1} + h \sum_{i=1}^{s} b_i k_i,
\eeq
where 
\beq 
k_i = f(x_{n-1} + c_i h, y_{n-1} + h \sum_{j<i} a_{ij} k_j). 
\eeq
Thus, the method is completely characterised by the matrix $a_{ij}$ (called the Runge-Kutta matrix), the numbers $b_i$ (called the weights) and the numbers $c_i$ (called the nodes). This is usually represented in the form of a table called the {\it Butcher tableau}. 
\beq
\begin{array}{c|ccccc}
0 & 0 & 0 &0 &\cdots  & 0 \\
c_2 & a_{21}& 0 &0 & \cdots & 0 \\
c_3 & a_{31} & a_{32} & 0 &\cdots & 0 \\
\vdots & \vdots & \vdots & \ddots& \cdots & 0 \\
c_2 & a_{s1}& a_{s2}& \cdots & a_{s s-1}& 0 \\
\hline
& b_1&b_2& b_3 & \cdots & b_s 
\end{array}
\equiv 
\begin{array}{c|c}
{\bf c} & {\bf A} \\
\hline
& {\bf b^T}
\end{array}
\eeq
The Butcher tableau of the forward Euler scheme is
\beq 
\begin{array}{c|c}
0&0\\
\hline
&1
\end{array}
\eeq
The Butcher tableau of the fourth-order RK integration is 
\beq
\begin{array}{c|cccc}
0&0&0&0&0\\
\frac{1}{2}&\frac{1}{2}&0&0&0\\
\frac{1}{2}&0&\frac{1}{2} &0&0\\
1&0&0&1&0\\
\hline
&\frac{1}{6}&\frac{1}{3}&\frac{1}{3}&\frac{1}{6}
\end{array}
\eeq

\subsubsection{Implicit Methods}
Explicit schemes are simple to implement, but often unstable and implicit schemes need to be employed. An implicit method usually involves solving a functional equation for $y_n$ at each time step. Although they are generally computationally intensive implicit schemes are generally more stable.  

The backward Euler scheme has the form 
\beq 
y_n = y_{n-1} + h f(x_n, y_n). 
\label{backeul}
\eeq 
and a general implicit RK method has the form 
\beq
y_n = y_{n-1} + h \sum_{i=1}^{s} b_i k_i,
\eeq
where 
\beq 
k_i = f(x_{n-1} + c_i h, y_{n-1} + h \sum_{j=i}^s a_{ij} k_j). 
\eeq
Note that the sum is not restricted to $j<i$ as was in the explicit case. Thus, the matrix $A$ in the Butcher tableau is no longer a lower triangular matrix and in general all entries are non-zero. 
The Butcher tableau for a implicit (backward) Euler scheme is 
\beq 
\begin{array}{c|c}
1&1\\
\hline
&1
\end{array}
\eeq
and for the implicit mid-point scheme given by 
\beq 
y_1 = y_0 + h f(x_0 + \frac{h}{2}, y_0 + \frac{h}{2} k_1)
\eeq
it is 
\beq 
\begin{array}{c|c}
1/2&1/2\\
\hline
&1
\end{array}
\eeq

\subsubsection{Stability of a RK scheme}
The stability of a numerical scheme is loosely defined as its ability to reproduce the qualitative behaviour of the exact analytic solution. For example, if the analytic solution converges to zero, it is expected that the numerical solution does the same. Consider the solution to the one-dimensional {\it linear autonomous} differential equation 
\beq 
\frac{dy}{dx} = \lambda y. 
\eeq
If $x_0$ is the starting point, the exact analytic solution at the $n+1$-th step is 
\beq 
y(x_0 + nh) = e^{\lambda nh} y(x_0). 
\eeq
The exact analytic solution is bounded if and only if $|e^{\lambda h}| \leq 1$, which is equivalent to ${\mathcal Re}(\lambda h ) \leq 0$. Since $h$ is always positive, this is equivalent to stating that the solution is bounded if and only if ${\mathcal Re}(\lambda) \leq 0$. Consider the numerical solution for the same system using the explicit Euler scheme. After the $n$-th step, the approximate solution is  
\beq 
y_{n} =  (1+ h \lambda)^n y_{0}. 
\eeq
For the numerical solution, the boundedness condition translates to $|1+\lambda h| \leq 1$. Thus, given the eigenvalue, this condition gives a minimum step size. The function $r(z) = 1+ z$ where $z= \lambda h$ is the stability function of the forward Euler scheme.  
For a multi-dimensional system ${\dot {\bf y}} = {\mathcal H} {\bf y}$, we can assume without loss of generality that ${\mathcal H}$ is diagonal and perform the analysis in the eigenbasis \citep{butcher}. The limits on the step size can be converted to limits on step sizes in a different basis using the basis transformation.

For a general RK scheme applied to autonomous differential equations, the stability function is (\citealt{butcher, harrier2}) 
\beq 
r(z) = 1 + z b^T(I - z A)^{-1} e = \frac{\det(I-zA+zeb^T)}{\det(I-zA)}, 
\label{stabfunc}
\eeq
where, $z = \lambda h$, $e = \{1, 1,1, \ldots\}^T$ is a s-dimensional vector and $b^T, A$ are to be read off from the Butcher tableau of the method. The stability condition is given by 
\beq 
|r(z)|\leq1. 
\eeq
The stability functions for some commonly used RK methods are given in table \ref{stabtab} in the text. 
The stability function for explicit methods is always a polynomial whereas for implicit methods it is always a ratio of rational functions of $z$. A more exhaustive list for implicit methods can be found in \cite{harrier2}, page 42. To compute the allowed step size $h$, one has to solve the equation 
$|r(z= h \lambda)| = 1$. 

\subsubsection{Non-Autonomous systems}
For a {\it linear non-autonomous} system, the stability function is generalized. For each stage at the $n$-th step of a RK method, define 
\bea
z_i &=& h \lambda(x_{n-1} + h c_i) \\
Z &=& {\rm diag}(z_1, z_2, \ldots z_s)\\
\label{stabfunc2} R(Z) &=& 1 + b^T Z(1-AZ)^{-1} e.
\eea 
 A RK method applied to a non-autonomous system is stable if, for all $z_1, z_2, \ldots, z_s$, such that $det(I-AZ) \neq0$, 
\beq 
|R(Z)| \leq 1. 
\eeq

When the system is autonomous, all $z_i$ are identical and $r(z) = R(z I)$, where $I$ is the identity matrix. For the Euler scheme, there is only one stage so $Z= z$ and $R(z) = r(z)$. The conditions to be solved for stability are 
\begin{enumerate}
\item  Explicit Euler scheme 
\beq 
|R(h,x)| = |1+h \lambda(x + h)| =1
\eeq
\item Implicit Euler scheme 
\beq 
|R(h,x)| = \left|\frac{1}{1- h\lambda(x+h)}\right|=1. 
\eeq
\item Implicit midpoint scheme
\beq
|R(h,x) | = \left|\frac{2+ h \lambda(x + h/2)}{2-h\lambda(x+h/2)}\right|=1
\eeq
\item 
Explicit RK-4 
\beq
R(h,x) = 1 + \frac{z_1}{6} + \frac{z_2}{3} + \frac{z_3}{3} + \frac{z_4}{6} + \frac{z_1z_2}{6} +  \frac{z_2 z_3}{6} +  \frac{z_3 z_4}{6} + \frac{z_1z_2 z_3}{12}  + \frac{z_2 z_3 z_4}{12} + \frac{z_1 z_2 z_3 z_4}{24}, 
\eeq
where the $z_i$s are functions of the eigenvalues evaluated at the sub-grid points defined by the nodes ($c_i$ values in the Butcher tableau): 
\bea
z_1 &=& h \lambda(x)\\
z_2 &=& h \lambda(x + \frac{h}{2})\\
z_3 &=& h \lambda(x + \frac{h}{2})\\
z_4 &=& h \lambda(x + h). 
\eea
\end{enumerate}
Thus, to solve for the step size along each eigendirection, one has to solve $|R(h,x)|=1$ at each point in the domain for the corresponding eigenvalue.

\subsection{Linear multistep methods}
Another important family of methods are the linear multistep methods. Here the $n$-th step can depend upon previous steps. There are two sub-families of these methods: Adams methods and Backward Differentiation Formula (BDF) methods.  
\begin{enumerate}
\item {\it Adams method}: The $k$-step explicit Adams method (also called Adams-Bashforth methods) is of the form 
\beq 
y_n = y_{n-1} + h \sum_{j=1}^k \beta_i f_{n-j}, 
\eeq
where 
\bea
\nonumber\beta_j &=& (-1)^{j-1} \sum_{i=j-1}^{k-1} \binom{l}{j-1} \gamma_l\\
\nonumber \gamma_l&=& (-1)^l \int_0^1 \binom{-s}{l} ds
\eea
The forward Euler method given by \eqnref{forwardEuler} corresponds to a Adams-Bashforth method with $k=1$ and $\beta =1$. 
The $k$-step implicit Adams method (also called Adams-Moulton) method has the form 
\beq 
y_n = y_{n-1} + h \sum_{j=0}^k \beta_i f_{n-j}, 
\eeq
The backward Euler method given by \eqnref{backeul}  corresponds to $k=1$ and $\beta_1=1$. 
The implicit trapezoidal rule given by 
\beq
y_{n} = y_{n-1} + \frac{h}{2} (f_n+ f_{n-1})
\eeq
corresponds to $k=1$ with $\beta_0 =\beta_1=\frac{1}{2}$. 

\item {\it Backward Differentiation Formula (BDF) method}: A $k$-step BDF method is given by 
\beq 
\alpha_0 y_n + \alpha_1 y_{n-1} + \ldots + \alpha_k y_{n-k} = h \beta_0 f_{n}, 
\eeq
where $\alpha_i$s and $\beta_0$ are derived by requiring that the error scale with a given order (order conditions). 
BDF1 is a first order method with $\alpha_0=1$, $\alpha_1=-1$ and $\beta_0=1$ and corresponds to the backward Euler scheme. 
BDF2 is a second order method which gives
\beq
y_n= \frac{4}{3} y_{n-1} -\frac{1}{3} y_{n-2} + \frac{2}{3}h f_n. 
\eeq
\end{enumerate}
Note that `linear' here refers to the method; in general, $f(y_n)$ could be a non-linear function of $y_n$.  
\subsubsection{Stability} 
A general linear $k$-step method has the general form \citep{petzold}
\beq 
\alpha_0 y_n + \alpha_1 y_{n-1} + \alpha_2 y_{n-2} + \ldots = h (\beta_0 f_n + \beta_1 f_{n-1} + \beta_2 f_{n-2} + \ldots).
\eeq
The choice $\alpha_0=1$ sets the overall scaling. For all Adams methods, $\alpha_j =0$ for $j>1$ and for all BDF methods, $\beta_j=0$ for $j>0$. 
Suppose $f_n=f(y_n) = \lambda y_n$; i.e., $\lambda$ is the eigenvalue of the 1D problem ${\dot y}= \lambda y$. This form then becomes 
\beq 
\sum_{k=0}^j (\alpha_j-h \lambda \beta_j) y_{n-j} =0. 
\eeq
This is a homogenous constant coefficient difference equation whose solutions can be expanded in terms of the roots $\xi_i$ of the polynomial 
\beq
\phi(\xi) = \sum_{k=0}^j \alpha_j \xi^{n-j} - h\lambda \sum_{k=0}^j \beta_j \xi^{n-j}.
\eeq
Usually the first sum is denoted as $\rho(\xi)$ and the second as $\sigma(\xi)$ so that $\phi(\xi) = \rho(\xi) - h \lambda \sigma(\xi)$. From the theory of difference equations, the solutions are stable if and only if the roots of the equation $\phi(\xi)=0$ satisfy 
\beq 
|\xi_{root}(z)|\leq1 \;\;\; {\rm for \; all \;roots},
\eeq
where $z = \lambda h$. They are absolutely stable if and only if $|\xi_{root}(z)| <1$. The stability polynomials for some methods are computed below. 
The stability is dictated by the behaviour in the complex $z$ plane. 
\begin{enumerate}
\item Forward Euler ($k=1$ explicit Adams) has $\alpha_0 =1, \beta_0=0, \alpha_1 = -1, \beta_1=1$. The stability polynomial is 
\beq
\phi(\xi) = 1 -(1+z) \xi^{-1}, 
\eeq
whose root is 
\beq 
\xi_{root}(z)  = 1+z.
\eeq
\item Backward Euler ($k=1$ implicit Adams, first order or BDF1) has $\alpha_0=1, \beta_0=1, \alpha_1 = -1, \beta_1=0$. The stability polynomial is 
\beq 
\phi(\xi)=(1-z)  -\xi^{-1},
\eeq
whose root is 
\beq
\xi_{root}(z) = \frac{1}{(1-z)}.
\eeq
\item Trapezoidal Rule ($k=1$ implicit Adams, second order) has  $\alpha_0=1, \beta_0 = 1/2, \alpha_1 = -1, \beta_1 =1/2$. The characteristic polynomial becomes 
\beq 
\phi(\xi)= (1-\frac{z}{2}) + (-1-\frac{z}{2})  \xi^{-1}, 
\eeq
whose root is 
\beq 
\xi_{root}(z) = \frac{2 + z }{2-z} 
\eeq 
\item BDF2 scheme has  $\alpha_0 = 1, \alpha_1 =-\frac{4}{3}, \alpha_2 = \frac{1}{3}, \beta_0 = \frac{2}{3}$. The stability polynomial becomes  
\beq
\phi(\xi) = \left(1-\frac{2 z}{3}\right) - \frac{4}{3}\frac{1}{\xi} + \frac{1}{3 \xi^2},
\eeq
whose roots are
\beq
\xi_{root}(z) = \frac{2 \pm \sqrt{1+ 2z}}{3-2z}. 
\eeq
The condition $|\xi_{root}|\leq1$ corresponding to root with the negative sign gives stability for all $z$. Thus, the constraining condition is given by the root corresponding to the positive sign.
\item BDF3 scheme has $\alpha_0 = 1, \alpha_1 =-\frac{18}{11}, \alpha_2 = \frac{9}{11}, \alpha_3 = -\frac{2}{11},  \beta_0 = \frac{6}{11}$. The stability polynomial becomes 
\beq
\phi(\xi)=(11-6z) \xi^3 -18 \xi^2 + 9 \xi + 2. 
\eeq
There are three roots. Numerically, we find that the complex roots always satisfy the stability condition. The constraining condition is given by the real root: 
\beq 
\xi_{root}(z) = \frac{6}{11-6z} + \frac{27 - 162 z}{9 (11-6z)f(z)} +\frac{ f(z)}{11-6z}, 
\label{BDF3formula}
\eeq
where $f(z) = (40 + 30 z + 36 z^2 + \sqrt{1573 + 1914 z + 864 z^2 - 3672 z^3 + 1296 z^4})^{1/3}$. In the text we use the notation $r(z) = \xi_{root}(z)$ to denote the stability function. 
\end{enumerate}

\capfigref{numstab} in the text shows the regions of stability in the complex $z$ plane for some popular methods. Generally, implicit methods are more stable than explicit methods. For the backward Euler and the BDF2 and 3 schemes, the stability criterion sets a lower bound on the step size for a positive eigenvalue. For example with a BDF2 scheme, $h_{min} = 4/\lambda$. A smaller step size will imply instability.  
For the BDF schemes, the size of the domain where the scheme is unstable increases with order. Thus, a higher order BDF scheme will converge faster, but the step size will have a greater restriction. A trade-off between stability and efficiency may be required while employing such schemes. 

\section{Detailed derivations of analytic forms in various limits}
\label{app:limits}
The system to solve is given by \eqnref{newsystem} given arbitrary initial conditions $y_{1,i}$ to $y_{5,i}$. We consider two limits: $\epsilon \ll1$ giving the super-horizon solutions and $\epsilon \gtrsim 1$ giving the sub-horizon solutions. Within each class we consider three epochs: radiation domination, when $\Omega_r\gg \Omega_m$, matter-radiation era when $\Omega_m$ and $\Omega_r$ are both non-zero and dark energy-matter era, when 
$\Omega_r\ll1$. This defines six separate regions where analytic forms can be obtained.  

\subsection{Super-horizon modes: $\epsilon \ll1$}
In the limit that $\epsilon\ll1$, neglecting all terms of order $\epsilon$ or higher, gives 
\beq
{\dot y}_1 =0; \;\; {\dot y}_2 =0;  \; \; {\dot y}_3 =0; \;\; {\dot y}_4 = -y_4 \;\;
\eeq
and 
\beq 
{\dot y}_5 = \frac{1}{2} \left[\Omega_m y_3 + 4 \Omega_r y_1 -(3 \Omega_m + 4 \Omega_r + 2) y_5\right]. 
\label{y5eqsuper} 
\eeq
The solutions to the first four variables are 
\beq 
y_1 = y_{1, i}; \; \; y_2= y_{2, i}; \; \; y_3 = y_{3, i}; \; \; y_4 = \frac{y_{4,i} a_i}{a}. 
\label{superhorifirst}
\eeq
 Substituting for $y_1$ and $y_3$ in \eqnref{y5eqsuper} gives 
\beq 
{\dot y}_5 = \frac{1}{2} \left[\Omega_m y_{3,i} + 4 \Omega_r y_{1,i} -(3 \Omega_m + 4 \Omega_r + 2) y_5\right]. 
\label{y5eqsuper1} 
\eeq

\subsubsection{Regions I and II: \emph{super-horizon modes in the radiation domination and radiation-matter eras}}
The solutions for both these regions can be obtained simultaneously. We solve \eqnref{y5eqsuper1} using an integrating factor $s$ defined as 
\beq 
{\dot s} = s \frac{3 \Omega_m +4 \Omega_r + 2}{2}. 
\label{eqnfors}
\eeq
In terms of this factor \eqnref{y5eqsuper1} becomes 
\beq 
\frac{d (y_5 \cdot s)}{d \ln a} = \frac{\Omega_m y_{3,i} + 4 \Omega_r y_{1,i}}{2} \cdot s. 
\eeq
The solution is \footnote{To prevent cumbersome notation, we have avoided using dummy indices under the integral sign. } 
\beq 
y_5 s= \int_{a_i}^a s \cdot \left(\frac{\Omega_m y_{3,i} + 4 \Omega_r y_{1,i}}{2}\right) d\ln a + c, 
\label{y5eqsuper2}
\eeq
with $c= y_{5,i} s_i$ where $s_i = s(a_i)$. 
To proceed further, one must solve for the integrating factor. From \eqnref{eqnfors} the solution for $s$ is 
\beq 
\ln s(a) =  \int \left(\frac{3 \Omega_m +4 \Omega_r + 2}{2} \right) d \ln a
\eeq
Using the derivative of $\Omega_m$ from \eqnref{omder} the integrand can be written as 
\beq 
\frac{3 \Omega_m +4 \Omega_r + 2}{2} = \frac{1}{2} \frac{ d \ln \Omega_m}{d \ln a} + \frac{5}{2}.  
\eeq
This gives 
\beq 
s(a) = \Omega_m^{1/2} a^{1/2}. 
\eeq 
Substituting for $s$ in \eqnref{y5eqsuper2} gives 
\beq 
a^{5/2} \Omega_m^{1/2} y_5 = \int_{a_i}^a a^{5/2} \Omega_m^{1/2} \cdot \left(\frac{\Omega_m y_{3,i} + 4 \Omega_r y_{1,i}}{2}\right) d \ln a  +  y_{5,i} \Omega_{m,i}^{1/2} a_i^{5/2}.
\label{y5eqsuper3}
\eeq
Now, define 
\beq 
x= \frac{a}{a_{eq,m-r}},  
\label{xaeq1}
\eeq
where $a_{eq,m-r} = \frac{\Omega_{r,0}}{\Omega_{m,0}}$. In the radiation-matter regime, 
\beq 
H^2 = H_0^2(\Omega_{m,0}a^{-3 }+ \Omega_{r,0} a^{-4}), 
\eeq
and using \eqnref{eq:omor}, the parameters in terms of $x$ are 
\beq
\Omega_r = \frac{1}{1+x}; \; \; \; \Omega_m = \frac{x}{1+x}.
\label{omegasinxaeq1}
\eeq
In terms of $x$ \eqnref{y5eqsuper3} becomes 
\bea
\frac{x^3}{\sqrt{1+x}} \cdot y_5 &=& \int_{x_i}^x \frac{x^3}{\sqrt{1+x}}  \left(\frac{y_{3,i}}{2}  \frac{x}{1+x} +  \frac{2 y_{1,i}}{1+x}  \right) d \ln x + \frac{y_{5,i}x_i^3}{\sqrt{1+x_i}}\\
&=& \frac{y_{3,i}}{2} \int_{x_i}^x \frac{x^3}{(1+x)^{3/2}}dx + 2y_{1,i} \int_{x_i}^x \frac{x^2}{(1+x)^{3/2}} dx + \frac{y_{5,i}x_i^3}{\sqrt{1+x_i}}\\
&=& \frac{y_{3,i}}{5} \left.\left\{ \frac{16+8x-2x^2+x^3}{\sqrt{1+x}}\right\}\right|_{x,i}^x + \frac{4 y_{1,i}}{3} \left.\left\{\frac{-8-4x+x^2}{\sqrt{1+x}} \right\}\right|_{x,i}^x + \frac{y_{5,i}x_i^3}{\sqrt{1+x_i}}
\eea
Thus, the full solution is 
\beq 
y_5(x) = \frac{\sqrt{1+x}}{x^3}\left( \frac{y_{3,i}}{5} \left.\left\{\frac{16+8x-2x^2+x^3}{\sqrt{1+x}} \right\}\right|_{x,i}^x + \frac{4 y_{1,i}}{3} \left.\left\{\frac{-8-4x+x^2}{\sqrt{1+x}} 
 \right\}\right|_{x,i}^x\right) +  y_{5,i}\left( \frac{x_i}{x}\right)^3 \sqrt{\frac{1+x}{1+x_i}}
\label{y5supersoln1}
\eeq
This solution is true for any initial conditions $y_{5,i}$, $y_{3,i}$ and $y_{1,i}$ (can,  in principle, be specified independently).  \\
{\it Approximations}\\
When $x_i\ll1$, the last term can be ignored and the terms in the curly brackets evaluated at $x_i$ reduce to just constants. Thus, the solution is 
\beq
y_5(x) \approx \frac{\sqrt{1+x}}{x^3} \left(\frac{-16 y_{3,i}}{5} + \frac{32 y_{1,i}}{3}\right) + 
\frac{y_{3,i}}{5} \left(\frac{16+8x-2x^2+x^3}{x^3}\right) + \frac{4 y_{1,i}}{3} \left(\frac{-8-4x+x^2}{x^3}\right)
\label{y5supersoln2}
\eeq
Further, when $y_{3,i} = 9/2 y_{5,i} $ and $y_{1,i}= 3/2 y_{5,i}$ as given by the adiabatic initial conditions (see \eqnref{newinit}), the solution becomes  
\beq
y_5(x) \approx \frac{y_{5,i}}{10} \left(\frac{16\sqrt{1+x}}{x^3} -\frac{16}{x^3} - \frac{8}{x^2}  + \frac{2}{x} + 9 \right). 
\label{y5supersoln3}
\eeq
This agrees with the solution obtained by  \cite{kodama_cosmological_1984}.
Note that \eqnrefs{y5supersoln1}, \eqnref{y5supersoln2} and \eqnrefbare{y5supersoln3} give the right limit at the initial time: when $x\ll1$, $y_5(x\ll1) \approx y_{5,i}$. 

\subsubsection{ Region III: \emph{super-horizon modes in the matter-dark energy era}}
In this limit, $\Omega_r \approx 0$ and \eqnref{y5eqsuper1} becomes 
\beq 
{\dot y}_5 = \left[\Omega_m y_3  -(3 \Omega_m + 2) y_5\right]. 
\label{y5mde1}
\eeq
Change the dependent variable from $a$ to $\Omega_m$ using \eqnref{omder}. This gives 
\beq 
{\dot y}_5 = \frac{d y_5}{d \Omega_m} \cdot 3 \Omega_m (\Omega_m-1), 
\eeq
and \eqnref{y5mde1} becomes 
\beq 
\frac{d y_5}{d \Omega_m}  + \frac{y_{3,i}}{6 (1-\Omega_m)} -\frac{(3\Omega_m + 2)}{\Omega_m(1-\Omega_m)} y_5 =0.
\eeq
The solution to this equation is in terms of hypergeometric functions. 
\beq
y_5(\Omega_m) = \frac{C \Omega_m^{1/3}}{(1-\Omega_m)^{5/6}} - \frac{y_{3,i}}{4}\frac{\Omega_m}{(1-\Omega_m)^{5/6}} \; \; {}_2F_1\left(\frac{2}{3}, \frac{1}{6},\frac{5}{3}, \Omega_m\right),  
\eeq
where the hypergeometric function (or series) is given by (\citealt{arfken}, 4$^{th}$ edition)
\beq
{}_2F_1(a, b, c, x) = \sum_{n=0}^\infty \frac{(a)_n (b)_n}{(c)_n} \frac{x^n}{n!}, 
\eeq
\beq
{\rm with } \; \; (a)_n = a(a+1)(a+2) \ldots (a+n-1) = \frac{(a+n-1)!}{(a-1)!}\; \; {\rm and} \;  (a)_0=1. 
\eeq
It is possible to recast the solution in terms of the variable $x$ defined as 
\beq 
x = \frac{a}{a_{eq, \Lambda-m}}, 
\eeq
where 
\beq 
a_{eq, \Lambda-m} = \left(\frac{\Omega_{m,0}}{\Omega_{\Lambda,0}}\right)^{1/3}. 
\eeq
This gives $\Omega_m = (1+ x^3)^{-1}$ and the solution becomes 
\beq 
y_5(x) = C \frac{\sqrt{1+x^3}}{x^{5/2}} -\frac{y_{3,i}}{4} \frac{1}{x^{5/2}(1+x^3)^{1/6}} \; \; {}_2F_1\left(\frac{2}{3}, \frac{1}{6},\frac{5}{3}, \frac{1}{1+x^3}\right). 
\eeq
The first term in the above expression is the homogenous solution and the second is the particular solution. Here $C$ is set by the initial conditions.
$C  = y_{5,i} \frac{x_i^{5/2}}{\sqrt{1+x_i^3}} + \frac{y_{3,i}}{4} \frac{1}{(1+x_i^3)^{2/3}} \; \; {}_2F_1\left(\frac{2}{3}, \frac{1}{6},\frac{5}{3}, \frac{1}{1+x_i^3}\right)$. When $x_i\ll1$, $C \rightarrow \frac{y_{3,i}}{4} \; \; {}_2F_1\left(\frac{2}{3}, \frac{1}{6},\frac{5}{3}, 1\right)$. The solutions above were obtained by ignoring the $\epsilon$ terms in the equations. It is possible to obtained refined approximations for $y_1$ to $y_4$ by substituting the above solutions in \cref{eq1,eq2,eq3,eq4} and integrating the resulting equations.  
This gives 
\begin{subequations}
\begin{align}
y_1(k,a) &= y_{1,i} - y_{2,i} \int_{a_i}^a \frac{\epsilon(k,a)}{3} d\ln a\\
 y_2(k,a) &= y_{2,i} +  y_{1,i}  \int_{a_i}^a  \epsilon(k,a) d\ln a - 2 \int_{a_i}^a y_5(a)  \epsilon(k,a) d\ln a \\
y_3(k,a) &= y_{3,i} -  y_{4,i} a_i \int_{a_i}^a  \frac{\epsilon(k,a)}{a} d\ln a, 
\end{align}
\end{subequations}
where $y_5(a) = y_5(x=a/a_{eq,m-r})$ given above. Deep in the radiation dominated era, $\epsilon \sim a$ and $y_5(a) \sim y_{5,i}$ and the solution is 
$y_2(k,a) \sim y_{2,i} +  (y_{1,i} - 2 y_{5,i}) (\epsilon(k,a)-\epsilon(k,a_i))$. In the other regimes, the integral is computed numerically. Consider \eqnref{eq4} for $y_4$. This can be re-written as 
\beq
\frac{d (a y_4)}{da} = -\epsilon(k,a) y_5(a). 
\eeq
Integrating once gives, 
\beq 
y_4(k,a) = \frac{y_{4,i} a_i}{a} -\frac{1}{a} \int_{a_i}^a \epsilon(k,a) y_5(a) da.
\eeq
In the radiation dominated era, the integral can be computed easily giving $y_4(k,a) \sim  \frac{y_{4,i} a_i}{a} -y_{5,i}\frac{\epsilon}{2 a^2}(a^2-a_i^2)$; for other regions we compute it numerically. 


\subsection{Horizon crossing ($\epsilon \sim 1$) and sub-horizon modes ($\epsilon >1$). }
To track the evolution of modes through horizon crossing and soon after, we need to solve the system \crefrange{eq1}{eq5} for non-zero $\epsilon$. 
The radiation and matter variables are coupled through $y_5$. In the radiation domination era, the coupled $y_1,y_2$ and $y_5$ system is solved first and the resulting $y_5$ sources the matter variables $y_3$ and $y_4$. In the radiation-matter and matter-dark energy eras the coupled $y_3,y_4$ and $y_5$ system is solved first and the resulting $y_5$ sources the radiation variables $y_1$ and $y_2$.

\subsubsection{Region IV: \emph{sub-horizon modes in the radiation domination era} }
The three equations that govern $y_1, y_2$ and $y_3$ are \eqnrefs{eq1}, \eqnrefbare{eq2} and \eqnref{eq5}. In addition, the algebraic equation for $y_5$ in the radiation domination regime becomes 
\beq
y_5 = \frac{4}{B_{rad}}\left(y_1+ \frac{1}{\epsilon} y_2\right), 
\label{y5algebraicrad}
\eeq
where $B_{rad} = 4 + 2 \epsilon^2/3$ and we have additionally assumed $\Omega_r \approx 1$ and $\Omega_m \approx 0$. As explained in the text, using the algebraic form for $y_5$ as a solution is justified for initial conditions set by inflation. Thus, there remain two degrees of freedom and we derive a second order differential equation for $y_5$ as follows. 

In the radiation domination limit, \eqnref{eq5} becomes. 
\beq 
{\dot y}_5 = 2 y_1 - \frac{B_{rad}+2}{2} y_5,    
\label{y5dotrad}
\eeq
Differentiate w.r.t. $\ln a$ and substitute \eqnref{eq1} to get 
\beq
{\ddot y}_5 = -\frac{2 \epsilon}{3} y_2 -\left(\frac{B_{rad}+2}{2}\right) {\dot y_5} - \frac{{\dot B}_{rad}}{2} y_5.
\eeq
Eliminate $y_1$ from \eqnrefs{y5algebraicrad} and \eqnrefbare{y5dotrad} to express $y_2$ as  
\beq 
y_2 = -\frac{\epsilon}{2}(y_5 + {\dot y}_5). 
\label{y2rad}
\eeq
Substitute for $y_2$ in \eqnref{y5dotrad} to get 
\beq
{\ddot y}_5 = {\dot y}_5 \left(\frac{\epsilon^2}{3} - \frac{B_{rad}+2}{2}\right) + y_5 \left(\frac{\epsilon^2}{3} - \frac{{\dot B_{rad}}}{2}\right). 
\eeq
Now in the radiation era, 
\beq
{\dot B}_{rad} \approx 4 \epsilon {\dot \epsilon}/3\;\;\;{\rm and\; from \; \eqnref{epsder}} \;\;\; {\dot \epsilon} = \epsilon \implies {\dot B}_{rad} = 4 \epsilon^2/3
\eeq
Substituting for $B_{rad}$ and ${\dot B}_{rad}$ above gives 
\beq 
{\ddot y}_5 + 3 {\dot y}_5 + \frac{\epsilon^2}{3}y_5 =0. 
\eeq
Since $\epsilon \sim a$ in the radiation era, 
\beq
\frac{d}{d\ln a} = \epsilon \frac{d}{d \epsilon} \;\;\;\; {\rm and} \;\;\; \frac{d^2}{d\ln a^2} = \epsilon \frac{d}{d\epsilon} + \epsilon^2 \frac{d^2}{d\epsilon^2}
\eeq
Thus in terms of $\epsilon$, the equation for $y_5$ in the radiation era is
\beq 
\epsilon^2 \frac{d^2 y_5}{d\epsilon^2} + 4  \epsilon \frac{d y_5}{d\epsilon} + \frac{\epsilon^2}{3}y_5=0.
\label{y5ineps}
\eeq
Define two new variables: the independent variable $x = \epsilon/\sqrt{3}$ and the dependent variable $u = y_5 x$. Substituting in \eqnref{y5ineps} gives the equation for $u$ and denoting derivatives w.r.t. $x$ with primes we get
\beq 
x^2 u'' + 2 x u' + (x^2-2) u =0, 
\eeq
whose solution is 
\beq 
u(x)  = a_1 J_1(x) + a_2 N_1(x), 
\eeq
where $J_1(x)$ and $N_1(x)$ are the Spherical Bessel functions of the first and second kind (the latter also called the spherical Neumann function). Thus, in the radiation limit when $\epsilon \gtrsim 1$, the solution for $y_5$ is 
\beq 
y_5(x) = a_1 \left(\frac{\sin x - x \cos x}{x^3}\right) -a_2\left(\frac{x \sin x +  \cos x}{x^3}\right).
\label{y5subhor}
\eeq
To compute $y_2$, use \eqnref{y2rad} and convert the dots (derivative w.r.t. $\ln a$) to primes (derivative w.r.t. $x$) to give 
\beq 
y_2 = -\frac{\sqrt{3}x}{2}\left(y_5(x) + x y_5'(x)\right). 
\eeq
Substituting from \eqnref{y5subhor} gives 
\beq 
\label{y2subhor}y_2(x) = -\frac{\sqrt{3}}{2 x^2} \left[ \left\{ 2  a_2 x + a_1 \left(x^2-2\right)\right\} \sin x + \left\{2 a_1 x-a_2
   \left(x^2-2\right)\right\} \cos x\right].
\eeq
Using \eqnref{eq2} and converting to $x$ variables gives 
\beq 
y_1 = \frac{1}{\sqrt{3}}y_2'(a) + 2 y_5(x). 
\eeq
Substituting for $y_2$ from \eqnref{y2subhor} gives 
\beq
\label{y1subhor} y_1(x) = - \frac{1}{2 x} \left[ (a_2 x -2 a_1 ) \sin x + (a_1 x + 2 a_2 ) \cos x \right].
\eeq

Once $y_5$ is known, the matter fields can be determined. Differentiating \eqnref{eq3} and substituting from \eqnref{eq4} gives the equation 
\beq 
{\ddot y}_3 + \left(1-\frac{{\dot \epsilon}}{\epsilon}\right) {\dot y}_3 - \epsilon^2 y_5 =0.
\label{y3ddot}
\eeq
In the radiation dominated limit, ${\dot \epsilon}/{\epsilon} \approx 1$ and the equation becomes 
\beq 
{\ddot y}_3 = \epsilon^2 y_5.  
\eeq
Integrating once gives 
\beq
{\dot y_3} = \int_{a_i}^a \epsilon^2 y_5 d \ln a +c. 
\eeq
Converting to the $x$ variable ($\epsilon = \sqrt{3} x$ and $ d\ln a = d \ln x$ in the radiation dominated era) gives
\beq
{\dot y_3}(x) = \frac{d y_3}{d \ln x} = \int_{x_i}^x 3 x y_5(x) dx + c = \frac{3}{x x_i} \left[ a_1(x \sin x_i - x_i \sin x) + a_2 (x_i \cos x - x \cos x_i)\right] + c, 
\label{y3dotrad} 
\eeq
where $c$ is the integration constant $c = {\dot y}_{3,i} = - \sqrt{3} x_i y_{4,i}$. This is

Using \eqnref{eq3} $y_4$ can be written in the $x$ variable as
\beq 
y_4(x)= -\frac{1}{\sqrt{3} x} {\dot y}_3. 
\eeq
Substituting from \eqnref{y3dotrad} gives 
\beq 
y_4(x) = - \frac{\sqrt{3}}{x^2 x_i} \left[ a_1(x \sin x_i - x_i \sin x) + a_2 (x_i \cos x - x \cos x_i)\right]  + \frac{y_{4,i} x_i}{x}.
\eeq
Integrating \eqnref{y3dotrad} once more gives $y_3 =\int_{a_i}^a  {\dot y}_3 d \ln a = \int_{x_i}^x  \frac{d y_3}{d \ln x} d\ln x$. Using \eqnref{y3dotrad} gives 
\beq y_3(x) = 3 a_1\left.\left(\frac{\sin x}{x} + \ln x \frac{\sin x_i }{x_i}  - \text{Ci}(x)\right)\right|_{x_i}^x
- 3 a_2 \left.\left( \frac{\cos x}{x} + \ln x \frac{\cos x_i }{x_i}  +  \text{Si}(x) \right)\right|_{x_i}^x -\sqrt{3} y_{4,i} x_i (\ln x - \ln x_i) + y_{3,i}, 
\eeq
where 
\beq
\text{Ci}(x) =-\int_x^\infty \frac{\cos x'}{x'} dx' \; \; \; {\rm and } \;\;\;  \text{Si}(x) = \int_0^x \frac{\sin x'}{x'}dx'. 
\eeq
$a_1$ and $a_2$ have to be set from initial conditions. There are two ways to set these constants. One choice is use the initial conditions $y_{1,i}$ and $y_{2,i}$ and solve $a_1$ and $a_2$. The other choice is to demand that $y_5$ is a constant at very early epochs, when $\epsilon \ll1$. In this limit, $a_2 = 0$ and $a_1 = 3 y_{5,i}$. Note that for $\epsilon\ll1$, this choice gives $y_{1,i} \approx 3/2 y_{5,i}$ and $y_{2,i} \approx -\epsilon_i/2 y_{5,i}$. 

Solutions of the E-B ystem in this regime have been derived analytically by \cite{hu_small-scale_1996}. They obtain a logarithmic growth for the matter overdensity $\delta$. This logarithmic growth is also reflected in our solution for $y_3$ (which is equal to $\delta + 3 \Phi$). 
The solutions derived above are valid for all $\epsilon$ in the radiation dominated era. No approximations made in the derivation invoke the magnitude of $\epsilon$. Thus, the solutions are also valid when $\epsilon\ll1$ and are expected to reduce to the super-horizon solutions for $x\ll1$.

\subsubsection{Region V: \emph{sub-horizon modes in the radiation-matter era}}
Here we first solve for the coupled system $y_3, y_4$ and $y_5$ and use the resulting solution to solve for the radiation variables. The three main equations are \eqnrefs{eq3}, \eqnrefbare{eq4} and \eqnrefbare{eq5}. These three are combined with the algebraic equation \eqnref{eq5II} to give a second order system. There are two main approximations made in this regime: (1) Although, $\Omega_r$ is not set to zero, perturbations in the radiation sector are ignored, and (2) $\epsilon$ is large enough such that $2 \epsilon^2/3 \gg  3 \Omega_m + 4 \Omega_r$.  

The starting point is \eqnref{y3ddot} which is valid at all epochs and scales (because it was derived directly from \eqnrefs{eq3} and \eqnrefbare{eq4}). We wish to express $y_5$ in terms of $y_3$ and its derivatives so that one can obtain a second order differential variable in $y_3$. This is done by combining \eqnrefs{eq5II}, where the radiation perturbations are ignored with \eqnref{eq3}. Ignoring the radiation perturbations, \eqnref{eq5II} becomes \footnote{Using this form for $y_5$ corresponds to ignoring the homogenous term. This is justified because it decays exponentially in this regime.} 
\beq 
y_5 = \frac{1}{B} \left[\Omega_m\left(y_3 + \frac{3 y_4}{\epsilon}\right)\right]. 
\label{y5mat}
\eeq 
Eliminate $y_4$ from the above by using \eqnref{eq3} to give $y_5$ in terms of $y_3$ and ${\dot y}_3$. Substitute the resulting $y_5$ in \eqnref{y3ddot} to give a second order equation for $y_3$:
\beq 
{\ddot y}_3 + \left(1-\frac{\dot \epsilon}{\epsilon} + \frac{3\Omega_m}{B} \right) {\dot y}_3 -\frac{\Omega_m}{B} y_3 \epsilon^2=0.
\label{y3mateq}
\eeq
Now, in the matter dominated eras, for most typical scales of interest, $\epsilon \gtrsim 1$ and hence the $\epsilon^2$ term dominates in the expression for $B$. Note that {\it this approximation is not valid in the radiation dominated era}. Setting $B \sim 2 \epsilon^2/3$, gives 
\bea
&&{\ddot y}_3 + \left(1-\frac{\dot \epsilon}{\epsilon} \right) {\dot y}_3 -\frac{3 \Omega_m}{2} y_3 =0,\\
\implies   &&{\ddot y}_3 +\left( 2 - \frac{3\Omega_m}{2} - 2\Omega_r \right) {\dot y}_3 - \frac{3}{2} \Omega_m y_3=0, 
\eea
where in the second equation is obtained by substituting for for ${\dot \epsilon}$ from \eqnref{epsder}. There is no $k$-dependence of $y_3$ in this regime and the only parameters in the equation are $\Omega_m$ and $\Omega_r$. This allows us to introduce the variable 
\beq 
x = \frac{a}{a_{eq,m-r}}, 
\eeq
as defined in \eqnref{xaeq1}. Using \eqnref{omegasinxaeq1} to substitute for the $\Omega$s, we get 
\beq 
y''_3 + \frac{3 x + 2}{2 x(1+x)} y'_3 - \frac{3}{2} \frac{y_3}{x (1+x)} = 0,  
\eeq
where the primes denote derivatives w.r.t. $x$. In the matter dominated era, when $x$ is large but before dark energy domination takes over, one mode grows as the scale factor i.e., linear in $x$. Thus, one solution has to be a polynomial of degree one in $x$ with $y''(x) =0$. This solution satisfies 
\beq 
\frac{y'_3}{y_3}  = \frac{3}{3 x+2}, 
\eeq 
whose solution is 
\beq 
y^{(1st)}_3 = 3 x + 2. 
\eeq
The superscript denotes the `first' solution. For a homogenous equation like the one above, the second solution can be constructed if one is known (\cite{arfken}, page 501). The second solution is 
\bea 
y^{(2nd)}_3&=& (3 x+ 2) \int \frac{\exp\left \{-\int  \frac{3 x' + 2}{2 x'(1+x')} dx'\right\}}{(3x + 2)^2} dx \\
&=& (3 x+ 2) \int \frac{dx}{x \sqrt{1+x} (3x+2)^2} \\
&=&-\frac{3}{4} \left[\left(x + \frac{2}{3} \right) \log \left(\frac{\sqrt{1+x}+1}{\sqrt{1+x}-1}\right) - 2 \sqrt{1+x}\right], 
\eea
where the integral was performed with the transformation $u= \sqrt{1+x}$.
Thus, the general solution is 
\beq 
y_3 = c_3 \left(x + \frac{2}{3} \right) + c_4 \left[\left(x + \frac{2}{3} \right) \log \left(\frac{\sqrt{1+x}+1}{\sqrt{1+x}-1}\right) - 2 \sqrt{1+x}\right], 
\eeq
and using $y_4 = -\epsilon^{-1} x y'_3$ gives
\beq 
y_4 = -\frac{1}{\epsilon(k, a_{eq} x)} \left\{ c_3 x + c_4 \left[x \log \left(\frac{\sqrt{1+x}+1}{\sqrt{1+x}-1}\right)  - \frac{2(1+3x)}{3\sqrt{1+x}} \right]  \right\}.
\eeq
The constants $c_1$ and $c_2$ can be set either by the initial conditions of $y_3$ and $y_4$ or by matching to the logarithmic solution for small scale modes in the radiation era. 

At first glance, it may seem inconsistent to keep the $\Omega_r$ in writing ${\dot \epsilon}$, but ignore the $\Omega_r$ term in \eqnref{y5mat}. 
Suppose this term is kept and the full expression for $y_5$ is substituted in \eqnref{y3ddot}. The term $\epsilon^2 y_5$ appearing in that equation then has a  contribution $\Omega_r y_1$ (assuming $B\sim \epsilon^2$). But note that, in the radiation dominated era, the solution for $y_1$ decays as $1/\epsilon$ (see \eqnref{y1subhor}). Thus, the contribution of the $\Omega_r y_1$ term is of $\mathcal{O}(\epsilon^{-1})$, whereas in ${\dot \epsilon}/\epsilon$ the $\Omega_r$ term is of $\mathcal{O}(1)$. Since, we have assumed $\epsilon \gg  1$ ignoring the former is justified. This point has also been discussed in a detailed footnote in \cite{weinberg_cosmological_2002}. 

Having computed $y_3$ and $y_4$, $y_5$ follows from \eqnref{y5mat}. $y_5$ is a source term for the coupled radiation system. To compute these solutions we will use the adiabatic approximation. First note that the radiation sector, for all values of $\epsilon$, can re-written in terms of two other variables 
\bea
\dot w &=& \frac{i \epsilon}{\sqrt{3}} w - \frac{2 i \epsilon}{\sqrt{3}} y_5 \label{rad1} \\
\label{rad2}
 \dot v &= & -\frac{i \epsilon}{\sqrt{3}} v + \frac{2 i \epsilon}{\sqrt{3}} y_5,
 \eea
where 
\bea 
w &=& y_1+ \frac{i}{\sqrt{3}}y_2 \\
v&=& y_1 - \frac{ i}{\sqrt{3}} y_2. 
\eea
The homogenous solutions for $w$ and $v$ and hence also for $y_1$ and $y_2$ are combinations of $\cos I(k,a)$ and $\sin I(k,a)$ where 
\beq 
I(k,a) = \int_{a_i}^{a} \frac{\epsilon(k,a')}{\sqrt{3}} d\ln a'.
\label{Ifunc}
\eeq 
The $y_5$ term decides the particular solution for this system and \eqnref{y5mat} gives $y_5$ in terms of $y_3$ and $y_4$. We invoke the adiabatic approximation to assume that in the radiation-matter regime when $\epsilon \gg 1$, the parameters $\Omega_m$ and $\Omega_r$ are approximately constants (see \S \ref{sec:adiabatic}). For the particular solution, we assume the ansatz: 
\bea
y_{p,1}(a) &=& a_3 y_3(a) + a_4 y_4(a),  \\
y_{p,2}(a) &=& b_3 y_3(a) + b_4 y_4(a),  
\eea
where $a_3, a_4$ are constants. 
Substitute this ansatz in \eqnrefs{eq1} and \eqnrefbare{eq2} and use \eqnrefs{eq3}, \eqnrefbare{eq4} and \eqnref{y5mat} to solve for $a_3,b_3,a_4$ and $b_4$. This gives 
\beq
a_3 = \frac{2 \Omega_m}{B+ 3 \Omega_m}, \;\;a_4=0, \; \; b_3=0, \;\;b_4 = \frac{6 \Omega_m}{B + 3 \Omega_m}. 
\label{eqa2b4}
\eeq
Thus, 
\bea
y_1(a) &=& c_1 \cos I(k,a) + c_2 \sin I(k,a) + a_3 y_3(a)\\ 
y_2(a) &=& \sqrt{3}\left(c_1 \sin I(k,a) -c_2\cos I(k,a)\right) + b_4 y_4(k,a)
\eea
Putting the initial conditions that $y_1(a_i) = y_{1,i}$ and ${\dot y_1}(a_i) = 3 y_{2,i}/\epsilon_i $ gives 
\beq 
c_1 = y_{1,i} - a_{3} y_{3,i} \; \; \; c_2 =-\frac{1}{\sqrt{3}}(y_{2,i} - b_4 y_{4,i})
\eeq
Note that the adiabatic approximation was not invoked in solving for the matter variables. Thus, having computed $y_1$, $y_2$, $y_3$ and $y_4$ is it possible to get a refined estimate of $y_5$ using \eqnref{eq5II}. This is more accurate than \eqnref{y5mat}. 
Some of the above solutions have been worked out in the past by \cite{meszaros_behaviour_1974,hu_small-scale_1996,weinberg_cosmological_2002}. 

\subsubsection{Region VI: \emph{sub-horizon modes in the matter-dark energy era}}
In this regime, $\Omega_r \approx 0$, but there is a possible dark energy component. Here we consider it to be the cosmological constant. In this case ${\dot \epsilon}  = -\epsilon(1 - 3\Omega_m/2)$ and 
\eqnref{y3mateq} becomes 
\beq 
{\ddot y}_3 +\left( 2 -\frac{3}{2} \Omega_m \right) {\dot y}_3 - \frac{3}{2} \Omega_m y_3=0. 
\eeq
Using \eqnrefs{Hderiv} and \eqnrefbare{omder}, 
 \bea
{\dot H} &=& -\frac{3 \Omega_m H}{2},\\ 
{\rm and}\; \;  {\ddot H} &=& \frac{9}{2} H  \Omega_m( 1-\frac{\Omega_m}{2} )
\eea
and it is easy to see that $y_3\sim H$ is one solution to the above equation. Thus, $y^{(1st)} \sim H$. 
Knowing the first solution, the second solution can be computed as 
\bea 
y_3^{(2nd)} &=& H \int \frac{\Omega_m^{1/2}}{a^{3/2}} H^2  da. \\
& \propto & H \int \frac{da}{(aH)^3}. 
\eea
Thus the full solution for $y_3$ is 
\beq 
y_3(a) = c_3 H + c_4 H \int \frac{da}{(aH)^3}. 
\label{y3mat}
\eeq
From \eqnref{y3mat}, $y_4(a)$ becomes 
\beq
y_4(a) = -\frac{1}{\epsilon} \left[ c_3  \frac{d H}{d\ln a}  + c_4 \left( \frac{1}{(aH)^2} + \frac{d H}{d \ln a} \int \frac{da}{(aH)^3} \right)\right].
\eeq
$c_3$ and $c_4$ are set by the initial conditions $y_{3,i}$ and $y_{4i}$. \\
{\it Recovering the usual linear growth equation}: 
When $\Omega_r \approx 0$ and $\epsilon \gg 1$, \eqnref{y5mat} becomes 
\beq 
y_5 = \frac{\Omega_m}{B} y_3,
\eeq
where $B = 3 \Omega_m + \frac{2}{3} \epsilon^2$. Substituting the definition $y_3 = \delta + 3 y_5$ gives 
\beq
y_5 \approx \frac{3 \Omega_m \delta}{2} \; \; {\rm or}\; \;  y_3 \approx \delta. 
\eeq
Substituting $y_5$ in terms of $y_3$ in \eqnref{eq5} and keeping terms to lowest order in $\epsilon$ gives ${\dot y}_5 \approx 0$ or $y_5$ is a constant. With these limits, \eqnrefs{eq3} and \eqnrefbare{eq4} become 
\bea
{\dot \delta} &=& -\epsilon y_4\\
{\dot y}_4 &=& -y_4 -\frac{3}{2} \frac{\delta \Omega_m} \epsilon.  
\eea
When these are combined we get 
\beq 
{\ddot \delta} + \left(2 - \frac{3}{2} \Omega_m\right) {\dot \delta} - \frac{3 \Omega_m}{2} \delta = 0. 
\eeq 
which is the usual linear growth equation. Note that the derivatives are w.r.t. $\ln a$ and $\Omega_m$ is the time-dependent matter density parameter (see \eqnref{eq:omor} in text). 

\subsection{Very small scales: $k \gtrsim 50$.  }
These modes have a very high value of $\epsilon$ even at very early times. We show that in this case, the radiation sector and matter sector can be decoupled. Each sector is solved separately and then combined to give the potential $y_5$. 

We note that the solution for $y_5$ can be written as 
\beq  
y_5(k,a) = C y_{5, hom}(k,a)  + y_{5, part}(k,a),  
\label{y5soln}
\eeq
where 
\bea
\nonumber y_{5, hom} (k,a) &=& \exp \left\{-\int_{a_i}^a  \frac{B+2}{2} d\ln a \right\} \\
\nonumber y_{5, part}(k,a) &=& \frac{1}{B} \left[ 4 \Omega_r \left(y_1 + \frac{1}{\epsilon} y_2\right) + \Omega_m \left(y_3 + \frac{3}{\epsilon} y_4 \right)\right] \\
\nonumber C & = & y_{5,i} - y_{5,part}(a_i). 
\eea
When $\epsilon$ is very large, the homogenous term in $y_5$ is exponentially suppressed and the 
particular solution is suppressed by $1/\epsilon^2$. Thus, the source terms in \eqnrefs{rad1} and \eqnrefbare{rad2} are suppressed by $1/\epsilon$. Thus only the homogenous solutions remain which are 
\bea
w(k,a) &\approx & w_i e^{i I(k,a)}\\
v(k,a) &\approx & v_i e^{-i I(k,a)}, 
\eea
where $I(k,a)$ is given by \eqnref{Ifunc}. Converting back to the $y_1, y_2$ variables gives 
\bea
\label{y1subhor1}y_{1}(k,a)  &=& y_{1,i}  \cos I(k,a) -  \frac{y_{2,i}}{\sqrt{3}} \sin I(k,a), \\
\label{y2subhor1}y_2(k,a) &=& y_{2,i}   \cos I(k,a) + \sqrt{3}y_{1,i} \sin I(k,a). 
\eea
In the matter sector, in \eqnref{eq4}, $y_5$ acts as a source term for $y_4$, which is also suppressed by $1/\epsilon$. Thus, only the homogenous term remains 
\beq
y_{4}(a) = y_{4,i} \frac{a_i}{a}.
\label{y4subhor1}
\eeq
Integrating \eqnref{eq3}, gives 
\beq 
y_3(k,a) = y_{3,i} - \int_{a_i}^a \epsilon(k,a) y_4(k,a)  d\ln a 
\eeq
In the radiation era, $\epsilon \sim a$. Substituting for $y_4$ from \eqnref{y4subhor1} and 
\beq 
y_{3}(k,a) = y_{4,i} \frac{ a_i}{a} \epsilon(k,a) \log \frac{a}{a_i} + y_{3,i}.
\label{y3subhor1}
\eeq
 Substituting for $y_1$ to $y_4$ in \eqnref{y5soln} gives, an estimate of $y_{5}$. 

These arguments hold true only in the radiation dominated era since the radiation fields evolve as $1/\epsilon$ and the matter fields grow at best as $\ln a$. $\epsilon \sim a$ in this regime and the gravitational potential $y_5$ decays as $\ln a/a^2$ as can be seen from \eqnref{y5soln}. However, in the matter dominated era, $\epsilon \sim \sqrt{a}$ and $y_3 \sim a$. So the gravitational potential $y_5$ does not decay, but instead goes to a constant $3\Omega_m y_3/(2 \epsilon^2)$ and the source terms involving $y_5$ in  \eqnrefs{rad1} and \eqnrefbare{rad2} actually grow as $\sqrt{a}$ and cannot be ignored.

\section{Including baryons and neutrinos}
\label{app:withbaryons}
Here we recast the full system (including baryons and massless neutrinos) in terms of the variables defined in \S \ref{sec:setup}.  
The equations governing the perturbations in the cold dark matter, massless neutrinos, photons, baryons 
are given by equation 43, 50, 64 and 67 in \citet{ma_cosmological_1995}.
We introduce the time variable `$\ln a$'; note that the derivative w.r.t. the conformal time $d/d\eta = aH d/d \ln a$. We also define three new parameters:  
\bea
v &=& \frac{\theta}{k},\\ 
 \epsilon_k &=& \frac{k}{aH},\\
\epsilon_c &=& \frac{a n_e \sigma_T}{aH},\\
\epsilon_H &=& \frac{(aH)^{-1}}{\eta}.
\eea
 Here $\epsilon_k$ is the same as the $\epsilon$ used elsewhere in the text. We use the subscript to distinguish it from $\epsilon_c$ and $\epsilon_H$. In terms of these parameters, the system is 

\begin{subequations}
\begin{align}
\label{fullsystem}
{\rm Cold} \; \;  {\rm Dark} \; \; {\rm Matter}: \; \;  \; \;
 \frac{ d \delta_c}{d \ln a} &= -\epsilon_k v_c + 3 \frac{d \phi}{d \ln a}, \\
\frac{ d v_c}{ d \ln a} &= -v_c + \epsilon_k \psi \\ 
\nonumber \\
{\rm Massless} \;\; {\rm Neutrinos}: \;\;\;\;
\frac{ d \delta_\nu}{d \ln a} &= -\frac{4}{3} \epsilon_k v_\nu + 4 \frac{d \phi}{d \ln a}, \\
\frac{ d v_\nu}{ d \ln a} &= \epsilon_k \left(\frac{\delta_\nu}{4} - \sigma_\nu +   \psi\right), \\
\frac{ d F_{\nu,l}}{ d \ln a} &= \frac{\epsilon_k}{2 l + 1}  \left[ l F_{\nu, l-1} - (l+1) F_{\nu, l+1}\right],\\ \\
\frac{ d F_{\nu,l_{max}}}{ d \ln a} &= \epsilon_k F_{\nu,l_{max}-1} -\epsilon_H (l_{max}+1)F_{\nu, l_{max}}.  \\  
\nonumber\\
{\rm Photons}: \; \; \; \; 
\frac{ d \delta_\gamma}{d \ln a} &= -\frac{4}{3} \epsilon_k v_\gamma + 4 \frac{d \phi}{d \ln a}, \\
\frac{ d v_\gamma}{ d \ln a} &= \epsilon_k \left(\frac{\delta_\gamma}{4} - \sigma_\gamma +   \psi\right)  + \epsilon_c (v_b - v_\gamma), \\
\frac{ d F_{\gamma 2}}{ d \ln a}  &= \frac{8}{15} \epsilon_k v_\gamma - \frac{3}{5} \epsilon_k F_{\gamma,3} - \epsilon_c \left(\frac{9}{5}\sigma_\gamma -\frac{1}{10}(G_{\gamma 0} + G_{\gamma 2})\right), \\ \\
\frac{ d F_{\gamma l}}{ d \ln a} &= \frac{\epsilon_k}{2 l + 1}  \left[ l F_{\gamma l-1} - (l+1) F_{\gamma l+1}\right] -\epsilon_c F_{\gamma l} ,\\ \\
\frac{ d G_{\gamma l}}{ d \ln a} &= \frac{\epsilon_k}{2 l + 1}  \left[l G_{\gamma l-1} - (l+1) G_{\gamma l+1}\right] -\epsilon_c \left[G_{\gamma l} -\frac{1}{2} \left(F_{\gamma 2} + G_{\gamma 0} + G_{\gamma 2} \right)\left(\delta_{l0} + \frac{\delta_{l2}}{5}\right)\right] ,\\ \\
\frac{ d F_{\gamma l_{max}}}{ d \ln a} &=\epsilon_k F_{\gamma l_{max}-1} - \epsilon_H (l_{max}+1) F_{\gamma l_{max}}  - \epsilon_c F_{\gamma l_{max}},\\ \\
\frac{ d G_{\gamma l_{max}}}{ d \ln a} &= \epsilon_k G_{\gamma l_{max}-1} - \epsilon_H (l_{max}+1) G_{\gamma l_{max}}  - \epsilon_c G_{\gamma l_{max}}. \\ 
\nonumber\\
{\rm Baryons}: \;\;\;\;
\frac{d \delta_b}{d \ln a} &=-\epsilon_k v_b + 3 \frac{d \phi}{d \ln a}, \\
\frac{ d v_b}{ d \ln a} &=-v_b + \epsilon_k ( c_s^2 \delta_b + \psi) + \epsilon_c  \frac{4 \Omega_\gamma}{3 \Omega_b} (v_\gamma - v_b). 
\end{align}
\end{subequations}
In this notation, the four coupled Einstein equations in the conformal Newtonian gauge become 
\begin{subequations}
\begin{align}
\label{timetime}\frac{d \phi}{d \ln a} + \psi + \frac{\epsilon_k^2}{3} \phi &= -\frac{1}{2} \sum_i \Omega_i \delta_i ,\\
\label{spacetime}\frac{d \phi}{d\ln a} + \psi &=  \frac{3}{2 \epsilon_k} \sum_i \Omega_i (1+w_i) v_i,\\
\label{spsptrace}\frac{d^2 \phi}{d\ln a^2} + \frac{d \phi}{d \ln a} \left(3 + \frac{ d \ln H}{d \ln a} \right) + \frac{d \psi}{d \ln a} + \left(3 + 2 \frac{d \ln H}{d\ln a}\right) \psi + \frac{\epsilon_k^2}{3} (\phi-\psi) &= \frac{3}{2} \sum_i w_i \delta_i \Omega_i, \\
\label{spsptraceless}\phi-\psi &= \frac{9}{2 \epsilon_k^2} \sum_i (1+w_i) \Omega_i \sigma_i.
\end{align}
\end{subequations}
Here sum over `$i$' denotes sum over the various components. 
The tight coupling regime is when $\epsilon_c \gg 1$. When the photon moments oscillate rapidly, $\epsilon_k \gg 1$; this is also the regime when the photons are free-streaming and the radiation is sub-dominant on scales of interest. Both are numerically stiff regimes. Traditional codes invoke the tight coupling approximation to treat the former and the free-streaming approximation to treat the latter.

\section{The reduced 4D system}
\label{app:4Dsystem}
\begin{figure}
\centering
\includegraphics[width=14cm]{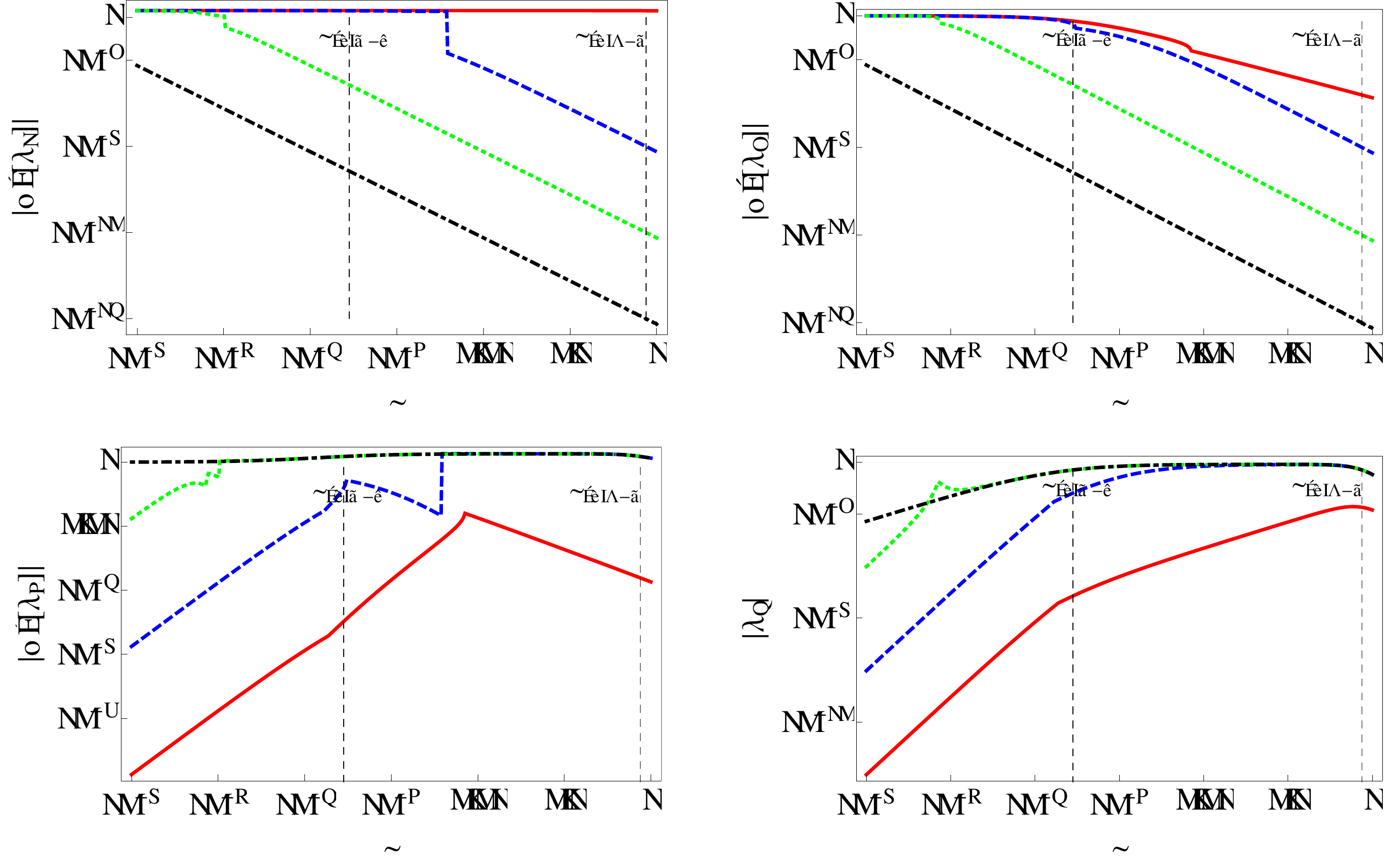}\\
\includegraphics[width=18cm]{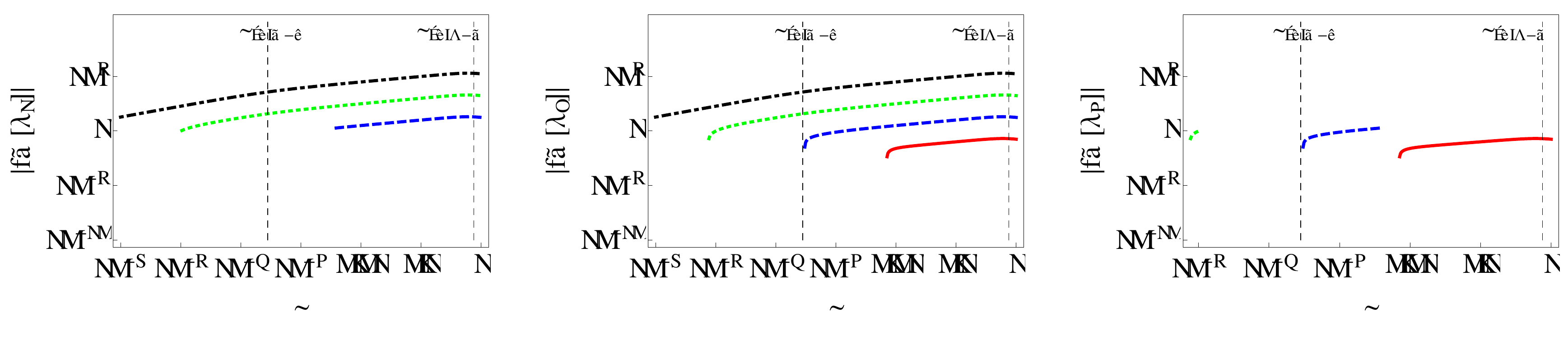}
\caption{Eigenvalues for the reduced four dimensional system. The colour coding is the same as that in \figref{eigenvalues}. The eigenvalue structure is different than that for the 5D system. In this case $\lambda_4$ is always real and hence no imaginary part. The imaginary part of $\lambda_3$ for the $k=100$ $h {\rm Mpc}^{-1}$ mode (black) is zero in the range of epochs shown. }
\label{4D}
\end{figure}

Substituting the algebraic equation for $y_5$, \eqnref{eq5II} in the \eqnrefs{eq1} to \eqnrefbare{eq4} gives a reduced four dimensional dynamical system for the variables $y_1$ to $y_4$. The time dependent eigenvalues for this system are shown in \figref{4D}. In stability analysis, the subset of eigenvalues with negative real part determines the step-size. Note that the absolute values of the real parts are of order unity or less, much smaller than the absolute value of the most negative real eigenvalue ($\lambda_1$) for the 5D system.
Thus, it is feasible that the 4D system may be numerically more stable. A more detailed analysis is required and whether this holds true when the full Botlzmann system is considered remains to be investigated. Both 4D and 5D systems show transitions to oscillations (sudden appearance of imaginary values), although the exact transition epoch may be slightly different in the two cases. 

\newpage
\bibliographystyle{mn2e.bst}
\bibliography{books,boltzcode}

\end{document}